\documentclass[acmsmall,natbib,screen,10pt,nonacm]{acmart}
\setcopyright{none}
\acmConference[arXiv]{arXiv.org e-Print archive}{\date}{e-Print}
\acmDOI{} 
\startPage{1}

\usepackage[utf8]{inputenc}
\usepackage{amsmath}
\usepackage{soul}
\usepackage{mathtools}
\usepackage{hyperref}
\usepackage{booktabs}
\usepackage{caption}
\usepackage{subcaption}
\usepackage{multicol}
\usepackage{xspace}
\usepackage{xfrac}
\usepackage{tikz}
\usetikzlibrary{matrix,backgrounds,arrows.meta}
\usepackage{cleveref}
\usepackage{sbnf}
\renewcommand\bnfcmt[1]{\textbf{(#1)}}
\usepackage{paper}
\usepackage{paralist}

\makeatletter
\let\@msm@th@eqref\eqref
\renewcommand{\eqref}[1]{%
  \begingroup
  \leavevmode
  \color{violet}%
  \hypersetup{linkbordercolor=[named]{violet}}%
  \@msm@th@eqref{#1}%
  \endgroup
}
\makeatother

\begin{document}

\title{A quantitative probabilistic relational Hoare logic}

\author{Martin Avanzini}
\orcid{0000-0002-6445-8833}
\affiliation{
  \institution{Centre Inria d’Université Côte d’Azur}
  \streetaddress{Route des Lucioles - BP 93}
  \city{Sophia Antipolis}
  \postcode{06902}
  \country{France}
}
\email{martin.avanzini@inria.fr}

\author{Gilles Barthe}
\orcid{}
\affiliation{
  \institution{MPI-SP}
  \streetaddress{Universit\"atstr 140}
  \city{Bochum}
  \postcode{44799}
  \country{Germany}
}
\email{gilles.barthe@mpi-sp.org}
\affiliation{
  \institution{IMDEA Software Institute}
  \streetaddress{Campus Montegancedo s/n}
  \city{Pozuelo de Alarcon}
  \postcode{28223}
  \country{Spain}
}
\email{gilles.barthe@mpi-sp.org}

\author{Davide Davoli}
\orcid{}
\affiliation{
  \institution{Centre Inria d’Université Côte d’Azur}
  \streetaddress{Route des Lucioles - BP 93}
  \city{Sophia Antipolis}
  \postcode{06902}
  \country{France}
}
\email{davide.davoli@inria.fr}

\author{Benjamin Grégoire}
\orcid{}
\affiliation{
  \institution{Centre Inria d’Université Côte d’Azur}
  \streetaddress{Route des Lucioles - BP 93}
  \city{Sophia Antipolis}
  \postcode{06902}
  \country{France}
}
\email{benjamin.gregoire@inria.fr}

\begin{abstract}
  We introduce \ERHL, a program logic for reasoning about relational
  expectation properties of pairs of probabilistic programs. \ERHL is
  quantitative, i.e., its pre- and post-conditions take values in the
  extended non-negative reals. Thanks to its quantitative assertions,
  \ERHL overcomes randomness alignment restrictions from prior logics,
  including \PRHL, a popular relational program logic used to reason
  about security of cryptographic constructions, and \APRHL, a variant
  of \PRHL for differential privacy. As a result, \ERHL is the first
  relational probabilistic program logic to be supported by
  non-trivial soundness and completeness results for all \emph{almost
  surely terminating} programs. We show that \ERHL\ is sound and
  complete with respect to program equivalence, statistical distance,
  and differential privacy. We also show that every \PRHL judgment is
  valid iff it is provable in \ERHL.  We showcase the practical
  benefits of \ERHL\ with examples that are beyond reach of \PRHL and
  \APRHL.
\end{abstract}



\maketitle

\repeatable{theorem}
\repeatable{lemma}
\repeatable{proposition}
\repeatable{corollary}
\section{Introduction}\label{s:intro}
\emph{Probabilistic Relational Hoare Logic} (\emph{\PRHL})~\cite{BartheGB09} is a
program logic to reason about relational properties of probabilistic
programs.
In \PRHL, judgments take the form
\[
  \RHT{}{\cond}{\cmd}{\cmdtwo}{\condtwo}\,,
\]
where $\cond,\condtwo : \Mem \times \Mem \to \{0,1\}$ are predicates on pairs of program memories $\Mem$, and
where $\cmd$ and $\cmdtwo$ are programs in a probabilistic, imperative language \PWHILELANG.
Informally, \PRHL\ judgments capture that execution of the two programs $\cmd$ and $\cmdtwo$ on
$\cond$-related inputs yield $\condtwo$-related output
distributions. The meaning of \PRHL\ judgments can be made formal
using a lifting from relations over states to relations over
distributions of states.  This lifting is defined formally using
probabilistic couplings, a fundamental tool from stochastic processes
theory and optimal
transport~\cite{Lindvall02,LevinPW09,Villani08}. Informally, the
notion of lifting quantifies existentially over all couplings and can
be used to combine two output distributions in such a way that their
combination satisfies a desirable property.

The coupling-based interpretation of \PRHL\ has one major advantage:
probabilistic reasoning is circumscribed to random assignments, and
all reasoning about other instructions is based on Boolean-valued
assertions. As a consequence, \PRHL\ proofs can leverage practical
verification techniques, including weakest verification calculus and
SMT-based reasoning. \PRHL\ and its variants are implemented in
several proof tools, including \tool{CertiCrypt}~\cite{BartheGB09},
\tool{EasyCrypt}~\cite{BartheGHB11}, the Foundational Cryptographic
Framework~\cite{DBLP:conf/post/PetcherM15},
\tool{CryptHOL}~\cite{DBLP:journals/joc/BasinLS20}, and
\tool{Iris}~\cite{DBLP:journals/pacmpl/TassarottiH19,GregersenAHTB24}.
These tools have been used extensively for mechanizing the security of
cryptographic constructions.

In spite of its success, \PRHL\ is incomplete, i.e.\, there exist
semantically valid \PRHL\ judgments that are not provable with the
\PRHL\ proof system. The main source of incompleteness is randomness
alignment: loosely speaking, the \PRHL\ proof system requires the
existence of a control-flow preserving bijection between random
assignments of the left and right program. As a consequence, it is
generally impossible to relate programs which perform a different
number of random assignments. For this reason, some tools like
\tool{EasyCrypt} combine \PRHL\ with other proof techniques, notably
eager/lazy sampling (which allows to move sampling around in programs)
and reification (which allows to replace a code snippet by a single
random assignment that samples a value from the semantics of the
snippet). Yet, the combination of \PRHL and these techniques remains
incomplete. Completeness results for \PRHL\ have remained elusive
since its introduction more than 15 years ago.

Furthermore, the incompleteness of \PRHL\ carries over to all its
extensions for reasoning about quantitative properties. Since
\PRHL\ cannot be used to reason about quantitative relational
properties, such as statistical distance, Kantorovich distance, or
differential privacy~\cite{DBLP:conf/tcc/DworkMNS06}, researchers have
developed extensions of \PRHL\ for specific properties. For instance,
\citet{BartheGHS17} introduce an extension of \PRHL, called \XPRHL,
that produces a so-called product program that computes for every pair
of inputs a coupling of the two output distributions. One can then use
expectation-based program logics on the product program to obtain an
upper bound on the statistical distance of the output distributions.
An alternative is to use \EPRHL~\cite{BartheEGHS18}, an extension of
\PRHL\ that targets relational expectation properties. Finally,
\citet{DBLP:conf/popl/BartheKOB12} use \APRHL, an extension of
\PRHL\ based on approximate couplings, to reason about differential
privacy. All these logics inherit the incompleteness from \PRHL---and
in some cases introduce further sources of incompleteness.

\begin{figure}
  $\begin{array}{rcl}
    w <* \{0,1\} \times \{0,1\} & \stackrel{?}{=} &
x <* \{0, 1\}; y <* \{0, 1\} ; w <- (x, y)\, \\
x<*A; \WHILE (x\in X) \DO x <* A &  \stackrel{?}{=} &
x <* A\setminus X
\end{array}$
\vspace{-2mm}
\caption{Both equivalences are not provable in \PRHL, due to
  constraints on randomness alignment. The first equivalence relates
  programs that do not perform the same number of samplings. The
  second equivalence relates rejection sampling in $A\setminus X$
  implemented as a loop with a uniform distribution over $A\setminus
  X$.}\label{fig:equiv}
\end{figure}

\paragraph*{Contributions}
This paper introduces \ERHL, a novel program logic for reasoning about
relational properties of probabilistic programs. The salient feature
of \ERHL\ is that it is a quantitative program logic, in the sense
that its assertions are mappings from pairs of states to (positive
extended) reals. Concretely, \ERHL manipulates judgments of the form
\[
  \RHT{Z}{\ex}{\cmd}{\cmdtwo}{\extwo}\,,
\]
where $Z$ is a type of \emph{auxiliary variables}, and where
$\ex,\extwo : Z \to \Mem \times \Mem \to \Rext$, the \emph{pre-} and \emph{post-expectation}, respectively.
The validity of \ERHL
judgments is based on probabilistic couplings. Informally, it states
that for all $z\in Z$ and memories $m_1$ and $m_2$, we have
$\ex~\mem_1~\mem_2 \geq \E{\mu_0}{\extwo z}$, where $\mu_0$ ranges over
all couplings of $\sem{\cmd}_{\mem_1}$ and $\sem{\cmdtwo}_{\mem_2}$
(we defer the exact definition of couplings to technical sections).

\ERHL\ is the first relational program logic for probabilistic programs
that is supported by non-trivial completeness theorems for
\emph{almost surely terminating} (\emph{\asterm}) programs.  Concretely, we show that
\ERHL\ is sound and complete for several important properties:
\begin{asparaitem}[-]
  \item \emph{Observational equivalence}. We show that \ERHL\ characterizes
  observational equivalence. This is in contrast with \PRHL,\ where
  strikingly simple examples of program equivalence cannot be
  proved due to randomness alignment constraints, see \Cref{fig:equiv}. 
  
  \item \emph{Statistical distance}. We show that \ERHL\ characterizes
  statistical distance between output distributions.

  \item \emph{Differential privacy}. We show that \ERHL\ captures both
  $\epsilon$- and $(\epsilon,\delta)$-differential privacy. This
  is in contrast with \APRHL, where simple
  examples such as randomized response~\cite{Warner65} and privacy
  amplification by
  subsampling~\cite{DBLP:conf/focs/KasiviswanathanLNRS08} cannot be
  proved.

  \item \emph{\PRHL validity}.
  We show that any valid \PRHL\ judgment, stated within \ERHL, can be proven.
  This is somewhat surprising, as validity in \PRHL\ asserts the existence of a coupling,
  whereas provability of a
  \ERHL\ judgment (implicitly) constructs a specific coupling.
  \ERHL\ does not increase proof complexity,
  any existing \PRHL\ proof can be embedded in \ERHL\ in a natural way.
\end{asparaitem}

Our completeness results are established in two steps. First, we prove
a completeness theorem for judgments of the form
\[
  \RHT{Z}{\ex}{\cmd}{\cmdtwo}{\extwo_1 \oplus \extwo_2}\,,
\]
where the post-expectation can be decomposed into the ``sum'' of two independent
expectations $\extwo_1$ and $\extwo_2$, depending only on the memory of $\cmd$ and
$\cmdtwo$, respectively.
This first theorem relies on two key facts:
\begin{inparaenum}[(a)]
  \item
  quantitative assertions provide the necessary expressiveness to
  encode the semantics of random samplings, and therefore eliminate
  issues with randomness alignment;
  \item
  validity in \ERHL, formulated in terms of \emph{$\star$-couplings},
  does not require programs to have the same termination behavior.
  This eliminates the need for termination conditions in asymmetric
  proof rules.
\end{inparaenum}
Second, we show that many \ERHL statements are provably equivalent to
a statement supported by our first completeness theorem. The existence
of equivalent statements of the required form is obtained by
Strassen's theorem~\cite{strassen1965existence}.

Then, we turn to adversarial computations. One main application of
\PRHL is reasoning about cryptography, which requires to reason about
adversaries. We extend the programming language with adversary calls
and introduce an adversary rule that can be viewed as a quantitative
counterpart to the up-to-bad adversary rule from \PRHL.

Finally, we illustrate the expressivity of \ERHL\ with several
examples that are out of scope of prior program logics, including
examples of program equivalences and differential privacy. We also
provide a few examples that were verified using prior logics, to
showcase features of the logic not covered by the other examples.

\paragraph*{Comparison with~\cite{BartheEGHS18,Aguirre:POPL:21}}
Our logic is not the first formalism to reason about quantitative
properties of probabilistic programs. Prior approaches include the
relational expectation logic \EPRHL\ by~\cite{BartheEGHS18} and the
relational pre-expectation calculus \RPE\ by~\cite{Aguirre:POPL:21}.

The latter is designed to reason about expected sensitivity, i.e.\,
about the expected distance between two outputs of the same program
run on different inputs. \RPE\ computes for a single program $C$ and a
relational pre-expectation $\extwo$ a pre-expectation $\ex$ such that
$\ex~\mem_1~\mem_2 \geq \min_{\mu_0}~\E{\mu_0}{\extwo}$, where $\mu_0$
ranges over all couplings of $\sem{\cmd}_{\mem_1}$ and
$\sem{\cmd}_{\mem_2}$. Our notion of validity generalizes their notion
of validity to the case of two programs.  \RPE\ is defined by
structural induction on the command $\cmd$. As already noted
by~\cite{Aguirre:POPL:21}, one cannot reason about expected
sensitively exactly and compositionally, because the expected
sensitivity of a program $\cmd;\cmd'$ cannot be established solely
from the expected sensitivity of its constituents---however, it can be
over-approximated, which guarantees soundness of \RPE. More
technically, the incompleteness is a consequence of the strong
constraints that \RPE\ imposes about randomness and control-flow
alignment. When applied to an instruction that samples from a
(constant) distribution $\mu$, \RPE\ returns $\min_{\mu_0}
\E{\mu_0}{\extwo}$, where $\mu_0$ ranges over all couplings of $\mu$
with itself. When applied to a conditional instruction $\IF \bexpr
\THEN \cmd \ELSE \cmdtwo$, $\RPE$ returns a pre-expectation that maps
pairs of memories for which the evaluation of $\bexpr$ disagree to
$\infty$. In contrast, \ERHL\ achieves completeness (for a class of
properties) by using one-sided rules to lift randomness and
control-flow alignment constraints and by carefully leveraging
Strassen's theorem. It would be reasonably straightforward to follow
the recipes laid out in our paper to incorporate both ingredients in
\RPE---in fact, an extension of $\RPE$ with one-sided rules is already
discussed in \cite[Section 8]{Aguirre:POPL:21}---and obtain a weakest
relational pre-expectation calculus with similar completeness
properties as \ERHL.

\EPRHL~\cite{BartheEGHS18} is a predecessor of \RPE. Their notion of validity is based on couplings, but otherwise, corresponds to ours.A minor difference is that their notion
of validity compares an affine function of the expected value of the
post-condition with the precondition, but said affine function is a
technical artifact that does not seem to be required in applications.
Contrary to \RPE, \EPRHL\ is defined by a proof system. However, the
proof system only uses 2-sided rules, so \EPRHL\ imposes similar
alignment constraints as \RPE.

Neither \EPRHL\ nor \RPE\ considers recursive procedures and
adversaries. Although it would be possible to extend both formalisms
with such constructs, some applications would also require handling
up-to-bad reasoning, thereby requiring a further extension.

\paragraph*{Comparison with \citet{AvanziniBGMV24}}
\citet{AvanziniBGMV24} defines a pre-expectation program logic for a
core probabilistic with recursive procedures. Like \ERHL, \EHL\ proves
upper-bounds of pre-expectation, i.e., functions taking values
into extended positive reals. \ERHL\ can be understood as a relational
counterpart of \EHL. Specifically, for each program construct, the
rule in \EHL\ gives rise to a corresponding one-sided and two-sided
rule in \ERHL. Moreover, the rule of consequence and the treatment of
recursive procedures in \ERHL\ is inspired from \EHL. A main novelty
of \ERHL\ is that it embeds Strassen's Theorem. Such an embedding is
not considered in \EHL, as \EHL\ is not a relational program logic.

\EHL\ is complete. The proof of completeness is by straightforward
induction on the structure of programs. The completeness of \EHL\ is
used implicitly in our completeness result. Specifically, there is a
derivation-preserving embedding of \EHL into \ERHL. Our one-sided
completeness is a by-product of the completeness of \EHL and of the
embedding.

\paragraph*{Organization of paper}
\Cref{s:language} defines the syntax and semantics of a core
probabilistic language. \Cref{s:couplings} introduces
$\star$-couplings and some auxiliary results.  \Cref{s:logic} presents
\ERHL and its soundness and completeness
theorems. \Cref{s:characterization} studies connections between \PRHL
and \ERHL---notable that \ERHL is complete for \PRHL judgments---
provides our characterizations of total variation, differential
privacy and Kantorovich-Rubenstein distances.  Then,
\Cref{s:extensions} presents extensions of the logic, noteworthy the
integration of procedure and adversary calls. It also presents a proof
of the PRF/PRP Switching Lemma within the logic. We conclude with
related work (\Cref{s:related}) and final remarks
(\Cref{s:conclusion}).
Some of the more involved proofs are relegated to the appendix.


\section{Language}\label{s:language}

We consider an imperative programming \texttt{pWhile} that evolves
probabilistically through the use of sampling instructions. 

\paragraph{Syntax} Let $\Var = \{\vx,\vy,\vz,\dots\}$ a set of \emph{variables}.
The set $\Stmt$ of statements is defined by the following syntax:
\begin{bnf}
  \cmd,\cmdtwo &
  \SKIP
  \bnfmid \vx <- \expr
  \bnfmid \vx <* \sexpr
  \bnfmid \cmd ; \cmdtwo
  \bnfmid \IF \bexpr \THEN \cmd \ELSE \cmdtwo
  \bnfmid \WHILE \bexpr \DO \cmd\,. & \\
 \end{bnf}
Here, $\expr \in \Expr$ is drawn from a set of \emph{expressions},
$\bexpr \in \BExpr$ is a \emph{Boolean} expression, and $\sexpr \in
\SExpr$ a \emph{sampling expression}.  The statements are mostly
standard.  With $\SKIP$ we denote the empty program, that acts as a
no-op.  Semantically, it is the unit to sequence composition
$\cmd;\cmdtwo$.  Syntactically, we sometimes write $\cmd \sep \cmdtwo$
if we intend to have $\SKIP$ as a unit also syntactically, that is
$\SKIP \sep \cmd = \cmd = \cmd \sep \SKIP$.  The statement $\vx <-
\expr$ gives the usual, deterministic assignment, whereas $\vx <*
\sexpr$ samples a value from $\sexpr$, and thereby makes the language
probabilistic.

\paragraph*{Monadic Denotational Semantics}
Semantics of imperative statements can be given in many ways. Here, we
endow the language with a denotational (monadic) style semantics,
lending itself better to the proofs of soundness and completeness of
our logic.
Since statements are probabilistic, we interpret them as functions from
states to (sub-)distributions of states, rather than as mere (partial)
state transformers.

For a countable set $A$, let $\Distr{A}$ denote the set of (discrete)
\emph{(sub-)distributions} over $A$, i.e., functions $d : A \to [0,1]$ with
$|d| \defsym \sum_{a \in A} d(a) \leq 1$.
We exclusively consider discrete
distributions, as it suffices for our work.
Distributions with $|d| = 1$ are called \emph{full} distributions.
The support $\supp(d)$ of $d \in
\Distr{A}$ is the set $\{ a \in A \mid d(a) > 0\}$.
The \emph{Dirac distribution} $\delta_{a}$ collects all mass on $a$, i.e.,
$\delta_{a}(b) \defsym 1$ if $a = b$
and $\delta_{a}(b) \defsym 0$ otherwise.
For two distributions $d_1 \in \Distr{A}$ and $d_2 \in \Distr{B}$, the
\emph{product distribution}
$d_1 \times d_2 \in \Distr{(A \times B)}$ is defined
by $(d_1 \times d_2)(a,b) = d_1(a) \cdot d_2(b)$ for all $a \in A$ and $b \in B$.

For an
event $\event \subseteq A$ we denote by $\Prob{d}{\event}$ the probability
$\sum_{e \in \event} d(e)$. This notation extends to predicates $P : A \to
\{0,1\}$ in the natural way.  Let $\Rext$ denote the
non-negative reals, adjoined with a top element $\infty$.  For a
function $f : A \to \Rext$, we denote by $\E{d}{f}$ the \emph{expected
value $\sum_{a \in A} d(a) \cdot f(a) \in \Rext$ of $f$ on $d$}.
Notice that when $f : A \to \{0,1\}$, then $\Prob{d}{f} = \E{d}{f}$.
We may sometimes write
$\E[a]{d}{e}$ rather than $\E{d}{\lambda a.\ e}$. 

The sub-distribution functor $\mathsf{D}$ forms a monad. The
\emph{unit} $\dunit{} : A\to\Distr{A}$ returns on $a \in A$ the Dirac distribution $\delta_{a}$.
The \emph{bind} $\dbind{} : \Distr{A} \to (A \to \Distr{B}) \to \Distr{B}$ is defined as
$\dbind{d}{f} \defsym \lambda b. \sum_{a \in \supp(d)} d(a) \cdot f \app a \app b : \Distr{B}$.
We may write $\dlet{a}{d}{f(a)}$ for $\dbind{d}{f}$.
With $\dfail : \Distr{A}$ we denote the sub-distribution with empty support.

We model \emph{memories} as mappings $m \in \Mem \defsym \Var \to \Val$ from variables to (a discrete set of) values $\Val$.
We write $m[\vx \mapped \val]$ for the memory obtained from $m$ by updating $\vx$ to $\val$.
\begin{figure}
  \centering
  \begin{tabular}{l@{\quad}l}
    \toprule
    $\cmd \in \Stmt$              & $\sem[]{\cmd}_\mem$                                                                                          \\[1mm]
    \midrule
    $\SKIP$                       & $\dunit{\mem}$                                                                                             \\[1mm]
    $\vx <- \expr$                & $\dunit{\mem[\vx \mapped \sem{\expr}_\mem]}$                                                               \\[1mm]
    $\vx <* \sexpr$               & $\dlet{\val}{\sem{\sexpr}_\mem}{\dunit{\mem[\vx \mapped \val]}}$                                           \\[1mm]
    $\cmd ; \cmdtwo$              & $\dlet{\mem'}{\sem[]{\cmd}_\mem}{\sem[]{\cmdtwo}_{\mem'}}$                                                     \\[1mm]
    $\ITE{\bexpr}{\cmd}{\cmdtwo}$ & $\begin{cases}
                                      \sem[]{\cmd}_\mem & \text{if $\sem[]{\bexpr}_\mem$,} \\
                                      \sem[]{\cmdtwo}_\mem & \text{otherwise.}
                                     \end{cases}$\\[1mm]
    $\WHILE\bexpr\DO\cmd$         & $\sup_{i \in \Nat} \sem[]{\WHILE<i>\bexpr\DO\cmd}_\mem$ \emph{where}                                                    \\[2mm]
    \multicolumn{2}{l}{
    \qquad
    $\begin{array}{l@{\,}l}
       \sem[]{\WHILE<0>\bexpr\DO\cmd}_\mem & \defsym \dfail \\[1mm]
       \sem[]{\WHILE<i+1>\bexpr\DO\cmd}_\mem & \defsym
                                             \begin{cases}
                                               \dlet{\mem'}{\sem[]{\cmd}_\mem}{\sem[]{\WHILE<i>\bexpr\DO\cmd}_{\mem'}} & \text{if $\sem{\bexpr}_\mem$,}\\
                                               \dunit{\mem} & \text{otherwise.}
                                             \end{cases}
     \end{array}$
    } \\[2mm]
    \bottomrule
  \end{tabular}
	\vspace{-6pt}
  \caption{Semantics of statements.}\vspace{-6pt}\label{fig:ds}
\end{figure}
We suppose that expressions $\expr \in \Expr$,
Boolean expressions $\bexpr \in \BExpr$ and
sampling expressions $\sexpr \in \SExpr$
are equipped with semantics $\sem{\expr}_{\mem} : \Val$,
$\sem{\bexpr}_{\mem} : \Bool$
and $\sem{\sexpr}_{\mem} : \Distr{\Val}$, respectively,
indexed by the memory $\mem \in \Mem$ in which these expressions are evaluated.
We assume that sampling instructions are \emph{lossless}, i.e. $\sem{\sexpr}_{\mem}$ is a
full distribution for all $\mem \in \Mem$.
The semantics of statements $\sem[]{\cmd}_\mem : \Distr{\Mem}$
can now be defined in a standard way, see \Cref{fig:ds}.

\paragraph*{Almost sure termination}
Statements may terminate with a probability that is strictly smaller
than one. We say that a statement is \emph{almost surely terminating}
(\emph{\asterm}) if it terminates with probability 1 on all inputs.


\section{Couplings}\label{s:couplings}
Before defining our coupling-based logic, we
review the basic definition and properties of couplings.
Recall that every sub-distribution $\mu \in \Distr{(A_1 \times
  A_2)}$ induces two marginals $\Pi_1(\mu)\in\Distr{A_1}$ and
$\Pi_2(\mu)\in\Distr{A_2}$ satisfying the clauses
$\Pi_1(\mu)(a_1)=\sum_{a_2} \mu(a_1,a_2)$ and
$\Pi_2(\mu)(a_2)=\sum_{a_1} \mu(a_1,a_2)$.
\begin{definition}[Coupling]\label{d:coupling}
  A coupling of two sub-distributions $d_1 \in \Distr{A_1}$ and $d_2
  \in \Distr{A_2}$ is a sub-distribution $\mu \in \Distr{(A_1 \times
    A_2)}$ such that $\Pi_i(\mu) = d_i$ for $i = 1,2$.  We write
  $\coupled{\mu}{d_1}{d_2}$ if $\mu$ is a \emph{coupling} between
  $d_1$ and $d_2$.
\end{definition}

Strassen's theorem~\cite{strassen1965existence} establishes a striking
equivalence between the existence of a coupling adhering to a
relation $\cond$, and a purely universal property. The following is a
variation of Strassen's theorem for discrete distributions, accounting
for a deficiency $\epsilon$.  Strassen's original theorem is the
instance with $\epsilon = 0$. Here, for a relation $\cond \subseteq A
\times B$ and $\event \subseteq A$, we use $\cond(\event) \defsym \{b \in B \mid
\exists a \in \event.\ a \mathrel{\cond} b\}$ for the image of $\event$ under $\cond$.
For a set $\eventtwo \subseteq B$, we
denote by $\setc{\eventtwo} \defsym B \setminus \eventtwo$ its complement when $B$ is clear from context.

\begin{proposition}[Strassen's Theorem (Discrete Version) with Deficiency]
  \label{p:Strassen}
  Let $d_1 \in \Distr{A}$ and $d_2 \in \Distr{B}$ be full distributions,
  let $\cond \subseteq A \times B$ be a relation
  and let $0 \leq \epsilon \leq 1$.
  The following two statements are equivalent:
  \begin{enumerate}
    \item\label{p:Strassen:ex} there exists a coupling $\mu$ of $d_1$ and $d_2$ with $\mu(\setc{\cond}) \leq \epsilon$;
    \item\label{p:Strassen:fa} for all $\event \subseteq A$, $d_1(\event) \leq d_2(\cond(\event)) + \epsilon$,
      or equivalently, $d_1(\event) + d_2(\setc{\cond(\event)}) \leq 1 + \epsilon$
  \end{enumerate}
\end{proposition}

Couplings are used to define the validity of \PRHL\ statements. One
consequence of this choice is that \PRHL\ judgements must always
relate programs that have the same probability of termination. This
entails that every reasoning about one single program requires that
the program is almost surely terminating. To avoid such requirements,
we use an alternative notion, called $\star$-coupling,  inspired from
prior work on differential privacy~\cite{DBLP:journals/lmcs/BartheEHSS19}.
For a set $A$, let $A^\star \defsym A \uplus \{\star\}$
and denote by $d^\star \in \Distr{A^\star}$
the full distribution extending $d \in \Distr{A}$ with $d^\star(\star) \defsym 1 - |d|$.
\begin{definition}[$\star$-coupling]
  A $\star$-coupling of two sub-distributions $d_1 \in \Distr{A_1}$ and
  $d_2 \in \Distr{A_2}$ is a full distribution $\mu \in
  \Distr{(A_1^\star \times A_2^\star)}$ such that $\Pi_i(\mu) =
  d_i^\star$ for $i = 1,2$.
  We write $\starcoupled{\mu}{d_1}{d_2}$ if $\mu$ is a
  \emph{$\star$-coupling} between $d_1$ and $d_2$.
\end{definition}
Thus, a $\star$-coupling between $d_1$ and $d_2$ is a coupling
between the extended distributions $d_1^\star$ and $d_2^\star$.
Since $d^\star = d$ whenever $d$ is a full distribution,
the following is immediate.
\begin{lemma}\label{l:coupling-vs-starcoupling}
Every coupling of
full distributions $d_1$ and $d_2$ is a $\star$-coupling, and conversely.
\end{lemma}


\section{Expectation based relational Hoare logic}\label{s:logic}
We now introduce the \emph{expectation based relational Hoare logic
\ERHLS} and state our two main meta-theoretical theorems, namely
soundness and the aforementioned completeness result.

\paragraph*{Judgments}
\ERHLS is a relational Hoare logic for reasoning about
\emph{quantitative} properties of probabilistic programs. As such, it
generalises pre- and post-conditions to real-valued
functions. Precisely, in \ERHLS judgments take the form
\[
  \RHT{Z}{\ex}{\cmd}{\cmdtwo}{\extwo} \,,
\]
where $Z$ is a non-empty type of \emph{logical} (or, \emph{auxiliary})
variables, \emph{assertions} $\ex,\extwo : Z \to \Mem \times \Mem \to
\Rext$ are functions, the \emph{pre-} and \emph{post-expectations},
respectively, and $\cmd,\cmdtwo$ are statements.

\begin{definition}[Validity]
  A judgment $\RHT{Z}{\ex}{\cmd}{\cmdtwo}{\extwo}$ is \emph{valid},
  written $\valid \RHT{Z}{\ex}{\cmd}{\cmdtwo}{\extwo}$,
  if \emph{for every} $z \in Z, \mem_1,\mem_2 \in \Mem$,
  \[
    \ex \app z \app (\mem_1,\mem_2) \geq \E{\mu}{\se{\extwo} \app z} \ \text{ \emph{for some} }\ \starcoupled{\mu}{\sem{\cmd}_{\mem_1}}{\sem{\cmdtwo}_{\mem_2}}\,,
  \]
  where $\se{\extwo}$ is the extension of $\extwo$ to a function $Z \to \Mem^\star \times
  \Mem^\star \to \Rext$ given by
  \[
    \se{\extwo} \app z \app (\mem_1,\mem_2) \defsym
    \begin{cases}
      \extwo \app z \app (\mem_1,\mem_2) & \text{if $\mem_1,\mem_2 \in \Mem$,} \\
      0 & \text{otherwise.}
    \end{cases}
  \]
\end{definition}
Since validity is witnessed by
$\star$-couplings, $\ERHL$ allows to relate programs that have different termination probabilities.
%
Validity can be equivalently stated by directly lifting the post-expectations from memories to distributions,
thereby highlighting the correspondence to the relational pre-expectation calculus \RPE\ by~\citet{Aguirre:POPL:21}, that is
defined in terms of such a lifting.
Let $\ke{\extwo} : Z \to \Distr{\Mem} \times \Distr{\Mem} \to \Rext$
be defined by $\ke{\extwo} \app z \app (d_1,d_2) \defsym \inf_{\starcoupled{\mu}{d_1}{d_2}} \E{\mu}{\se{\extwo} \app z}$.
\begin{lemma*}{lemma:charvalidity}
  The following two statements are equivalent.
  \begin{enumerateenv}
    \item\label{l:charvalidity:valid} $\valid \RHT{Z}{\ex}{\cmd}{\cmdtwo}{\extwo}$; and
    \item\label{l:charvalidity:alt} for every $z \in Z, \mem_1,\mem_2 \in \Mem$,
    $\ex \app z \app (\mem_1,\mem_2) \geq \ke{\extwo} \app z \app (\sem \cmd_{\mem_1}, \sem \cmdtwo_{\mem_2})$.
  \end{enumerateenv}
\end{lemma*}

\paragraph*{Syntactic conventions}
To avoid notational overhead, we will represent assertions as real-valued expressions, using the following conventions.
We use $\lft{\expr}$ and $\rght{\expr}$ to state that program variables occurring in expression $\expr$
refer to the first and second memory supplied to the assertion, respectively.
In particular, if $x \in \Var$ is a program variable, then $\lft{x}$
refer to the value of $x$ in $\cmd$, and $\rght{x}$ refers to $x$ in $\cmdtwo$.
When notationally convenient, we may identify relations $\cond \subseteq A^1 \times \cdots \times A^n$ with their characteristic functions
$\cf{\cond} : A^1 \times \cdots \times A^n \to \{0,1\}$, and thereby use them freely as assertions.
To incorporate classical reasoning into \ERHLS,
we use the \emph{guard operator} $\bexpr \gd \ex$ as syntactic sugar for  $\IF \bexpr  \THEN \ex \ELSE \infty$.
To prove validity of a judgment
$\RHT{Z}{\bexpr \gd \ex}{\cmd}{\cmdtwo}{\bexprtwo \gd \extwo}$
one can assume that input states are related by $\bexpr$, but dually,
one has to establish that $\bexprtwo$ holds certainly.
The type $Z$ is usually a product type $Z = Z_1 \times \cdots \times Z_n$, modelling
$n$ logical variables. In this case, we may write
$\RHT{z_1 \in Z_1,\dots,\ z_n \in Z_n}{\ex}{\cmd}{\cmdtwo}{\extwo}$
to make the logical variables explicit.
We omit $Z$ from judgments, when not relevant. 

\begin{figure}
  \small
  \begin{rules}[Two-sided rules]
    \Infer[erhls][Skip]
    { \derivable \RHT{Z}{\ex}{\SKIP}{\SKIP}{\ex}}
    {}
    \quad
    \Infer[erhls][Asgn]
    { \derivable \RHT{Z}{\ex[\lft\vx \mapped \lft\expr,\rght\vy \mapped \rght\exprtwo]}{\vx <- \expr}{\vy <- \exprtwo}{\ex}}
    {}
    \\[1mm]
    \Infer[erhls][Sample]
    { \derivable
      \RHT{Z}
      {\E[(\val_1,\val_2)]{\mu}{\ex[\lft\vx \mapped \val_1,\rght\vy \mapped \val_2]}}
      {\vx <* \sexpr}
      {\vy <* \sexprtwo}
      {\ex}}
    {\forall \mem_1,\mem_2.\ \coupled{\mu \app (\mem_1,\mem_2)}{\sem{\sexpr}_{\mem_1}}{\sem{\sexprtwo}_{\mem_2}}}
    \\[1mm]
    \Infer[erhls][If]
    { \derivable \RHT{Z}{\lft{\bexpr} \leftrightarrow \rght{\bexprtwo} \gd \ex}
      {\IF \bexpr \THEN \cmd_1 \ELSE \cmd_2}
      {\IF \bexprtwo \THEN \cmdtwo_1 \ELSE \cmdtwo_2}
      {\extwo}}
    {
       \derivable \RHT{Z}{\lft{\bexpr} \land \rght{\bexprtwo} \gd \ex}{\cmd_1}{\cmdtwo_1}{\extwo}
      &  \derivable \RHT{Z}{\lnot \lft{\bexpr} \land \lnot \rght{\bexprtwo} \gd \ex}{\cmd_2}{\cmdtwo_2}{\extwo}
    }
    \\[1mm]
    \Infer[erhls][While]
    { \derivable
      \RHT{Z}
      {\lft{\bexpr} \leftrightarrow \rght{\bexprtwo} \gd \ex}
      {\WHILE \bexpr \DO \cmd}
      {\WHILE \bexprtwo \DO \cmdtwo}
      {\neg \lft{\bexpr} \land \neg \rght{\bexprtwo} \gd \ex}}
    {  \derivable
      \RHT{Z}
      {\lft{\bexpr} \land \rght{\bexprtwo} \gd \ex}
      {\cmd}
      {\cmdtwo}
      {\lft{\bexpr} \leftrightarrow \rght{\bexprtwo} \gd \ex}
    }
    \\[1mm]
    \Infer[erhls][Seq]
    { \derivable \RHT{Z}{\ex}{\cmd_1 \sep \cmd_2}{\cmdtwo_1 \sep \cmdtwo_2}{\extwo}}
    {
       \derivable \RHT{Z}{\ex}{\cmd_1}{\cmdtwo_1}{\exthree}
      &  \derivable \RHT{Z}{\exthree}{\cmd_2}{\cmdtwo_2}{\extwo}
    }
  \end{rules}
  \vspace{-2mm}
  \begin{rules}[One-sided rules]
    \Infer[erhls][Asgn-L]
    { \derivable \RHT{Z}{\ex[\lft\vx \mapped \lft\expr]}{\vx <- \expr}{\SKIP}{\ex}}
    {}
    \qquad
    \Infer[erhls][Sample-L]
    { \derivable \RHT{Z}{\E{v \leftarrow \lft\sexpr}{\ex[\lft\vx \mapped \val]}}
      {\vx <* \sexpr}
      {\SKIP}
      {\ex}}
    {}\\[1mm]
    \Infer[erhls][If-L]
    { \derivable \RHT{Z}{\ex}
      {\IF \bexpr \THEN \cmd_1 \ELSE \cmd_2}
      {\SKIP}
      {\extwo}}
    {
       \derivable \RHT{Z}{\lft{\bexpr} \gd \ex}{\cmd_1}{\SKIP}{\extwo}
      &  \derivable \RHT{Z}{\lnot \lft{\bexpr} \gd \ex}{\cmd_2}{\SKIP}{\extwo}
    }
    \quad
    \Infer[erhls][While-L]
    { \derivable \RHT{Z}{\ex}{\WHILE \bexpr \DO \cmd}{\SKIP}{\lnot \lft{\bexpr} \gd  \ex}}
    {
       \derivable \RHT{Z}{\lft{\bexpr} \gd \ex}{\cmd}{\SKIP}{\ex}
    }
  \end{rules}
  \vspace{-2mm}
  \begin{rules}[Logical rules]
    \Infer[erhls][Conseq]
    { \derivable \RHT{Z}{\ex}{\cmd}{\cmdtwo}{\extwo}}
    {
      \begin{lalign}[t]
        \\
        \derivable \RHT{Z'}{\ex'}{\cmd}{\cmdtwo}{\extwo'}
      \end{lalign}
      &
        \forall \mem_1\,\mem_2\,d_1\,d_2.\
        \begin{lalign}[t]
          \bigl(\forall z'.\ \ex' \app z' \app (\mem_1,\mem_2) \geq \inf_{\starcoupled{\mu}{d_1}{d_2}} \E{\mu'}{\se{\extwo'} \app z'}\bigr) \\
          \imp \bigl(\forall z.\ \ex \app z \app (\mem_1,\mem_2) \geq \inf_{\starcoupled{\mu}{d_1}{d_2}} \E{\mu}{\se{\extwo} \app z}\bigr)
        \end{lalign}
    }
    \\[1mm]
    \Infer[erhls][Nmod-L]
    {  \derivable \RHT{Z}{\ex}{\cmd}{\cmdtwo}{\extwo} }
    {
      x \not\in \modV{\cmd}
      &
       \derivable
      \RHT
      {\val \in \Val,\ Z}
      {\lft{\vx} = \val \gd \ex}
      {\cmd}{\cmdtwo}
      {\extwo[\lft{\vx} \mapped \val]}
    }
    \\[1mm]
    \InferEq[erhls][Strassen]
    { \derivable \RHT{z\in Z}{\ex}{\cmd}{\cmdtwo}{\cf{\setc\cond_z}} }
    {
      \derivable
      \RHT{M \subseteq \Mem,\ z\in Z}
      {1 + \ex}
      {\cmd}
      {\cmdtwo}
      {\lft{\cf{M}} + \rght{\cf{\setc{\cond_z(M)}}}}
    }
    [\!\raisebox{2pt}{\text{$\cmd$ and $\cmdtwo$ \asterm}}]
\end{rules}
  \vspace{-3mm}
  \caption{Core rules of \ERHLS. Although not presented here, all depicted one-sided ``left''-rules, as well as Rule~\ref{erhls:Nmod-L} come with
  symmetric ``right''-rules.}\label{fig:erhls}
\end{figure}

\subsection{Deductive system}
As usual for relational logics, the deductive system underlying
$\ERHLS$ consists of \emph{two-sided}, \emph{one-sided} and
\emph{logical rules}.  The rules are presented in \Cref{fig:erhls}.

\emph{Two-sided} rules apply to structurally equal programs.
Rule~\ref{erhls:Skip} and Rule~\ref{erhls:Asgn} directly reflect the
statements' semantics within pre- and post-expectations.
Rule~\ref{erhls:Sample} compares two sampling instructions.
The pre-expectation is computed explicitly from a coupling of the two sampling operations $\sexpr$ and $\sexprtwo$.
The next two rules deal with conditionals and loops. These rules are
\emph{synchronous}, in the sense that the compared statements
evaluate in lock step.
In Rule~\ref{erhls:If}, the classical pre-condition $\lft{\bexpr} \leftrightarrow
\rght{\bexprtwo}$ imposes that the evaluation of the two
statements descend either both into the then-branch, or both into the
else-branch.
Thus, it is sufficient to compare the two branches
synchronously. Similarly, Rule~\ref{erhls:While} enforces
that the two while-statements execute in lockstep through
the classical assertion $\lft{\bexpr} \leftrightarrow \rght{\bexprtwo}$. On the quantitative
part of the assertion, the 
rule lifts an invariant $\ex$ on the bodies to one on the two while-statements.
After completion, the rule asserts that the two guards are
falsified, while in the premise, one can assume that the two guards
hold. The final two-sided rule, Rule~\ref{erhls:Seq}, deals with
sequentially composed commands. It is reminiscent of the corresponding
rule found in classical Hoare Logic that facilitates compositional
reasoning.
Through the equivalence $\cmdtwo = \SKIP \sep \cmdtwo = \cmdtwo \sep \SKIP$,
it enables asynchronous reasoning through the application of one-sided rules.

Such \emph{one-sided rules}, presented next in \Cref{fig:erhls}, deal with
judgments of the form
\[
  \RHT{Z}{\ex}{\cmd}{\SKIP}{\extwo}\,.
\]
Although not explicitly shown in the Figure, symmetric rules $\RHT{Z}{\symt{\ex}}{\SKIP}{\cmd}{\symt{\extwo}}$
are also present in the logic. Here, $\symt{\exthree} \defsym \lambda z\ (\mem_1,\mem_2).\ \exthree \app z \app (\mem_2,\mem_1)$ is
the assertion symmetric to $\exthree$.
Conceptually, the one-side rules all follow their two-sided counterparts.
Rules~\ref{erhls:Asgn-L} and~\ref{erhls:Sample-L}
explicitly compute the pre-expectation based on the post-expectation,
rule~\ref{erhls:If-L} performs case analysis on the conditional's guard,
and Rule~\ref{erhls:While-L} relies on an invariant on the loop's bodies.

The next set of rules, the \emph{logical rules}, deal with proof
structure.  As standard in Hoare logics, also \ERHLS contains a
\emph{weakening rule}, or \emph{rule of consequence}. In its simplest
form, this rule permits strengthening of the post-, and weakening of
the pre-expectation:
\[
  \Infer[erhls][Weaken]
  { \derivable \RHT{Z}{\ex}{\cmd}{\cmdtwo}{\extwo}}
  {\ex \geq \ex' &  \derivable \RHT{Z}{\ex'}{\cmdtwo}{\cmd}{\extwo'} & \extwo' \geq \extwo}
\]
Our Rule~\ref{erhls:Conseq} is inspired by the
adaptation complete rule of~\citet{Nipkow2002}. It is more
more general than the usual weakening rule, and permits a change on logical variables.
Adapting to our notations, Nipkow's rule for partial correctness
of classical Hoare Logic
reads as
\[
  \Infer{\derivable Z: \HT{P}{\cmd}{Q}}
  {\derivable Z': \HT{P'}{\cmd}{Q'}
    & \forall \mem\,\mem'.\ (\forall z'.\ P' \app z' \app \mem \imp Q' \app z' \app \mem') \imp (\forall z.\ P \app z \app \mem \imp Q \app z \app \mem')
  }
\]
In our quantitative, relational logic, the input memory $\mem$ turns into a pair of input memories $\mem_1,\mem_2$, the output memory $\mem'$ into a pair of distributions $d_1, d_2$.
The two inner implications reflect the validity of premise and conclusion,
via \Cref{lemma:charvalidity}.



Rule~\ref{erhls:Nmod-L} internalizes variables that remain constant throughout evaluation---the variables not in the set of \emph{variables $\modV{\cmd}$ modified by $\cmd$}%
\footnote{
  To be precise, a variable $\vx$ is modified by $\cmd$ if $\mem \app \vx \not=\mem' \app \vx$
  for some initial memory $\mem$ and final memory $\mem' \in \supp(\sem{\cmd}_{\mem})$.}%
as logical ones.
This rule is relevant in practice as it permits deriving rules akin to the \emph{frame rule} from separation logic, or the \emph{rule of constancy}, in Hoare logics, as explained in a moment.

Rule~\ref{erhls:Strassen} incorporates Strassen's theorem with
deficiency (see \Cref{p:Strassen}) into $\ERHL$.
In the rule, we denote by $\cf{S}$ the characteristic function associated
with a set $S$; $\cond_z$ refers to a family of relations on
memories, potentially parametric in $z$.
The double line indicates that the rule can be applied in both directions.
By the \asterm assumptions,
the $\ERHL$ judgments witness the existence of a coupling on output distributions,
thereby enabling the applications of Strassen's theorem.
In the premise, corresponding to the universal statement \eqref{p:Strassen:fa} of \Cref{p:Strassen},
the set of memories $M$ is implicitly universally quantified.
The conclusion, on the other hand, is a direct encoding of statement \eqref{p:Strassen:ex}.
The significance of this rule lies in its ability to
prove an upper-bound on a relational property,
in terms of an upper-bounds on sum of probability of two one-sided events.

\subsection{Derived rules}

\begin{figure}
  \small
  \begin{rules}[WP-style rules for sampling instructions]
    \Infer[erhls][WP-Sample]
    {
      \derivable \RHT{Z}{\ex}{\cmd;\vx <* \sexpr}{\cmdtwo;\vy <* \sexprtwo}{\extwo}
    }
    {
      \coupled{\mu}{\sem{\sexpr}}{\sem{\sexprtwo}} ~~
      \derivable \RHT{Z}{\ex}{\cmd}{\cmdtwo}{\E[(v_1,v_2)]{\mu}{\extwo[\lft\vx \!\mapped v_1,\rght\vy \!\mapped v_2]}}
    }
    ~
    \Infer[erhls][WP-Sample-L]
    {
      \derivable \RHT{Z}{\ex}{\cmd;\vx <* \sexpr}{\cmdtwo}{\extwo}
    }
    {
      \derivable \RHT{Z}{\ex}{\cmd}{\cmdtwo}{\E[v]{\lft\sexpr}{\extwo[\lft\vx \!\mapped v]}}
    }
  \end{rules}
  \begin{rules}[Derived logical rules \vspace{-3mm}]
      \Infer[erhls][Inst]
      {\derivable \RHT{Z}{\ex[z\mapped\expr]}{\cmd}{\cmdtwo}{\extwo[z\mapped \expr]}}
      {\expr \perp \modV{\cmd} \cup \modV{\cmdtwo} & \derivable \RHT{Z'}{\ex}{\cmd}{\cmdtwo}{\extwo}}
      \qquad
      \Infer[erhls][Ctx]
      {\derivable \RHT{Z}{F[\ex]}{\cmd}{\cmdtwo}{{F[\extwo]}}}
      {
        \begin{lalign}
          F[\cdot] \text{concave, non-decreasing} \\
          F \perp \modV{\cmd} \cup \modV{\cmdtwo}
          \qquad \derivable \RHT{Z}{\ex}{\cmd}{\cmdtwo}{\extwo}
        \end{lalign}
      }
    \end{rules}
  \vspace{-2mm}
  \caption{Derived rules for practical applications.}
  \label{fig:erhls:usable}
\end{figure}

In this section, we introduce derived rules that ease reasoning and
will subsequently used in examples. Our first set of rules supports
backward reasoning style akin to \emph{weakest pre-condition (WP)
reasoning}. Specifically, these rules reason on the final instruction
in a sequence, and combine the rule specific to this instruction with
Rule~\ref{erhls:Seq}. As an example, \Cref{fig:erhls:usable} lists a
two-sided Rule~\ref{erhls:WP-Sample} and the left-sided
Rule~\ref{erhls:WP-Sample-L} for sampling instructions.  Similar rules
can be derived for other instructions. WP reasoning is commonly used
to automate program verification. User intervention can be restricted
to explicit weakening steps (e.g., applications of
Rule~\ref{erhls:Conseq}) and the treatment of sampling instructions
and while loops, where the coupling and invariant have to be supplied,
respectively.

The second set of rules is made of logical rules that streamline
reasoning in practice.  Rule~\ref{erhls:Inst} in
\Cref{fig:erhls:usable} allows replacing logical variables with
a constant expression $E$ that evaluates to $z'\in Z'$, and
whose logical variables are taken from $Z$.
Rule~\ref{erhls:Ctx} allows placing assertions
under a context $F[\cdot]$.  It encompasses for instance a \emph{frame
rule}---taking $F[\cdot] = \bexpr \gd \Box$ for predicate $\bexpr
\perp \modV{\cmd}, \modV{\cmdtwo}$--- or the \emph{constant
propagation law} in pre-expectation calculi
\cite{OlmedoKKM16}---taking $F[\cdot] = \Box + \expr$ where $\expr$ is
constant.
Soundness of Rule~\ref{erhls:Ctx} is derived via Jensen's reverse inequality, which explains the side condition on
$F$.\footnote{
I.e., $F[p \cdot r_1 + (1-p) \cdot r_2] \geq p \cdot F[r_1] + (1-p) \cdot F[r_2]$
and $r_1 \geq r_2$ implies $F[r_1] \geq F[r_2]$.
}

As all of these rules are derived from the core rules
from~\Cref{fig:erhls}, we may use them freely in the examples that we
discuss next.

\subsection{Examples}
This section illustrates the logic at work with two examples---while
simple, these examples are out of scope of prior approaches.

\subsubsection{Two samplings vs. one sampling}
Our first example shows that \ERHLS\ does not impose random alignment constraints. As a simple instance, consider
proofing equivalence of
\[
  \cmd \defsym w <* \{0,1\} \times \{0,1\} \quad \text{\emph{and}} \quad \cmdtwo \defsym x <* \{0, 1\}; y <* \{0, 1\} ; w <- (x, y)\,,
\]
as mentioned in the introduction.
As we will show in \Cref{c:equiv}, the equivalence of $\cmd$ and $\cmdtwo$
can be established by deriving the following judgment:
\[
  \derivable z \in \RHT{\{0,1\}^2}{1} \cmd \cmdtwo {(z = \lft w) + (z \neq \rght w)}\,.
\]
By exploiting linearity of expectation, and weakest pre-expectation reasoning on $\cmd$
via Rule~\ref{erhls:WP-Sample-L}, we reduce our goal to:
\[
  \derivable z \in \RHT{\{0,1\}^2}{1}{\SKIP}{\cmdtwo}{\sum_{a, b\in \{0, 1\}} \frac{1}{4}{(z = (a, b))}+(z \neq \rght w)}.
\]
Reasoning similarly on $\cmdtwo$, this goal further reduces to:
\[
  \derivable z \in \RHT{\{0,1\}^2}{1}{\SKIP}{\SKIP}{\sum_{a, b\in \{0, 1\}} \frac{1}{4}{[z = (a, b)]}+\sum_{a, b\in \{0, 1\}} \frac{1}{4}{[z \neq (a, b)]}}.
\]
Now it is not difficult to see that the post-expectation sums up to 1 independently of $z \in \{0,1\}^2$, thus the goal can be discharged with an application of Rule~\ref{erhls:Skip}.

\subsubsection{Rejection sampling}
\renewcommand{\ud}[1]{#1}
\newcommand{\rs}{\cmd_{\mathsf{RS}}}
Next, we demonstrate a non-trivial application of \ERHLS in order to prove functional correctness of a simple rejection sampling algorithm
\[
  \rs \defsym x<*A; \WHILE x \in X \DO x <* A \,,
\]
by proving it equivalent to $x <* A\setminus X$, that samples the value of $x$ uniformly from $A \setminus X$.
We fix $X \subset A$, thereby the reference implementation is well-defined and $\rs$ is
\asterm.
Similar to before, as later justified by \Cref{c:equiv}, we prove the equivalence by showing:
\[
  \derivable
  \RHT{a \in A}
  {1}
  {\rs}{x <* X\setminus A}
  {(a = \lft{x}) + (a \neq \lft{x})} \,.
\]

The central proof step involves the reasoning around the loop within $\rs$. As a preparatory step,
we prove that it satisfies the invariant $\ex(x)$, given by
\[
  \ex(x) \defsym \IF \lft x \notin X \THEN a = \lft x \ELSE \frac{a \in A\setminus X}{|A\setminus X|} \,.
\]
Precisely, we have
\[
  \derivable \RHT{a \in A}{\ex(\lft{x})}{\WHILE x \in X \DO x <* A}{\SKIP}{\lft{x} \not\in X \gd \ex(\lft{x})} \,,
\]
via an application of Rule~\ref{erhls:While-L}. The premise
$\derivable
\RHT{a \in A}{x \in X \gd \ex(x)}{x <* A}{\SKIP}{\ex(x)}$
is easily discharged via Rule~\ref{erhls:Sample-L}
and a weakening step, for the latter exploiting the classical pre-condition $x \in X$
and the first identity in
\begin{equation}
    \label{eq:rsid}
    \E[v]{\ud{A}}{\ex(v)} = \frac{a \in A \setminus X}{|A \setminus X|} = \E[v]{\ud{A\setminus X}}{a = v} \,.
\end{equation}
By observing $\lft{x} \not\in X \gd \ex(\lft{x})\ge a = \lft{x}$, and using the second identity in \eqref{eq:rsid}, the preparatory step on the loop allows us to establish
\[
  \derivable
  \RHT{a \in A}{\E[v]{\ud{A\setminus X}}{a = v}}{\rs}{\SKIP}{a = \lft{x}} \,,
\]
and consequently
\[
  \derivable
  \RHT{a \in A}
  {\E[v]{\ud{A\setminus X}}{a = v} + \E[v]{\ud{A\setminus X}}{a \neq v}}
  {\rs}{\SKIP}
  {(a = \lft{x}) + \E[v]{\ud{A\setminus X}}{a \neq v}} \,,
\]
by an application of Rule~\ref{erhls:Ctx}.
Using this as a premise to the right-sided version of Rule~\ref{erhls:WP-Sample-L},
we derive
\[
  \derivable
  \RHT{a \in A}
  {\E[v]{\ud{A\setminus X}}{a = v} + \E[v]{\ud{A\setminus X}}{a \neq v}}
  {\rs}{x <* X\setminus A}
  {(a = \lft{x}) + (a \neq \rght{x})} \,.
\]
It is now sufficient to see that the pre-expectation is
identical to $1$.


\subsection{Soundness and completeness}\label{s:sound-complete}
We are ready to present our two main meta-theoretical results, viz, soundness of \ERHL and our completeness
result.
\begin{theorem*}[Soundness]{t:soundness}
  Any derivable judgment is valid, i.e.,
  \[
     \derivable \RHT{Z}{\ex}{\cmd}{\cmdtwo}{\extwo}
    \Imp
     \valid \RHT{Z}{\ex}{\cmd}{\cmdtwo}{\extwo}
    \,.
  \]
\end{theorem*}
Conversely, \emph{completeness} (in the sense of Cook) states that
every valid judgment is derivable. As for other relational Hoare
logics, completeness is not obvious. Certainly, the two-sided fragment
is incomplete, because two-sided rules require programs to have
identical structure. A further difficulty with coupling-based logics
is that the notion of validity existentially quantifies over
couplings, and thus it is not clear if the proof system can always be
used to construct (implicitly) a coupling, when it exists.

In the following, we prove a partial completeness result, in three steps. We
first start with a completeness result on one-sided judgments, i.e.,
where one of the programs is $\SKIP$. Then, we prove a limited completeness
result, for a specific class of post-conditions. Finally, we use
Strassen's theorem to extend this class of post-conditions to a more
general class that encompasses the applications of \ERHL\ discussed in
\Cref{s:characterization}.
\paragraph*{One-sided completeness}
For brevity, we present completeness for left-sided judgments.
The central step behind one-sided completeness is given by the following auxiliary lemma,
stating that a form of \emph{most-general one-sided judgment} can be derived.
\begin{lemma}\label{l:mgu}
  $\derivable
  \RHT{Z}
  {\lambda z\,(\mem_1,\mem_2).\ \E{\sem{\cmd}_{\mem_1} \times \delta_{\mem_2}}{\ex \app z}}
  {\cmd}
  {\SKIP}
  {\ex}$,
  for any post-expectation $\ex$.
\end{lemma}
The lemma is provable by induction on the structure of $\cmd$.
One-sided completeness, as expressed by the following theorem, is now a direct consequence.
Indeed, from the mentioned judgment, any valid one can be derived by
one application of Rule~\ref{erhls:Conseq}.
\begin{theorem*}[One-sided completeness]{t:complete-oneside}
  Any valid one-sided judgment is derivable, i.e.,
  \[
    \valid \RHT{Z}{\ex}{\cmd}{\SKIP}{\exftwo}
    \Imp
    \derivable \RHT{Z}{\ex}{\cmd}{\SKIP}{\exftwo}
    \,.
  \]
\end{theorem*}
Dual, completeness for right-sided judgments, $\RHT{Z}{\ex}{\SKIP}{\cmd}{\exftwo}$ can be obtained.
\paragraph*{A two-sided completeness result.}
With Rule~\ref{erhls:Seq}, two dual one-sided judgments can always be combined to a two-sided one:
\[
  \Infer*[erhls][Seq]{
    \derivable \RHT{Z}{\ex}{\cmd}{\cmdtwo}{\extwo}
}
{
  \derivable \RHT{Z}{\ex}{\cmd}{\SKIP}{\exthree}
  & \derivable \RHT{Z}{\exthree}{\SKIP}{\cmdtwo}{\extwo}
}
\]
Through this observation, \Cref{l:mgu} can be extended to two-sided judgments as follows.
\begin{lemma*}{l:completeness}
  $\derivable
  \RHT{Z}
  {\lambda z\,(\mem_1,\mem_2).\ \E{\sem{\cmd}_{\mem_1} \times \sem{\cmdtwo}_{\mem_2}}{\ex \app z}}
  {\cmd}
  {\cmdtwo}
  {\ex}$,
  for any post-expectation $\ex$.
\end{lemma*}
The pre-expectation is a reification of the $\star$-coupling
implicit to the above proof, viz, the product coupling.
Our main completeness result specializes the post-expectation to a combination
$\ex \oplus \extwo$
of two separate functions $\ex, \extwo$, depending only on memories of the left and right
program statement, respectively.
\begin{theorem*}{t:completeness}
  Let $\cmd, \cmdtwo$ be \asterm programs, let $\extwo,\exthree : Z \to \Mem \to \Rext$,
  and let $\oplus : \Rext \times \Rext \to \Rext$ be any operator
  that commutes with the expectation in the following sense:
  \[
    \E{\mu}{\ex \oplus \extwo} = \E{\mu}{\ex} \oplus \E{\mu}{\extwo}\,.
  \]
  The following three statements are equivalent:
  \begin{enumerate}
    \item\label{t:completeness:derivable} $\derivable \RHT{Z}{\ex}{\cmd}{\cmdtwo}{\lft{\extwo} \oplus \rght{\exthree}}$;
    \item\label{t:completeness:valid} $\valid \RHT{Z}{\ex}{\cmd}{\cmdtwo}{\lft{\extwo} \oplus \rght{\exthree}}$; and
    \item\label{t:completeness:form} for all $z \in Z$, $\mem_1,\mem_2 \in \Mem$, $\ex \app z \app (\mem_1,\mem_2) \geq \E{\sem{\cmd}_{\mem_1}}{\extwo \app z} \oplus \E{\sem{\cmdtwo}_{\mem_2}}{\exthree \app z}$.
  \end{enumerate}
\end{theorem*}
Here, the commutativity property ensures that
\[
  \E{\sem{\cmd}_{\mem_1}}{\extwo \app z} \oplus \E{\sem{\cmdtwo}_{\mem_2}}{\exthree \app z} = \E{\sem{\cmd}_{\mem_1} \times \sem{\cmdtwo}_{\mem_2}}{\lft{\extwo} \oplus \rght{\exthree}} \,.
\]
Via this identity, the theorem is in essence an application of soundness (\Cref{t:soundness}) (implication $\eqref{t:completeness:derivable} \Rightarrow \eqref{t:completeness:valid}$) and
\Cref{l:completeness} (implication $\eqref{t:completeness:form} \Rightarrow \eqref{t:completeness:derivable}$).
\asterm is required only for the implication $\eqref{t:completeness:valid} \imp \eqref{t:completeness:form}$
to hold.

\paragraph*{Discussion} One potential drawback of our completeness theorem is that it requires to prove almost sure termination by external means, whereas it would be arguably more elegant to prove it \emph{within} the logic. Indeed, there exist many logical formalisms to prove almost sure termination, see e.g.\,~\cite{DBLP:journals/pacmpl/McIverMKK18}. However, our interpretation of judgments does not lend itself to support such proofs, as it entails upper bounds rather than lower bounds on expected outcomes of the programs. Furthermore, designing practical methods for reasoning about lower bounds poses challenges of its own, see e.g.\,~\cite{DBLP:journals/pacmpl/HarkKGK20}, which are beyond the scope of the current paper. However, as a point of comparison, we note that \PRHL\ also requires external means to prove termination when applying one-sided rules for loops.


\section{Applications}\label{s:characterization}
This section presents several applications of our results. The first result concerns \emph{completeness of \ERHL\ w.r.t.\, \PRHL}.
Second, it presents the \emph{characterisation of program
equivalence} within \ERHL\ that was already
used in the examples of the previous section.
Finally, it presents completeness of \ERHL\ wrt.\ several
quantitative properties of interest: \emph{statistical distance},
\emph{differential privacy}, and the \emph{Kantorovich distance}.
%
All these results fundamentally rely on our completeness theorem (\Cref{t:completeness}), and
Strassen's theorem, as embodied by Rule~\ref{erhls:Strassen}.

\subsection{Completeness w.r.t.\, \PRHL\ judgments}\label{s:prhl}
In \PRHL, judgments take the form
\[
  \RHT{}{\cond}{\cmd}{\cmdtwo}{\condtwo}\,,
\]
where $\cond,\condtwo : \Mem \times \Mem \to \{0,1\}$ are predicates on pairs of memories.
Such a judgment asserts that on $\cond$-related inputs,
$\cmd$ and $\cmdtwo$ yield $\condtwo$-related
outputs. Assertions are lifted to distributions via couplings.
\begin{definition}[Semantics of \PRHL Judgments]
  \[
    \valid[\PRHL] \RHT{}{\cond}{\cmd}{\cmdtwo}{\condtwo}
    \Defiff  \sem{\cmd}_{\mem_1} \rlift{\condtwo} \sem{\cmdtwo}_{\mem_2}
    \quad \text{for all} \quad \mem_1 \mathrel{\cond}\mem_2\,,
  \]
  where $d_1 \rlift{\condtwo} d_2$ if there exists a coupling $\mu$ of $d_1$ and $d_2$ with
$\mem_1 \mathrel{\condtwo} \mem_2$ for all $(\mem_1,\mem_2) \in \supp(\mu)$.
\end{definition}

\begin{figure}
  \small
  \begin{rules}
        \Infer[prhl][Skip]
    {\derivable[\PRHL] \RHT{}{\cond}{\SKIP}{\SKIP}{\cond}}
    {\strut}
    \qquad
    \Infer[prhl][Sample]
    {\derivable[\PRHL] \RHT{}{\forall \val \in \supp(\sexpr).\ \cond[\vx,\vy \mapped \val,f(\val)]}
      {\vx <* \sexpr}
      {\vy <* \sexprtwo}
      {\cond}}
    {\strut \text{$f : \supp(\sexpr) \to \supp (\sexprtwo)$ is a bijection}}
     \\[2mm]
    \Infer[prhl][If]
    {\derivable[\PRHL] \RHT{}{\cond}
      {\IF \bexpr \THEN \cmd_1 \ELSE \cmd_2}
      {\IF \bexprtwo \THEN \cmdtwo_1 \ELSE \cmdtwo_2}
      {\condtwo}}
    {
      \valid \cond \to \lft\bexpr \leftrightarrow \rght\bexprtwo
      &
       \derivable[\PRHL] \RHT{}{\lft\bexpr \land \cond}{\cmd_1}{\cmdtwo_1}{\condtwo}
       &
       \derivable[\PRHL] \RHT{}{\lnot\lft\bexpr \land \cond}{\cmd_2}{\cmdtwo_2}{\condtwo}
     }\\[2mm]
   \Infer[prhl][Seq]
    {\derivable[\PRHL] \RHT{}{\cond}{\cmd_1;\cmdtwo_1}{\cmd_2;\cmdtwo_2}{\condtwo}}
    {\derivable[\PRHL] \RHT{}{\cond}{\cmd_1}{\cmd_2}{\condthree}
      &
      \derivable[\PRHL] \RHT{}{\condthree}{\cmdtwo_1}{\cmdtwo_2}{\condtwo}}
    ~
    \Infer[prhl][While]
    {
      \derivable[\PRHL]
      \RHT{}
      {\cond}
      {\WHILE \bexpr \DO \cmd}{\WHILE \bexprtwo \DO \cmdtwo}
      {\neg \lft\bexpr \land \cond}
    }
    { \valid \cond \to (\lft\bexpr \leftrightarrow \rght\bexprtwo)
      & \derivable[\PRHL]
      \RHT{}
      {\lft{\bexpr} \land \cond}
      {\cmd}
      {\cmdtwo}
      {\cond}
    }
  \end{rules}
  \vspace{-3mm}
  \caption{Two-sided rules of \PRHL.}
  \label{f:prhl}
\end{figure}
As for \ERHL, \PRHL\ consists of a set of two-sided, one-sided and logical proof
rules. The two-sided rules are given in \Cref{f:prhl}. All the rules
are based purely on Boolean reasoning, except for Rule~\ref{prhl:Sample},
where one has to establish a bijection on the random assignments. The remaining
rules preserve this bijection alignment. They are reminiscent of the rules one
expects from classical Hoare-logic, e.g., one has a compositional
rule for sequential statements \ref{prhl:Seq},
conditionals are reasoned through case-analysis \ref{prhl:If},
and proofs of
while loops \ref{prhl:While} are facilitated through an invariant $\cond$ on the loops body.
Additional assumptions on guards enforce a synchronous control flow, but through
a corresponding case rule
programs with asynchronous control flow can be dealt with.
As for \ERHLS, corresponding one-sided rules exist.
An important difference is that the one-sided rule for loops imposes a termination condition:
\[
  \Infer[prhl][While-L]
  {\derivable[\PRHL] \RHT{}{\cond}{\WHILE \bexpr \DO \cmd}{\SKIP}{\lnot \lft\bexpr \land  \cond}}
  {
    \RHT{}{\lft\bexpr \land \cond}{\cmd}{\SKIP}{\cond}
    & \vDash \cond \to \lft\cond_1
    & \cond_1 \vDash \WHILE \bexpr \DO \cmd \text{ \asterm}
  }
\]
The side-condition is essential to soundness of the rule, as the termination probability of
the two programs has to coincide.

As a first step to our completeness result, we define a semantic preserving
embedding of \PRHL judgments into \ERHL.
Due to the fact that validity in \ERHL is defined in terms of upper-bounds, our
embedding of \PRHL\ is defined by contra-position:
\begin{lemma}[Semantic Embedding of \PRHL]
  \label{l:prhl-embedding}
  \[
    \validP \RHT{}{\cond}{\cmd}{\cmdtwo}{\condtwo}
    \Imp
    \valid \RHT{}{\cf{\setc{\cond}}}{\cmd}{\cmdtwo}{\cf{\setc{\condtwo}}}\,.
  \]
  The converse holds when $\cmd$ and $\cmdtwo$ are \asterm.
\end{lemma}
\begin{proof}
  For the direct implication, consider initial memories $\mem_1$ and $\mem_2$.
  If these are not $\cond$-related, then
  any $\star$-coupling $\mu$ of the output distributions (e.g., the product $\star$-coupling)
  proves
  \[
    (\cf{\setc{\cond}})(\mem_1,\mem_2) \geq \E{\mu}{\se{\cf{\setc{\condtwo}}}} \,
  \]
  since the right-hand side is trivially bounded by $(\cf{\setc{\cond}})(\mem_1,\mem_2) = 1$.
  On the other hand, if the memories are $\cond$-related, validity of the \PRHL judgment
  yields a coupling $\nu$ of the output distributions whose support is related by $\condtwo$,
  i.e., $\E{\nu}{\cf{\setc{\condtwo}}} = 0$. Extending this coupling,
  by placing the missing mass $1 - |\nu|$ on $(\star,\star)$, yields a $\star$-coupling
  $\mu$ that adheres again to the above inequality.

  For the converse implication, observe that for $\cond$-related inputs,
  validity of the \ERHL judgment yields a $\star$-coupling $\mu$ for which
  $\E{\mu}{\se{\cf{\setc{\condtwo}}}} = 0$.
  When $\cmd$ and $\cmdtwo$ are \asterm, $\mu$ is also a coupling witnessing
  the validity of the \PRHL judgment.
\end{proof}

It is worth noting that under this interpretation, every $\PRHL$ rule can derived
as a consequence of the corresponding $\ERHL$ rule. In this sense, \ERHL constitutes
a conservative extension of \PRHL. Any \PRHL proof can be turned, mechanically, into
a corresponding \ERHL proof.
More interestingly, through Strassen's Theorem,
every valid \PRHL\ judgment is provable in \ERHLS.
\begin{theorem}[Characterisation of \PRHL Judgments]
  \label{t:prhl-complete}
  For \asterm programs $\cmd$ and $\cmdtwo$,
  \[
    \validP \RHT{}{\cond}{\cmd}{\cmdtwo}{\condtwo}
    \Iff
    \derivable \RHT{}{\cf{\setc{\cond}}}{\cmd}{\cmdtwo}{\cf{\setc{\condtwo}}}
  \]
\end{theorem}
\begin{proof}
  For the direct implication, suppose $\validP \RHT{}{\cond}{\cmd}{\cmdtwo}{\condtwo}$.
  Taking its semantic embedding (\Cref{l:prhl-embedding}),
  an application of Strassen's theorem with deficiency (\Cref{p:Strassen})
  yields
  \[
    \valid \RHT{M\subseteq \Mem}{1+\cf{\setc{\cond}}}{\cmd}{\cmdtwo}{\cf{M}+\cf{\setc{\condtwo(M)}}} \,.
  \]
  This judgment falls into the fragment for which \ERHL\ is complete (\Cref{t:completeness}).
  An application of Rule~\ref{erhls:Strassen} yields then $\derivable \RHT{}{\cf{\setc{\cond}}}{\cmd}{\cmdtwo}{\cf{\setc{\condtwo}}}$.

  The converse implication follows more directly, using soundness (\Cref{t:soundness}) and
  \Cref{l:prhl-embedding}.
\end{proof}

\subsection{Program equivalence}
Recall that two
programs are equivalent if their semantics coincide, i.e. map
identical inputs to identical output distributions.

\begin{corollary}[Characterisation of Program Equivalence]\label{c:equiv}
  Let $\cmd$ and $\cmdtwo$ be two \asterm programs. The following statements are equivalent:
  \begin{enumerate}
  \item \(
    \derivable \RHT{\mem \in \Mem}{\cond \gd 1}{\cmd}{\cmdtwo}{\lambda \mem \,(\mem_1,\mem_2).\ (\mem = \mem_1) + (\mem \neq \mem_2)}
    \);
  \item
    \(
    \sem{\cmd}_{\mem_1} = \sem{\cmdtwo}_{\mem_2}
    \)
    for all $\mem_1 \mathrel{\cond} \mem_2$.
  \end{enumerate}
\end{corollary}
\begin{proof}
  It is not difficult to reason that two full distributions $d_1$ and $d_2$ over $A$ are
  equal iff
  \[
    1 \geq \Prob[x]{d_1}{a = x} + \Prob[x]{d_2}{a \not= x} \text{ for all $a \in A$} \,.
  \]
  Using that \asterm program statements yield full output distributions, the
  statement is thus a direct consequence of \Cref{t:completeness}.
\end{proof}
Observe how the direct application of \PRHL-completeness would require to prove a judgment that is parameterized by sets of memories, rather than a single one.
The corollary can also be cast into a rule for program equivalence. Actually, the following
rule is slightly more permissive, and allows reasoning about the value that two expressions take.
\[
    \InferEq[erhls][Equiv]
    {\ctx \derivable \RHT{Z}{P \gd 0}{\cmd}{\cmdtwo}{\lft{\expr} \neq \rght{\exprtwo}}}
    {
      \ctx \derivable \RHT{\val \in \Val,\ Z}{P \gd 1}{\cmd}{\cmdtwo}{(\lft\expr = \val) + (\rght\exprtwo \neq \val)}
    }[\text{$\cmd$ and $\cmdtwo$ \asterm}]
\]

\subsection{Statistical distance}
It is often the case that two programs are almost equivalent. In
order to capture almost equivalence precisely, it is often useful to
consider the statistical, or total variation distance, distance
between output distributions.

\begin{definition}[Total Variation Distance]\label{d:tv}
The \emph{total variation distance} between two full distributions of equal weight
$d_1,d_2 \in \Distr{A}$ is defined as
\[
  \TV(d_1,d_2)
  \defsym \sup_{\event \subseteq A} |d_1(\event) - d_2(\event)|
  = \frac{1}{2} \sum_{a \in A} |d_1(a) - d_2(a)|
  = \inf_{\coupled{\mu}{d_1}{d_2}} \Prob{\mu}{\neq} \,.
\]
\end{definition}
The last equality is often known as the fundamental lemma of
couplings. Based on this lemma, we can characterize the total
variation distance between output distributions directly within \ERHL.
\begin{corollary*}[Characterisation of Total Variation]{c:tv}
  Let $\cmd$ and $\cmdtwo$ be two \asterm programs.
  The following are equivalent:
  \begin{enumerate}
    \item\label{c:tv:deriv} $\derivable \RHT{}{\cond \gd \ex}{\cmd}{\cmdtwo}{\cf{\neq}}$; and
    \item\label{c:tv:bounded} $\TV(\sem{\cmd}_{\mem_1}, \sem{\cmdtwo}_{\mem_2}) \leq \ex \app (\mem_1,\mem_2)$ for all $\mem_1 \mathrel{\cond} \mem_2$.
  \end{enumerate}
\end{corollary*}
\begin{proof}
  The implication $\eqref{c:tv:deriv} \imp \eqref{c:tv:bounded}$ follows by soundness (\Cref{t:soundness}), and the fundamental lemma of coupling.
  For the converse implication, observe that by the fundamental lemma, \eqref{c:tv:bounded} implies validity
  $\valid \RHT{}{\cond \gd \ex}{\cmd}{\cmdtwo}{\cf{\neq}}$,
  and consequently
  \[
    \valid \RHT{\eventm\subseteq \Mem}{\cond \gd 1 + \ex}{\cmd}{\cmdtwo}{\cf{\eventm} + \cf{\setc{\eventm}}}
  \]
  by Strassen's theorem with deficiency (\Cref{p:Strassen}).
  By \Cref{t:completeness} this valid judgment is derivable, which allows us
  to conclude \eqref{c:tv:deriv} by one application of Rule~\ref{erhls:Strassen}.
\end{proof}

\subsection{Differential privacy}
Differential privacy ~\cite{Dwork} is a mathematically rigorous
framework to design and analyze privacy-preserving statistical
algorithms. The strength of differential privacy is to provide strong
individual guarantees through probabilistic noise. The main idea of
differential privacy is that running a program on two adjacent inputs,
typically structured inputs that only differ in one item, yields two
distributions that are similar w.r.t.\, specific notion of similarity.
There exist many notions of differential privacy. In this work we
adopt the original notion of \emph{$(\epsilon, \delta)$-differentially
privacy}:

\begin{definition}[Differential privacy~\cite{Dwork}]\label{def:dp}
  A program $\cmd$ is \emph{$(\epsilon, \delta)$-differentially private}
  on memories related by $\cond$,
  if
  \[
    \forall \event \subseteq \Mem.\ \Prob{\sem{\cmd}_{\mem_1}}{\event} \leq \exp(\epsilon) \cdot \Prob{\sem{\cmd}_{\mem_2}}{\event} + \delta \quad \text{for all} \quad \mem_1 \mathrel{\cond} \mem_2\,.
  \]
\end{definition}
\noindent
For \asterm programs,
\ERHLS\ characterizes the set of $(\epsilon,\delta)$-differentially privacy, in the following way.
\begin{corollary*}[Characterisation of Differential Privacy]{c:dp}
  Let $\cmd$ be \asterm.
  Then
  \[
    \derivable
    \RHT{\event \subseteq \Mem}
    {\cond \gd \delta + \exp(\epsilon)}
    {\cmd}
    {\cmd}
    {\lft{\cf{\event}} + \exp(\epsilon) \cdot \rght{\cf{\setc{\event}}}}
    \,.
  \]
  if and only if $\cmd$ is \emph{$(\epsilon, \delta)$-differentially private}
  on memories related by $\cond$.
\end{corollary*}
\begin{proof}
  For \asterm $\cmd$, $(\epsilon, \delta)$-differential privacy can be rephrased as
  \[
     \delta +\exp(\epsilon) \geq \Prob{\sem{\cmd}_{\mem_1}}{\event} + \exp(\epsilon) \cdot \Prob{\sem{\cmd}_{\mem_2}}{\setc{\event}} \,,
   \]
   for all $\event \subseteq \Mem$ and $\cond$ related inputs $\mem_1$, $\mem_2$.
   The claim is thus a direct corollary to \Cref{t:completeness}. 
 \end{proof}

\subsubsection{Example: Randomized response}
\newcommand{\rr}{\cmd_{\mathsf{RR}}}
Randomized response~\cite{Warner65} is a simple mechanism to conduct surveys
and data collection, while maintaining the privacy of individual respondents.
Suppose each individual of a population has a private
property and the goal of an interviewer is to measure the percentage
of population that has this property. If each individual reveals to
the interviewer whether it has the property or not, their privacy is
compromised. In randomized response, instead, the individuals each toss a coin.
Based on the output of the coin toss, an individual either reveals the
information, or answer randomly to the question. Randomized
response can be used to make an algorithm differentially private.
As a simple example, consider
\[
  \rr \defsym x <* \{0, 1\}; \IF x=1 \THEN y <*\{0, 1\} \ELSE y <- b\,,
\]
which, depending on the outcome $x$ of a tossed coin,
either fills the variable $y$ with a random bit,
or it stores in $y$ a private bit $b$.
Privacy of $\rr$ cannot be proven in \APRHL. Informally, this is
because \APRHL\ features rules for basic mechanisms and rules for
composing privacy budgets, but has no suitable primitive rule to
reason about the kind of sampling behavior performed in $\rr$.

We use \ERHL to derive an $\epsilon$ for which $\rr$ is $\epsilon$-differentially private.
To this end, guided by \Cref{c:dp}, we derive
\begin{equation}
  \label{eq:rr}
  \tag{\dag}
  \derivable
  \RHT[t]{\event \subseteq \{0,1\}}
  {
    \frac{1}{2} \cdot \E[y]{\{0,1\}}{(y \in \event) + \exp(\epsilon)\cdot(y \notin \event)}
    + \frac{1}{2} \cdot ((\lft b \in \event) + \exp(\epsilon) \cdot (\rght b \notin \event))
  }
  {\rr}
  {\rr}
  {(\lft y \in \event) + \exp(\epsilon) \cdot (\rght y \notin \event)}
\end{equation}
by reasoning backward using only the two-sided rules. All sampling instructions
are coupled via the the identity coupling.
By an application of Rule~\ref{erhls:Weaken},
\[
  \derivable
  \RHT{\event \subseteq \{0,1\}}
  {\exp(\epsilon)}
  {\rr}
  {\rr}
  {(\lft y \in \event) + \exp(\epsilon)\cdot (\rght y \notin \event)}\,.
\]
follows for $\exp(\epsilon) \geq 3$, i.e., $\rr$ is $(\log 3)$-differentially private.
To see this, observe that the pre-expectation in \eqref{eq:rr} is identical to
\[
  \ex = \exp(\epsilon) + \frac{1}{2} \cdot \left((1 - \exp(\epsilon))  \cdot \E[y]{\{0,1\}}{y \in \event} + (\lft{b} \in \event) - \exp(\epsilon) \cdot (\rght{b} \in \event) \right)
  \,.
\]
The application of Rule~\ref{erhls:Weaken} requires $\exp(\epsilon) \geq \ex$, or equivalently,
\[
  (\exp(\epsilon) - 1)  \cdot \E[y]{\{0,1\}}{y \in \event} \geq (\lft{b} \in \event) - \exp(\epsilon) \cdot (\rght{b} \in \event) \,,
\]
for all $\event \subseteq \{0,1\}$.
The only case where this inequality does not follow trivially is when $\event$ is a singleton set. In this case, $\E[y]{\{0,1\}}{y \in \event} = \frac{1}{2}$, the right-hand side is at most $1$,
from which we infer that $\exp(\epsilon) \geq 3$ is indeed sufficient (and necessary).
Note that this bound is tight~\cite{DBLP:journals/fttcs/DworkR14}.

\subsubsection{Example: Privacy amplification by subsampling}
\label{sec:subsampling}
\newcommand{\filter}[2]{#1 \ominus #2}
\newcommand{\dbsim}{\lesssim}
The subsampling mechanism is broadly employed in the design of differentially private mechanisms such as differentially private machine learning algorithms.
As it does not rely on any feature of the algorithm it is applied to, it is also suitable for those applications where the target algorithm is only available as a black box.

In a nutshell, the subsampling mechanism amplifies the privacy guarantees of a differentially private algorithm $\cmd[A]$ by applying it to a small randomly sampled subset of records of an initial database.
The effectiveness of the subsampling mechanism depends on how the subset is sampled. One of the simplest ways is to employ the so-called \emph{Poisson Filter}. With this filter, each record in the initial dataset can be included in the final one---where the differentially private algorithm $\cmd[A]$ runs---with some probability $\gamma$. In our language, we can describe this construction as follows:
\[
  \cmd_{\cmd[A]} \defsym U <* \CALL{PoissonFilter}(|d|, \gamma); d <-\filter{d}{U}; \cmd[A]
\]
Here $d$ is the database, $\CALL{PoissonFilter}(n, \gamma)$ samples a Poisson filter with bias $\gamma$ for $n$ entries as a bitmask $U$, and  $\filter{a}{U}$ is the database obtained by masking elements
with a distinguished element $\bot$ according to $U$.

Let $d_1 \dbsim d_2$ denote that databases $d_1$ and $d_2$ coincide, up to possibly one entry set to $\bot$ in $d_1$.
Based on \Cref{c:dp},
it can be shown that if $\cmd[A]$ is $(\epsilon, \delta)$-differentially private algorithm on $\dbsim$-related databases,
then the subsampling amplified version $\cmd_{\cmd[A]}$ is
$(\epsilon_\gamma, \delta_\gamma)$-differentially private on $\dbsim$-related entries,
for $\epsilon_\gamma \defsym {\log(1+\gamma(\exp(\epsilon)-1))}$ and
$\delta_\gamma \defsym \delta \cdot \gamma$~\cite{smith:privacysample}.

We outline the proof.
Wlog.\ we suppose $\cmd[A]$ depends only on the initial database, i.e., it can be attributed a semantic $\sem{\cmd[A]}(d) : \Distr{\Mem}$.
By an application of \Cref{l:completeness} on $\cmd[A]$ and WP-reasoning via two-sided rules, using the identity coupling on the sampling of the filter $U$,
we derive
\[
  \derivable
  \RHT[t]
  {\event \subseteq \Mem}
  { \lft{d} \dbsim \rght{d} \gd
    \E[U]
    {\op{PoissonFilter}(n,\gamma)}
    { \Prob{\sem{A}(\filter{\lft{d}}{U})}{\event}
      + \exp(\epsilon_\gamma)\cdot\Prob{\sem{A}(\filter{\rght{d}}{U})}{\event}}
  }
  {\cmd_{\cmd[A]}}{\cmd_{\cmd[A]}}
  {\lft{\cf{\event}} + \exp(\epsilon_\gamma) \cdot \rght{\cf{\setc{\event}}}}\,
\]
Weakening the pre-expectation, reasoning identical to \cite[Theorem 29]{Steinke}
using the DP-assumption on $\cmd[A]$,
now yields
\[
  \derivable
  \RHT {\event \subseteq \Mem}  {\lft{d} \dbsim \rght{d} \gd \gamma_\delta + \exp(\epsilon_\gamma)}
  {\cmd_A}{\cmd_A}
  {\lft{\cf{\event}} + \exp(\epsilon_\gamma) \cdot \rght{\cf{\setc{\event}}}}\,,
\]
establishing our claim through \Cref{c:dp}. Note that the provided amplification bound is tight.

\subsection{Kantorovich-Rubinstein distance}

\newcommand{\metric}{\delta}

The \emph{Kantorovich-Rubinstein distance}---also known as
\emph{Wasserstein distance}---measures how far two probability
distributions $d_1, d_2 \in \Distr A$ are apart,
lifting a metric $\delta$ on $A$ to a metric $W_p^\delta$ on $\Distr A$.
\begin{definition}[Kantorovich distance]
  \label{d:krmetric}
  Let $\metric : A \times A \to \Rext$ be a function.
  For $p\in [1 ,\infty)$, the Wasserstein $p$-distance between two full distributions $d_1, d_2 \in \Distr A$ is defined as
  \[
    W_{p}^\metric(d_1, d_2 )
    \defsym \inf _{\coupled{\mu}{d_1}{d_2}}\left(\E[(a_1, a_2)]  \mu {\metric(a_1,a_2)^{p}}\right)^{\frac 1 p}.
  \]
\end{definition}
Notice that a direct consequence of this definition is that \ERHL can prove upper bounds to the \emph{ Kantorovich-Rubinstein distance} between two programs, as it is stated by the following proposition.

\begin{lemma}
  \label{rem:krmetric}
  If
   \(
     \derivable \RHT{}{ \cond \gd \ex}{\cmd}{\cmdtwo}{\metric^p},
  \)
  then
  \(
  \ex \app (\mem_1, \mem_2)^{\frac 1 p} \ge
  W_{p}^\metric(\sem \cmd_{\mem_1}, \sem \cmdtwo_{\mem_2})
  \)
  for all $\mem_1 \mathrel{\cond} \mem_2$.
\end{lemma}
While \Cref{rem:krmetric} establishes only an upper-bound, we can
characterize the Kantorovich distance for the case $p=1$
through the \emph{Kantorovich-Rubinstein duality}.

\begin{theorem}[Kantorovich-Rubinstein duality~\cite{kantorovich1958space}]
  \label{thm:krduality}
  For a metric $\delta : A \times A \to \Rext$, 
  \[
    W_{1}^\metric(d_1, d_2 )
    = \inf_{\mathclap{\coupled{\mu}{d_1}{d_2}}}\ \E{\mu}{\metric}
    = \sup_{f \in L_1}(\E{d_1}{f}-\E{d_2}{f}),
  \]
  where the $\sup$ ranges over all 1-Lipschitz functions
  $L_1 \defsym \{f\in \Real^\Mem \mid ||f||_L\leq 1\}$.
\end{theorem}
A direct consequence of \Cref{thm:krduality} is the following sound
and complete characterisation of the $W_1^\metric$ distance between
pairs of \asterm programs, provided the metric is bounded.

\begin{theorem*}[Characterisation of Kantorovich distance]{thm:kmdcomplete}
  Let $\cmd$ and $\cmdtwo$ be two \asterm\ programs.
  and let $\metric: \Mem \times \Mem \to [0, h]$ be a metric, where $0 \leq h < \infty$. 
  The following statements are equivalent:
  \begin{enumerate}
  \item\label{thm:kmdcomplete:deriv} $\derivable \RHT{f \in L_1} { \cond \gd \ex+h} \cmd\cmdtwo  {\lft{f} - \rght{f} + h}$; and
  \item\label{thm:kmdcomplete:metric}
    $W_1^\metric(\sem \cmd_{\mem_1},\sem \cmdtwo_{\mem_2}) \leq \ex (\mem_1, \mem_2)$ for all $\mem_1 \mathrel{\cond} \mem_2$.
  \end{enumerate}
\end{theorem*}
\noindent
In \eqref{thm:kmdcomplete:deriv}, $f$ is a universally quantified $1$-Lipschitz function. $1$-Lipschitzness can be used in the proofs, e.g., via Rule~\ref{erhls:Conseq}.
Note that since the difference between two applications of a $1$-Lipschitz function $f$ is bounded by $h$,
the post-expectation in~\eqref{thm:kmdcomplete:deriv} remains non-negative, independent of $f$.

\subsubsection{Example: Algorithmic stability of Stochastic Gradient Descent}
\label{sec:sgd}
\newcommand{\loss}{\ell}
\newcommand{\grad}{\nabla}
\newcommand{\diffone}{\simeq_1}
\newcommand{\sgd}{\cmd_{\mathsf{SGD}}}

\emph{Algorithmic stability} expresses the tendency of a learning mechanism to produce models that have similar performance on unseen examples
and on the training dataset~\cite{Bousquet}.
We revisit algorithmic stability of \emph{Stochastic Gradient Descent (SGD)},
a classic testbed for verification of relational expected properties,
modeled as follows:
\[
  \sgd \defsym\pw{
    \var{w} <- \var{w}_0; \var{t} <- 0 ;\\
    \WHILE \var{t} < T
    \DO{
      \var{s} <* \var{S};\\
      \var{g} <- \grad(\loss(\var{s}))(\var{w});\\
      \var{w} <- \var{w} - \var{\alpha_t} \cdot \var{g};\\
      \var{t} <-\var{t}+1
    }
  }
\]
This program runs on a
set of examples $S\in A^n$ and ``learns'' a sequence of parameters
$w\in \Real^d$, repeatedly updating them in order to minimize a \emph{loss
function} $\loss: A \to \Real^d\to [0, 1]$.

\citet{Hardt} has shown that SGD is
$\left(\frac{2L^2}{n}\sum_{t=0}^{T-1}\alpha_t\right)$\emph{-uniformly
  stable} for suitable loss-functions $\loss$ and step-sizes $\alpha_t$:
for all examples $a\in A$, we have
\[
  | \E[\mem]{\sem \sgd_{\mem_S}}{\loss(a,\mem(w))} - \E[\mem]{\sem \sgd_{\mem_{S'}}}{\loss(a,\mem(w))} | \leq \frac{2L^2}{n}\sum_{t=0}^{T-1}\alpha_t,
\]
\noindent
where $\mem_S$ and $\mem_{S'}$ are memories storing
training sets $S$ and $S'$, respectively, that differ in exactly
one example, denoted by $\mem_S \diffone \mem_{S'}$ in the following.

\citet{Aguirre:POPL:21} have proven this statement in their pre-expectation calculus for probabilistic sensitivity.
We re-cast the proof into \ERHLS. As a first step,
a crucial observation is that the absolute distance of two expectations
$|\E{d_1}{f_1} - \E{d_2}{f_2}|$, such as the one expressing algorithmic stability, is bounded from above itself
by the Kantorovich distance $W^\delta_1$ given by the metric $\delta(a_1,a_2) = |f_1(a_1) - f_2(a_2)|$ (see e.g. \cite[Theorem 2.4]{Aguirre:POPL:21}).
By \Cref{rem:krmetric}, a sufficient condition for stability of $\sgd$
is thus given by the \ERHLS\ judgment
\[
  \RHT{a \in A}{\diffone \gd \frac{2L^2}{n}\sum_{t=0}^{T-1}\alpha_t}{\sgd}{\sgd}{\ |{\loss(a, \lft w)} - {\loss(a, \rght w)}|}\,.
\]
To derive this judgment, one can now proceed by reasoning identical to \cite{Aguirre:POPL:21}, using the two-sided rules of \ERHLS.
The invariant computed in \cite{Aguirre:POPL:21} serves essentially as an invariant in the application of Rule~\ref{erhls:While}.

Algorithmic stability of SGD was also established using \EPRHL~\cite{BartheEGHS18}. In comparison to \EPRHL's judgments, those of \ERHL are only meant to reason about expected values, instead of reasoning about sensitivities \emph{and} relational properties. This renders \ERHL's proof system closer to that of \APRHL and thus simpler than \EPRHL's one. 


\section{Extensions of the logic}\label{s:extensions}
In this section, we extend our programming language and logic with
recursive procedures and adversary calls, and illustrate how the
resulting logic can be used to reason about cryptographic proofs.
Throughout this section, \textbf{we exclude Rule~\ref{erhls:Strassen} from the calculus}. The technical reason is discussed after \Cref{t:procs}.

\subsection{Procedure calls}

We first consider the extension of our language with
statically scoped, recursive procedures.

\paragraph*{Extensions of the language} To model procedures,
we introduce the set $\Op = \{\fn,\fntwo,\dots\}$ of \emph{procedure names},
and we assume that the set of variables $\Var$ is
partitioned into local variables $\LVar$ and global variables $\GVar$.
Global variables should be understood as implicit input and output to procedures,
while local ones are statically scoped. Because of this distinction
each memory $\mem$ can be seen as the disjoint union of a global
memory $\gmem\in \GMem$ with domain $\GVar$ and
a local memory $\lmem\in \LMem$ with domain $\LVar$. 
The set $\Stmt$ of statements now also includes the
instruction $\vx <* \CALL\fn(\expr)$, which executes the procedure $\fn$
with argument $\expr$ and assigns its return
value to $\vx$. Zero or more than one argument can be passed to procedures as tuples.
For simplicity, we require that
$\vx$ is a local variable---this will simplify the rule for procedure
calls in our logic---but nothing prevents us from writing
$\vx <* \CALL\fn(\expr); \vy <- \vx$ if we want to assign the result
to a global variable $\vy$.

\emph{Procedures} are declared through \emph{procedure definition} of the form
$\PROC\fn(\vx) \cmd;\RET{\expr}$, 
where $\vx \in \LVar$ is the \emph{formal parameter},
$\cmd \in \Stmt$ the \emph{body}
and $\expr \in \Expr$ the \emph{return expression} of $\fn$.
A \emph{program} $\prog$ is a finite sequence of
(mutually exclusive) \emph{procedure definitions}.
Throughout the following, we keep the program $\prog$ fixed,
and we assume that it always contains the identity procedure $\PROC{id}(x) \RET x$,
lifting the no-op statement $\SKIP$ to the level of procedures.
This simplifies the treatment of one-sided judgments related to procedures.

\paragraph*{Semantics}

The interpretation of commands now
also depends on a
\emph{procedure environment}
\[
  \env : \Op \to (\GMem\times\Val) \to \Distr{(\GMem\times\Val)}\,,
\]
attributing to each $\fn \in \Op$ its
semantics
$\sem[\env]{\fn}_{\gmem,\val} \defsym \env \app \fn \app (\gmem,\val)$. 
The semantics of a procedure invocation $\vx <* \CALL\fn(\expr)$,
evaluates $\fn$ on the global portion of the memory and the
argument. After the execution of $\fn$,
the global memory is updated and the return value is stored in $\vx$, captured in
the semantics as follows:
\[
  \sem[\env]{\vx <* \CALL\fn(\expr)}_\mem \defsym \dlet{(\gmem',r)}{\sem[\env]{\fn}_{\gmem, \sem{\expr}_\mem}}{\dunit{(\gmem' \uplus \lmem[\vx \mapped r])}}.
\]
Apart from carrying a procedure environment, the semantics of
statements depicted in~\Cref{fig:ds} remains otherwise identical.

Procedures can be mutually recursive. To formalize this,
for each $n \in \Nat$ we first define procedure environments $\env<n>$
that reflect the semantics of procedures
up to $n$ unfoldings of recursive calls, failing when this threshold has been reached:
\label{def:env}
\begin{align*}
    \env<0> \app \fn
    & \defsym \lambda (\gmem,\val).\ \dfail \\
    \env<n+1> \app \fn
    & \defsym \lambda (\gmem,\val).\
      \dlet{\mem'}
      {(\sem[{\env<n>}]{\cmd}_{\gmem\uplus\lmemz[\vx \mapped \val]})}
      {\dunit{(\gmem',\sem{\expr}_{\mem'})}} \\
    & \qquad \qquad \text{ \emph{where} } (\PROC\fn(\vx)\cmd;\RET{\expr}) \in \prog.
\end{align*}
In the case $n \not= 0$,
the local memory is initialized to an \emph{initial memory} $\lmemz$, assigning
to each variable $\vx \in \LVar$ an (arbitrary but fixed) default value, and
the formal parameter $\vx$ is bound by the argument. Upon completion of the procedure's body $\cmd$, the return
expression is evaluated and returned, together with the potentially modified global memory.
The full semantics of a procedure are captured by letting the threshold $n$ be arbitrary large,
as captured by the environment. $\env_\prog \defsym \sup_{n \in \Nat} \env<n>$
associated to $\prog$. As we keep $\prog$ fixed in the following,
it is justified to abbreviate $\sem[\env_\prog]{\cdot}$ by $\sem{\cdot}$ in the following.

\paragraph{Extension of $\ERHL$} To deal with procedures, we endow our logic with a \emph{logical contexts} $\ctx$,
i.e., sets of auxiliary judgments of the form
$\RHT{Z}{\exf}{\fn}{\fntwo}{\exftwo}$ that relate pairs of procedures.
Following the semantics of procedures,
assertions $\exf, \exftwo$ are drawn from $Z \to (\GMem \times \Val) \times (\GMem \times \Val) \to \Rext$.
In the pre-expectation $\exf$, the two arguments of type $\GMem \times \Val$ refer
to the global memory and the formal parameter of $\fn$ and $\fntwo$, respectively,
while in the post-expectation $\exftwo$ they refer to the modified global memory
and return value.
Despite the change in type, validity is defined in line with that of ordinary judgments.
Specifically, a judgment on procedures is \emph{valid},
in notation $\valid \RHT{Z}{\exf}{\fn}{\fntwo}{\exftwo}$, if
for every $z \in Z, i_1,i_2 \in \GMem \times \Val$,
\[
   \exf \app z \app (i_1,i_2) \geq \E{\mu}{\se{\exftwo} \app z} \text{ for some } \starcoupled{\mu}{\sem{\fn}_{i_1}}{\sem{\fntwo}_{i_2}}.
\]

\paragraph*{Procedure rules}

\begin{figure}
  \begin{rules}
    \Infer[erhls][Call]
    {\ctx \derivable
      \RHT{Z}
      {\exf[\lft{\varg}\mapped\lft\expr,\rght{\varg}\mapped\rght\exprtwo]}
      {\vx <* \CALL{\fn}(\expr)}
      {\vy <* \CALL{\fntwo}(\exprtwo)}
      {\exftwo[\vres_1\mapped\lft\vx,\vres_2\mapped\rght\vy]}
    }
    {
      (\RHT{Z}{\exf}{\fn}{\fntwo}{\exftwo}) \in \ctx
    }
    \\[3mm]
    \Infer[erhls][Call-L]
    {\ctx \derivable
      \RHT{Z}
      {\exf[\lft\varg\mapped\lft\expr,\rght\varg\mapped\rght\vy]}
      {\vx <* \CALL{\fn}(\expr)}
      {\SKIP}
      {\exftwo[\lft\vres\mapped\lft\vx,\rght\vres\mapped\rght\vy]}
    }
    {
      (\RHT{Z}{\exf}{\fn}{\op{id}}{\exftwo}) \in \ctx
    }
    \\[3mm]
  \Infer[erhls][Proc]
    {\ctx \derivable \RHT{Z}{\exf}{\fn}{\fntwo}{\exftwo}}
    {
      \begin{array}{c}
        (\PROC{\fn}(\vx)\cmd;\RET{\expr}) \in \prog
        \qquad (\PROC{\fntwo}(\vx)\cmdtwo;\RET{\exprtwo}) \in \prog
        \\[1mm]
        \initz \defsym \forall \vy \in \LVar.\, \vy \neq \vx \imp \lft{\vy} = \lmemz(\vx) = \rght{\vy} \\[1mm]
        \ctx \derivable
        \RHT{Z}
        {\initz \gd \exf[\lft{\varg}\mapped\lft{\vx}, \rght{\varg}\mapped\rght{\vx}]}
        {\cmd}{\cmdtwo}
        {\exftwo[\lft{\vres}\mapped\lft{\expr}, \rght{\vres}\mapped\rght{\exprtwo}]}
      \end{array}
    }
    \\[3mm]
    \Infer[erhls][Ind][\textsc{ProcInd}]
    {\ctx \derivable \RHT{Z}{\ex}{\cmd}{\cmdtwo}{\extwo}}
    {
      \begin{array}{c}
        \ctxtwo = (\RHT{Z_i}{\exf_i}{\fn_i}{\fntwo_i}{\exftwo_i})_{i=1,\dots,k} \\[1mm]
        \forall 1 \leq i \leq k.\ \ctx,\ctxtwo \derivable \RHT{Z_i}{\exf_i}{\fn_i}{\fntwo_i}{\exftwo_i}
        \quad \ctx,\ctxtwo \derivable \RHT{Z}{\ex}{\cmd}{\cmdtwo}{\extwo}
      \end{array}
    }
  \end{rules}
  \vspace{-3mm}
  \caption{Procedure rules.}\label{fig:procedures}
\end{figure}

To support reasoning about procedure calls, all the deduction rules
of $\ERHL$ are extended to carry an additional logical context $\ctx$,
e.g. the rule for sequencing
statements takes now the form:
\[
    \Infer[erhlexample][Seq]
    {\ctx \derivable \RHT{Z}{\ex}{\cmd_1 \sep \cmd_2}{\cmdtwo_1 \sep \cmdtwo_2}{\extwo}}
    {
      \ctx \derivable \RHT{Z}{\ex}{\cmd_1}{\cmdtwo_1}{\exthree}
      & \ctx \derivable \RHT{Z}{\exthree}{\cmd_2}{\cmdtwo_2}{\extwo}
    }
\]

Additional rules for procedures are in \Cref{fig:procedures}.
Rule~\ref{erhls:Call} is the two-sided rule for
procedure calls, lifting the auxiliary judgment
$\RHT{Z}{\exf}{\fn}{\fntwo}{\exftwo}$ from the logical context to
a corresponding pair of calls. In the conclusion of the rule, the
arguments $\expr$ and $\exprtwo$ are supplied to the pre-expectation
$\exf$, conversely, in the post-expectation $\exftwo$ the return
parameters are substituted by the assigned variables $\vx$ and $\vy$,
respectively.  Note that $\exf$ and $\exftwo$, parameterised by
definition only on global memories, are implicitly lifted in the
conclusion to functions on general memories. Rule \ref{erhls:Call-L}
follows its two-sided counterpart. Although
not explicitly shown, the deductive system also includes the
symmetrical version of Rule \ref{erhls:Call-L}. 

Rule~\ref{erhls:Proc} lifts an assertion relating the bodies
of two procedures to a judgment on the procedures. 
In the pre-expectation of the premise, the classical condition
reflects that upon invocation, the local memory is initialised to $\lmemz$,
the substitution in the post-expectation relates the result parameter to the two return values.
The rule is only interesting in combination with Rule~\ref{erhls:Ind}, which enables inductive
reasoning on procedures. This rule internalizes judgments
on procedures $\ctxtwo$ within the logical context, once all of them have been proven,
via Rule~\ref{erhls:Proc}.

The following theorem establishes the soundness
of $\ERHL$ discussed in this section.
\begin{theorem}[Soundness and Completeness with Procedures]\label{t:procs}
  \begin{enumerateenv}
    \item\label{t:procs:sound} \emph{Soundness:}  Any derivable judgment is valid, i.e.,
    \[
      \derivable \RHT{Z}{\ex}{\cmd}{\cmdtwo}{\extwo}
      \Imp
      \valid \RHT{Z}{\ex}{\cmd}{\cmdtwo}{\extwo}
      \,.
    \]
    \item\label{t:procs:complete} \emph{Completeness:} Any valid judgment in the form mentioned in \Cref{t:completeness} is derivable, i.e.,
    \[
      \valid \RHT{Z}{\ex}{\cmd}{\cmdtwo}{\lft{\extwo} \oplus \rght{\exthree}}
      \Imp \derivable \RHT{Z}{\ex}{\cmd}{\cmdtwo}{\lft{\extwo} \oplus \rght{\exthree}}
      \,.
    \]

  \end{enumerateenv}
\end{theorem}
The proof of soundness in the presence of recursive procedures is significantly more involved.
It proceeds in two steps. First, we establish soundness wrt. approximated semantics, permitting
recursion up-to depth $n \in \Nat$. This intermediate result, proven by induction on $n$ and side-induction
on the structure of the derivation of the judgment,
allows us to take care of by Rule~\ref{erhls:Ind}, embodying inductive reasoning within \ERHLS.
In effect, this first step yields a family of $\star$-couplings $(\mu_n)_{n \in \Nat}$ that witness validity of the derived judgment
under approximated semantics $\sem[\env<n>]{\cdot}$.
As a second step, we then show that this sequence of $\star$-couplings converges, in a certain sense, to a $\star$-coupling
$\mu$ that witnesses validity of the derived judgment under full semantics.

Alas, the first step fails for Rule~\ref{erhls:Strassen}, due to the \asterm side-conditions,
inherited from Strassen's theorem (\Cref{p:Strassen}).
Since \asterm does not extend to the approximated semantics---an \asterm program may fail to evaluate to a full distribution under approximated semantics when the recursion threshold $n$ is reached---Strassen's theorem cannot be applied.
This does not mean that Rule~\ref{erhls:Strassen} is unsound, in general, in the presence of recursive procedures.
However, one has to be careful in combination with inductive proofs through Rule~\ref{erhls:Ind},
which would require additional book-keeping and thereby complicate the logic. As Rule~\ref{erhls:Strassen} is not
strictly necessary for our completeness results---Strassen's theorem can always be applied at the level of validity---we refrained from including the rules in the logic extended to procedures.

The completeness result can be proven following the outline given in \Cref{s:sound-complete},
through one-sided completeness. Crucial, most-general (one-sided) judgments (\Cref{l:mgu})
remain derivable under a most-general, one-sided logical context $\ctx$.
As the judgments in $\ctx$ are themselves derivable, \Cref{l:mgu} is recovered by one application
of Rule~\ref{erhls:Ind}. From there, the proof proceeds identical to the fragment without procedures.

\subsection{Adversaries}

\newcommand{\Adv}{\mathsf{Adv}}
\newcommand{\AOp}{\mathsf{AdvFun}}
\newcommand{\adv}[1][]{\op{\mathcal{A}}_{#1}}
\newcommand{\advtwo}[1][]{\op{\mathcal{B}}_{#1}}
\newcommand{\orcl}{\fn}
\newcommand{\orcltwo}{\fntwo}
\newcommand{\vorcl}{\textit{orcl}}
\newcommand{\ADV}[1]{\CALL{\adv[#1]}}
\newcommand{\ADVTWO}[1]{\CALL{\advtwo[#1]}}
\newcommand{\advenv}{\zeta}
\newcommand{\writeV}[1]{\mathsf{Write}_{\uncolor{codefn}{#1}}}

In this section, we consider an extension of our language and logic with
adversary calls, i.e., invocations to arbitrary procedures that have access to
global variables and an oracle.

\paragraph*{Extension of the language}
To extend the language, we permit \emph{adversary calls} $\vx <- \ADV{\orcl}(\expr)$, where $\adv$
is drawn from a set $\Adv = \{\adv,\advtwo,\dots\}$ of \emph{adversary names}.
Each adversary call is parameterised by an \emph{oracle}, i.e., a pre-defined procedure $\orcl \in \Op$,
accessible to $\adv$ via a dedicated variable $\vorcl$.%
\footnote{In practice, we permit $\adv$ to be parameterised by more than one oracle. Here, the restriction helps us to avoid notational overhead. }
Adversaries $\adv$ refer to arbitrary procedures,
granted access to a subset $\GVar_{\adv} \subseteq \GVar$ of global variables
which are all read-only, except for a distinguished set of \emph{writable variables}
$\writeV{\adv} \subseteq \GVar_{\adv}$.

\paragraph*{Semantics} To model adversarial code in the semantics, we
index the interpretation of program statements by a second environment, the \emph{adversary environment} $\advenv$.
This environments maps each $\adv \in \Adv$ to a declaration
\[
  \advenv \app \adv = \PROC{\adv}(\vx) \cmd;\RET{\expr}
  \,.
\]
Procedure calls in the body $\cmd$ are confined
to \emph{oracle calls} $x <* \CALL{\vorcl}(\expr)$, the
oracle itself is determined at invocation time of the adversary.
We require that adversary environments are consistent with accessible and writable variables, i.e.,
the body $\cmd$---except during the invocation of the oracle---%
accesses only variables in $\GMem_{\adv}$, and modifies only those
within $\writeV{\adv}$.
For instance, if the adversary executes an
instruction $\vx <- \expr$, then $\vx \in \writeV{\adv}$ and $\expr$ mentions only variables in $\GMem_{\adv}$.
Notice that the memory content of a global variable $\vx \not\in \writeV{\adv}$ may change
during an invocation of $\adv[\fn]$, precisely when it is modified while executing the supplied oracle $\fn$.
The semantics of an adversary $\adv[\fn]$ parameterised by oracle $\fn$ now follows that of an ordinary (non-recursive) procedure,
with the code taken from the adversary environment, and the oracle $\vorcl$ effectively set to $\fn$:
\begin{align*}
  \sem[\env,\advenv]{\adv[\orcl]}_{\gmem,\val}
  & \defsym
      \dlet{\mem'}
       {\sem[{\env[\vorcl \mapped \env \app \orcl],\advenv}]{\cmd}_{\gmem\uplus\lmemz[\vx \mapped \val]}}
       {\dunit{(\gmem',\sem{\expr}_{\mem'})}}
    \text{ \emph{where} } \advenv \app \adv = \PROC{\adv}(\vx)\cmd;\RET{\expr}\,.
\end{align*}
Similar to before, we write $\sem[\advenv]{\cdot}$ for $\sem[\env,\advenv]{\cdot}$ when $\env = \env_\prog$.
\paragraph*{Extension of \ERHLS}
To extend the logic for programs with adversarial code, the notion of judgment can remain identical, apart from
the fact that program statements may now contain adversarial calls. However, judgments will now be \emph{valid}
if validity in the original sense holds \emph{independent} of the adversarial code, that is,
$\valid \RHT{Z}{\ex}{\cmd}{\cmdtwo}{\extwo}$ states that
\emph{for all adversary environments $\advenv$},
for all $z \in Z$, $\mem_1,\mem_2 \in \Mem$, there
exists
$\starcoupled{\mu}{\sem[\advenv]{\cmd}_{\mem_1}}{\sem[\advenv]{\cmdtwo}_{\mem_2}}$ with
$\ex \app z \app (\mem_1,\mem_2) \geq \E{\mu}{\se{\extwo} \app z}$.
Similar, validity for procedure declarations is defined by quantifying
over all adversary environments.

\newcommand{\BAD}{\expression{B}}
\begin{figure}
  \small
  \begin{rules}
    \Infer[erhls][Adv-L]
    { \ctx \derivable \RHT{Z}{\exf}{\adv[\orcl]}{\op{id}}{\exf} }
    { \exf \perp \writeV{\adv}
      & \exf \perp \{\lft{\varg},\rght{\varg},\lft{\vres},\rght{\vres}\}
      & \ctx \derivable \RHT{Z}{\exf}{\orcl}{\op{id}}{\exf}
    }
    \\[1mm]
    \Infer[erhls][Adv]
    { \ctx \derivable
      \RHT{Z}
      {\exfthree_\varg}
      {\adv[\orcl]}
      {\adv[\orcltwo]}
      {\exfthree_\vres}
    }
    {
      \begin{array}{c}
        \exf,\exftwo,\BAD \perp \writeV{\adv}\\[1mm]
        \exf,\exftwo \perp \{ \lft{\varg},\rght{\varg},\lft{\vres},\rght{\vres} \} \\[1mm]
        \exfthree_\vx \defsym \IF \BAD \THEN \exftwo \ELSE (\eqmem{\vx,\GVar_{\adv}} \gd \exf)
      \end{array}
      \ 
      \begin{array}{c}
        \ctx \derivable \RHT{Z}{\BAD \gd \exftwo}{\orcl}{\op{id}}{\BAD \gd \exftwo} \\[1mm]
        \ctx \derivable \RHT{Z}{\BAD \gd \exftwo}{\op{id}}{\orcltwo}{\BAD \gd \exftwo} \\[1mm]
        \ctx \derivable\RHT{Z}{\neg \BAD \land \eqmem{\varg,\GVar_{\adv}} \gd \exf}{\orcl}{\orcltwo}{\exfthree_\vres}
      \end{array}
    }
  \end{rules}
  \vspace{-3mm}
  \caption{Adversary rules.}
  \label{fig:erhls:adversary}
\end{figure}

\paragraph*{Adversary rules}\Cref{fig:erhls:adversary} gives our adversarial rules.
The first rule is the one-sided rule covering the cases where the adversary is
invoked on the left, a symmetric right rule also present, although not shown in the figure. 
The assumptions placed on $\exf$, with $\exf \perp V$ indicating that $\exf$
does not depend on variables $V$,%
\footnote{
  Formally, for $\expr \in \Expr$ an expression on $\Mem$ we define $\expr \perp V$ iff
  $\sem{\expr}_{\mem_1} = \sem{\expr}_{\mem_2}$ for all $\mem_1,\mem_2$ that may differ only on $\Var \setminus V$.
  This notation extends naturally to assertions.
}
establish that $\exf$ remains invariant when executing the adversary. The premise on the oracle $\orcl$ enforces that invariance is retained when the adversary invokes $\orcl$. 

The second rule in \Cref{fig:erhls:adversary} covers
the invocation of identical adversaries $\adv[\orcl]$ and $\adv[\orcltwo]$ parameterised by different
oracles. Similar in spirit to the one-sided rule, 
it lifts an invariant relating the two oracles to one that relates $\adv[\orcl]$ and $\adv[\orcltwo]$.
To see this more clearly,
in simplified form where a designated \emph{bad event} $\BAD$ never occurs, the rule reads as
\[
  \Infer
  { \ctx \derivable
    \RHT{Z}
    {\eqmem{\varg,\GMem_{\adv}} \gd \exf}
    {\adv[\orcl]}
    {\adv[\orcltwo]}
    {\eqmem{\vres,\GMem_{\adv}} \gd \exf}
  }
  {
    \exf \perp \writeV{\adv}
    & \exf \perp \{ \lft{\varg},\rght{\varg},\lft{\vres},\rght{\vres} \}
    &
    \ctx \derivable
    \RHT{Z}
    {\eqmem{\varg,\GMem_{\adv}} \gd \exf}
    {\orcl}{\orcltwo}
    {\eqmem{\vres,\GMem_{\adv}} \gd \exf} \\[1mm]
  }
\]
The predicate $\eqmem{x}$ is short form for $\lft\vx = \rght\vx$, similar $\eqmem{V}$ asserts
that all variables in $V$ coincide this way. 
Its usage in the rule keeps the control flow of the two compared adversary invocations synchronised,
during execution of the adversarial code and when invoking the oracle.
Again, the premise on $\exf$ enforces that $\exf$ remains invariant when executing the adversary code.
By the last premise, invariance is retained. In this simplified form,
as the control flow is identical on both sides, it is justified to consider only synchronized
calls to the two oracles. No one-sided assumptions are needed. 

To make the logic usable in practice, the rule incorporates reasoning
up-to a bad event $\BAD$. Once this event has been reached, the control flow of the two sides are allowed to diverge.
Additional one-sided premises provide an invariant $\exftwo$ that has to hold. 

The main theorem of this section states that our adversary rules are sound.
\begin{theorem*}[Soundness of Adversary Rules]{t:adversary}
  The adversary rules in \Cref{fig:erhls:adversary} are sound.
\end{theorem*}

\subsection{Example: PRF/PRP switching lemma}

\newcommand{\coll}{\mathit{coll}}
\newcommand{\vc}{\var{c}}
\newcommand{\vl}{\var{l}}
\newcommand{\sz}[1]{|#1|}
\newcommand{\NIL}{\constr{[\,]}}
\newcommand{\CONS}{\mathrel{\constr{::}}}
\newcommand{\QQ}{\constr{Q}}
\newcommand{\UNDEF}{\constr{\bot}}
\newcommand{\prp}{\op{prp}}
\newcommand{\prf}{\op{prf}}
\newcommand{\gprp}{\op{game\_prp}}
\newcommand{\gprf}{\op{game\_prf}}
\begin{figure*}[t]
  \centering
  \small
  \[
    \pw{
      \VAR \vl;\\
      \\[-3mm]
      \PROC\prf(\vx) {
        \IF \sz\vl \geq \QQ \THEN \vy <- \UNDEF \\
        \ELSEIF \vx \in \CALL{dom}(\vl)
        \THEN \vy <- \vl(\vx) \\
        \ELSE \pw{
          \vy <* A;\\
          \vl <-(\vx, \vy) \CONS \vl;
        }\\
        \RET{\vy}
      }
      \\[-3mm]
      \PROC\gprf() {
        \vl <- \NIL;\\
        \vx <- \ADV{\prf}();\\
        \RET \vx
      }
    }
    \qquad\qquad
    \pw{
      \VAR \vl;\\
      \\[-3mm]
      \PROC\prp(\vx) {
        \IF \sz\vl \geq \QQ \THEN y <- \UNDEF \\
        \ELSEIF \vx \in \CALL{dom}(\vl)
        \THEN \vy<-\vl(\vx) \\
        \ELSE \pw{
          \vy <* A\setminus \CALL{codom}(\vl);\\
          \vl <-(\vx, \vy) \CONS \vl;
        }\\
        \RET{\vy}
      }
      \\[-3mm]
      \PROC\gprp() {
        \vl <- \NIL;\\
        \vx <- \ADV{\prp}();\\
        \RET \vx
      }
    }
  \]
  \vspace{-3mm}
  \caption{Functions $\gprf$ and $\gprp$ model a game  where an adversary $\adv$ is meant to distinguish between a function $\prf$ and $\prp$, implementing a random function, and permutation, respectively.}
  \label{fig:prp/prf}
\end{figure*}

The \emph{\emph{PRP/PRF} switching lemma} is an important
cryptographic result used to show that pseudorandom
functions and pseudorandom permutations are computationally
indistinguishable~\cite{Bellare2008}. It
states that an adversary $\adv$
can distinguish a random function from a
random permutation over $n$ elements,
using at most $\QQ$ oracle queries,
with probability at most $\frac {\QQ({\QQ-1})}{2 \cdot n}$.
The corresponding games used in a game-based cryptographic proof are given
in \Cref{fig:prp/prf}. The oracles $\prf$ and $\prp$, implementing
a random function and random permutation over $A$, respectively,
are represented internally as an associate list $\vl$.
The two oracles differs only
in the sampling instruction. They stop
answering once $\QQ \in \Nat$ distinct queries have been performed. 

The switching lemma can be proven in \ERHL by proving a bound on the total variation
of the two games via \Cref{c:tv}, precisely, we have
\[
  \derivable
  \RHT{}
  {\frac{\QQ(\QQ-1)}{2|A|}}
  {\gprp}
  {\gprf}
  {\lft \vres \neq \rght \vres},
\]
assuming that the adversary $\adv$ has no access to the internal
representations $\vl$, i.e., $\GVar_{\adv} = \emptyset$.  The crux of
the underlying derivation lies in the application of the adversary
rule~\ref{erhls:Adv}, instantiated as follows.  The bad event $\BAD$
states that a \emph{collision} occurred in $\prf$, i.e., that
$\vl(\vx_1) = \vl(\vx_2)$ for some $\vx_1 \neq \vx_2$, causing $\prf$
and $\prp$ to diverge.  In case of a collision, we use the invariant
$\exftwo \defsym 1$.  Finally, the invariant $\exf$ covering the case
where no collision occurred (yet) is given by $\eqmem{\vl} \gd\sum_{0
  \leq i < \QQ-\lft{n}} \frac{i+\lft{n}}{|A|}$, for $n \defsym
\sz{\vl}$ the number of distinct queries carried out so far. Since the
two associated lists are initially empty, this invariant is initially
identical to $\frac{\QQ(\QQ-1)}{2|A|}$, as demanded by the proof.

Under this instantiation, the two one-sided premises of
rule~\ref{erhls:Adv} are easily discharged, the remaining two-sided
proof obligation
\begin{gather*}
  \derivable
  \RHT{}
  {\neg \BAD \land \eqmem{\varg, \vl} \gd\sum_{0 \leq i < \QQ-{\lft{n}}} \frac {i+\lft{n}} {|A|}}
  {\prf}{\prp}
  {          \IF \BAD \THEN 1 \ELSE (\eqmem{\vres, \vl} \gd \sum_{0 \leq i < \QQ-{\lft{n}}} \frac {i+\sz {\lft \vl}} {|A|})
}
\end{gather*}
can be derived using only two-sided rules.  The only non-trivial part
of the proof is where the inner else branches are compared. There, one
proceeds by weakest precondition reasoning.  When comparing the two
sampling instruction, elements from $A$ on the left not in
$\CALL{codom}(\lft{\vl})$ are coupled with identical elements on the
right with probability $\frac{1}{|A|}$.  The total missing mass on the
remaining elements is divided, with equal probability $\frac{1}{|A|
  \cdot |A \setminus \CALL{codom}(\lft{\vl})|}$ on the right.  Using
this coupling, the final weakening step follows from simple
arithmetical reasoning on pre-expectations.

The PRP/PRF lemma has been proved using multiple systems, including
\tool{Clutch}~\cite{GregersenAHTB24} and \PRHL.  A key difference is
that the entire proof is conducted directly within \ERHL, whereas
previous approaches use \PRHL or a variant to prove conditional
equivalence, and program
semantics (or a non-relational logic) to upper bound the probability
of the ``bad'' event, i.e. the complement of the event conditioning
equivalence.


\section{Related work}\label{s:related}

\paragraph*{Expectation-based program logics}
Seminal work by Kozen~\cite{Kozen:JCSC:85} uses pre-expectations for
reasoning about probabilistic programs. \citet{MM05}
introduce a weakest pre-expectation calculus for reasoning about the
combination of probabilities and non-determinism. Their calculus is
exact, but variants of the calculus can be used to prove upper or
lower bounds on expected values. Reasoning about upper or
lower bounds rather than exact values eases verification. However,
reasoning about lower bounds is significantly harder than reasoning
about upper bounds, see e.g.~\cite{DBLP:journals/pacmpl/HarkKGK20}.
In general, these works focus on positive pre-expectations, that take
values in the (extended) non-negative reals. \citet{KK:LICS:17}
explore how to generalize pre-expectation calculi to mixed-signed
expectations; informally, the resulting calculus internalizes a proof
that the expectation of the absolute value of the pre-expectation is
well-defined.

\paragraph*{Coupling-based program logics and calculi}
\citet{BartheGB09} develop the foundations of \PRHL, and prove in the
Coq proof assistant its soundness w.r.t.\ a denotational semantics of
programs. \citet{DBLP:journals/corr/BartheEGHSS15} establishes the
connection with probabilistic couplings. It follows from the embedding
of \PRHL\ into \ERHL\ that all examples in \PRHL\ can be encoded into
\ERHL.

Several works explore couplings for higher-order probabilistic
languages~\citet{DBLP:conf/esop/ABBBG18,DBLP:journals/pacmpl/TassarottiH19,DBLP:journals/pacmpl/VasilenkoVB22,GregersenAHTB24}
develop coupling-based logics for higher-order languages. It would be interesting to explore extensions of \ERHL\ to the higher-order setting.

BlueBell~\cite{BaoDF24} is a novel program logic that combines
relational and non-relational reasoning for probabilistic
programs. Specifically, BlueBell features a joint conditioning
modality that supports reasoning about distribution laws,
probabilistic couplings and their interactions. These interactions
alleviate some of the constraints on randomness alignment, and allow
proving some of the simple examples that are not provable in \PRHL.
On the other hand, tracking precisely distribution laws of program
variables is generally challenging---similar to proving lower bounds
on expected values.

\paragraph*{Coupling-based logics for differential privacy}
\APRHL~\cite{DBLP:conf/popl/BartheKOB12} is an extension of \PRHL\ to
reason about differential privacy. \APRHL manipulates judgments of the
form $\ARHT{}{\cond}{\cmd}{\cmdtwo}{\condtwo}{\epsilon,\delta}$, where
$\epsilon>0$ and $0\leq \delta <1$ are the usual parameters of
differential privacy. The rules are similar to those of \PRHL, except
that they are instrumented to track the privacy budget through the
derivation. Over the years, the expressiveness of \APRHL was enhanced
with the introduction of additional
rules~\cite{DBLP:conf/lics/BartheGGHS16,DBLP:conf/ccs/BartheFGGHS16}.
These rules provide useful additional reasoning principles that could
be used to verify advanced examples such as the Sparse Vector
Technique. However, basic examples such as randomized response and
privacy amplification remain out of scope for these extended
logics. Moreover, the repeated addition of new (and at times \emph{ad
hoc}) proof rules is rather unsatisfactory both from foundational and
practical perspectives. This suggests that it would be beneficial to
provide a tight integration between \APRHL\ and \ERHL. Unfortunately,
the interpretation of \APRHL\ uses different notions of approximate
couplings, requiring a careful analysis of existing proof rules to
determine if they can be embedded in \ERHL. Broadly speaking, there
exist three notions of approximate couplings: 1-couplings, which are
similar to usual couplings and use a single witness, 2-couplings which
use two witnesses, and 0-couplings, which formulate couplings as a
generalization of Strassen's theorem. 0-couplings are used for
instance in~\cite{DBLP:journals/entcs/Sato16}; 1-couplings are used
for instance in~\cite{DBLP:conf/popl/BartheKOB12}; 2-couplings are
used for instance
in~\citet{DBLP:conf/icalp/BartheO13,DBLP:journals/pacmpl/AlbarghouthiH18,DBLP:journals/lmcs/BartheEHSS19,justin:phd}. All
rules based on 0-couplings and on the 2-witness version of
$\star$-coupling
from~\cite{DBLP:journals/lmcs/BartheEHSS19,justin:phd} are admissible
in \ERHL---in the latter case, this follows from the equivalence with
the 0-coupling definition proved in~\cite{justin:phd}. It follows that
all examples in these systems, including Above Threshold, can be
encoded into \ERHL.

\paragraph*{Coupling-based logics for Kantorovich distance}
\ERHL can be used to establish upper bounds on Kantorovich
distance. It subsumes other approaches, like the relational
pre-expectation calculus \RPE\ by~\cite{Aguirre:POPL:21}, and the
relational expectation logic \EPRHL\ by~\cite{BartheEGHS18}. One major
difference with \RPE\ is that the latter selects for each sampling
instruction the coupling that minimizes the Kantorovich distance.
Since the Kantorovich distance is not compositional, \RPE\ does not always
produce tight bounds. Moreover, \RPE\ imposes strong constraints on
randomness alignment---similar to \PRHL.

\paragraph*{Coupling-based logics for quantum programs}
One important application of \PRHL\ is proving security of
post-quantum cryptographic constructions. With the advent of
post-quantum cryptography, i.e.\, classical algorithms that are secure
against quantum adversaries, there is a strong incentive to extend
coupling-based reasoning to quantum programs. Examples of such logics
have been developed
in~\cite{DBLP:journals/pacmpl/Unruh19,DBLP:journals/pacmpl/BartheHYYZ20,DBLP:conf/ccs/BarbosaBFGHKSWZ21}. To
our best knowledge, the completeness of these logics has not been
investigated.

\paragraph*{Automated proofs for probabilistic programs}
There is a large body of work that develops fully automated tools for
(dis)proving properties of probabilistic programs. These tools use a
broad range of probabilistic tools, notably martingales, probabilistic
game semantics, and are not limited to using couplings. In some
restricted settings, it is possible to achieve completeness and
decidability, see e.g.\, \cite{DBLP:conf/cav/KieferMOWW12} for
equivalence and
\cite{DBLP:conf/lics/BartheCJS020,DBLP:conf/ccs/ChadhaS0B23} for
differential privacy. However, the majority of works aims for
incomplete methods for (approximate)
equivalence~\cite{DBLP:chatterjeePLDI24}, Kantorovich
distance~\cite{DBLP:journals/pacmpl/WangFCDX20}, and differential
privacy~\cite{DBLP:conf/ccs/BichselGDTV18,DBLP:conf/ccs/DingWWZK18}.


\section{Conclusion}\label{s:conclusion}
\ERHL\ is a relational program logic for expectation-based properties
of probabilistic programs. \ERHL\ subsumes many prior coupling-based
logics, and is the first relational probabilistic program logic to
achieve soundness and completeness for non-trivial properties, and in
particular program equivalence, statistical distance, differential
privacy, and Kantorovich metric. It remains an open problem whether
\ERHL\ is complete.




\begin{acks}
This research was supported
by the
\grantsponsor{ANR}{\textit{Agence Nationale de la Recherche} (French National Research Agency)}{https://anr.fr/}
as part of the France 2030 programme (\grantnum{ANR}{ANR-22-PECY-0006})
and through the project "Probabilistic Program Semantics" (\grantnum{ANR}{ANR-19-CE48-0014}).
\end{acks}


\newpage
\appendix
\section{Properties of the Semantics}
We start with some observations of the expectation of $\ex$ w.r.t. to program
semantics. 
\begin{proposition}\label{p:sem-props}
	The following properties hold:
	\begin{enumerate}
		\item \label{p:sem-props:E-continuous}
		\emph{continuity:} $\E{\sup_i d_i}{f} = \sup_i (\E{d_i}{f})$ for every
		$\omega$-chain $d_0 \leq d_1 \leq d_2 \leq \dots$\,;
		\item \emph{monotony:} \label{p:sem-props:E-mono}
		$f \leq g \Imp \E{d}{f} \leq \E{d}{g}$;
		\item \emph{bind:}\label{p:sem-props:E-bind}
		$\E{\dbind{d}{h}}{f} = \E{d}{\lambda a. \E{h \app a}{f}}$;
		\item \emph{unit:}\label{p:sem-props:E-unit}
		$\E{\dunit{a}}{f} = f \app a$.
	\end{enumerate}
\end{proposition}
The first point is a discrete version of Lebesgue's Monotone Convergence
Theorem~\citep[Theorem 21.38]{Schlechter96},
implying monotony. The bind equation can be proven by unfolding and re-arranging
sums. The unit law follows by definition.
These laws are sufficient to prove the characterisation of
$\E{\sem{\cmd}_{m}}{\ex}$ given in \Cref{fig:evt}, which we tacitly employ throughout the proofs of
soundness and completeness.
\begin{figure}
	\centering
	\begin{tabular}{l@{\quad}l}
		\toprule
		$\cmd \in \Stmt$               & $\E{\sem{\cmd}_\mem}{\ex}$                                \\[1mm]
		\midrule
		$\SKIP$                       & $\ex \app
		\mem$                                                     \\[1mm]
		$\vx <- \expr$               & $\ex \app \mem[\vx \mapped
		\sem{\expr}_\mem]$                     \\[1mm]
		$\vx <* \sexpr$             & $\E{\sem{\sexpr}_\mem}{\lambda \val. f \app \mem[\vx \mapped \val]}$ \\[1mm]
		$\vx <* \CALL\fn(\expr)$    & $\E{\sem{\fn}_{\gmem} \app \sem{\expr}_\mem}{\lambda (\gmem',r).\ \ex \app (\gmem' \uplus \lmem)[\vx \mapped r]}$\\[1mm]
		$\cmd_1 ; \cmd_2$           & $\E{\sem{\cmd_1}_\mem}{\lambda \mem'.\E{\sem{\cmd_2}_{\mem'}}{\ex}}$ \\[1mm]
		$\ITE{\bexpr}{\cmd_1}{\cmd_2}$ & $\begin{cases}
			\E{\sem{\cmd_1}_\mem}{\ex}     & \text{if $\sem{\bexpr}_\mem$,} \\
			\E{\sem{\cmd_2}_\mem}{\ex}  & \text{otherwise.}
		\end{cases}$
		                                   \\[1mm]
		$\WHILE\bexpr\DO\cmdtwo$       & $\sup_{i \in \Nat} \E{\sem{\WHILE<i>\bexpr\DO\cmdtwo}_\mem}{\ex}$ \\[1mm]
		\bottomrule
	\end{tabular}
	\caption{Structural expectation rules.}\label{fig:evt}
\end{figure}
\begin{proposition}
  All equalities depicted in \Cref{fig:evt} hold.
\end{proposition}
Formally this proposition can be proven inductively. 
The only equality worth mentioning is that given w.r.t. to while loops, all
remaining ones follow directly from \Cref{p:sem-props}.
The crucial step in the proof of this equality lies in proving
$F_{f}^{(i)} \app \mem = \E{\sem{\WHILE<i>\bexpr\DO\cmd}_\mem}{\ex}$,
which can be easily verified by induction on $i$.
From there, the equality follows essentially by continuity of expectations
(\Cref{p:sem-props}.\eqref{p:sem-props:E-continuous}).

\section{Alternative characterization of Validity}

A remarkable property of $\ERHL$ is the characterization of validity in terms
of \emph{Kantorovich Lifting}, as stated below:


\begin{lemma}[Lemma 3.3 of \cite{Aguirre:POPL:21}]
  \label{lemma:Aguirre3.3}
  let $d_1, d_2\in \Distr{A}$ be two sub-distributions of countable support
  with the same weight and let $\ex: A\times A \to \Rext$. There is a
  coupling $\coupled \mu {d_1} {d_2}$ such that:
  \[
    \E{\mu}{\ex} = \inf_{\coupled {\nu} {d_1} {d_2}} {\E{\nu}{\ex}}.
  \]
\end{lemma}

\again{lemma:charvalidity}
\begin{proof}
  The implication $\eqref{l:charvalidity:valid} \Rightarrow \eqref{l:charvalidity:alt}$
  is a direct consequence of the definition of
  validity and of the infimum. For the inverse implication, we observe
  \[
    \ke{\extwo} \app z \app (\sem \cmd_{\mem_1}, \sem \cmdtwo_{\mem_2}) =
    \inf_{\coupled{\mu}{\sem[\star]{\cmd}_{\mem_1}} {\sem[\star]{\cmdtwo}_{\mem_2}}} \E{\mu}{\se{\extwo}\app z}\,.
  \]
  \Cref{lemma:Aguirre3.3} guarantees a coupling that establishes the infimum.
\end{proof}

\section{Proof of Soundness}
\newcommand{\leqc}{\sqsubseteq}
In this section we prove soundness of \ERHLS in the presence of procedure calls.
That is, we prove \eqref{t:procs:sound} in \Cref{t:procs}. We also show soundness of the
Strassen rule, and thereby \Cref{t:soundness}.

With an eye on proofs, we parameterise validity of assertions by a procedure
environment
\begin{align*}
  & \env \valid \RHT{Z}{\ex}{\cmd}{\cmdtwo}{\extwo} \\
  & \qquad \defiff \forall z \in Z, \mem_1,\mem_2 \in \Mem.\
    \exists \starcoupled{\mu}{\sem[\env]{\cmd}_{\mem_1}}{\sem[\env]{\cmdtwo}_{\mem_2}}.\
    \ex \app z \app (\mem_1,\mem_2) \geq \E{\mu}{\se{\extwo} \app z}\,,\\
  & \env \valid[\mu] \RHT{Z}{\exf}{\fn}{\fntwo}{\exftwo} \\
  & \qquad \defiff \forall z \in Z, i_1,i_2 \in \GMem \times \Val.\
    \exists \starcoupled{\mu}{\sem[\env]{\fn}_{i_1}}{\sem[\env]{\fn}_{i_2}}.\ 
    \ex \app z \app (i_1,i_2) \geq \E{\mu}{\se{\extwo} \app z}\,.
\end{align*}
When $\env = \env_\prog$, the environment associated with $\prog$, we may omit $\env$ from the left-hand side;
recovering the expected notion of validity.

We will justify inductive reasoning within \ERHLS through a Scott-inductive argument, in the sense
that validity is a Scott-continuous predicate. This in particular means that we will reason inductively
that validity holds if we interpret procedures by their finite approximations $\env<n>$,
e.g,
\begin{equation}
  \label{eq:approx-sound}
  \tag{Approximated Soundness}
  \derivable \RHT{Z}{\ex}{\cmd}{\cmdtwo}{\extwo}
  \quad\text{ implies }\quad
  \env<n> \valid \RHT{Z}{\ex}{\cmd}{\cmdtwo}{\extwo} \text{ for all $n \in \Nat$.}
\end{equation}
Lifting this result from approximations requires that we provide a 
$\star$-coupling $\mu$ relating the two program statements in their full semantics, in such a way
that the expected value of $\extwo$ 
remains bounded w.r.t.\ this coupling. To this end, we employ two theorems.
The first, a discrete version of \cite[Theorem 5.19]{Villani08}, is used to witness the existence of $\mu$.
Its proof can be found in \cite[Theorem 16]{Aguirre:POPL:21}.
\begin{proposition}[Convergence of Discrete Couplings]\label{p:coupling-conv}
  Let $(d_{1,n})_{n \in \Nat}$ and $(d_{2,n})_{n \in \Nat}$ be two sequence of (sub)distributions
  that converge pointwise to $d_1$ and $d_2$, respectively.
  Let $(\coupled{\mu_i}{d_{1,n}}{d_{2,n}})_{n \in \Nat}$ be a sequence of couplings.
  Then there exists an infinite subsequence of $(\mu_n)_{n \in \Nat}$ that converges to a coupling
  $\coupled{\mu}{d_1}{d_2}$.
\end{proposition}
The theorem speaks about couplings, but extends to $\star$-couplings through the lifting $(\cdot)^\star$.
The second theorem is a discrete version of Fatou's lemma with converging measures and uniform $f$.

\begin{lemma}[Fatou's lemma with converging measures, discrete version]\label{l:fatou}
  Let $(d_n)_{n \in \Nat}$ be a sequence of distributions on $A$
  that converges point-wise to $d$, and
  let $f : A \to \Rpos$ integrable w.r.t.\ all $d_n$, i.e., $\E{d_n}{f} < \infty$.
  Then
  \[
    \E{d}{f} \leq \liminf_{n \to \infty} \E{d_n}{f} \,.
  \]
\end{lemma}

The following lemma, a combination of the previous two, is used to justify our reasoning by approximations.
The premise of the implication will be instantiated by the validity statement on approximations,
the conclusion will then yield validity w.r.t. full semantics. 
\begin{lemma}\label{l:approx-lifting}
  Let $k \in \Rext$ and $f : A \times B \to \Rext$. 
  Let $(d_{1,n})_{n \in \Nat}$ and $(d_{2,n})_{n \in \Nat}$ be two sequence of (sub)distributions
  that converge pointwise to $d_1 \in \Distr{A}$ and $d_2 \in \Distr{B}$, respectively.
  Then
  \[
    \forall n.\ \exists \starcoupled{\mu_n}{d_{1,n}}{d_{2,n}}.\ \E{\mu_n}{\se{f}} \leq k
    \Imp \exists \starcoupled{\mu}{d_1}{d_2}.\ \E{\mu}{\se{f}} \leq k \,.
  \]
\end{lemma}
\begin{proof}
  Suppose that for all $n \in \Nat$ there is $\starcoupled{\mu_n}{d_{1,n}}{d_{2,n}}$
  with $\E{\mu_n}{\se{f}} \leq k$.
  As the distributions $d_{1,n}$ and $d_{2,n}$ are pointwise converging to $d_1$ and $d_2$,
  respectively, their extensions to full distributions 
  $d_{1,n}^\star \in \Distr{A^\star}$ and $d_{2,n}^\star \in \Distr{A^\star}$ are pointwise converging to $d_1^\star$ and $d_2^\star$, respectively.
  \Cref{p:coupling-conv} yields a coupling of
  $d_1^\star$
  and $d_2^\star$, i.e.
  $\starcoupled{\mu}{d_1}{d_2}$,
  as the pointwise limit of a subsequence $(\mu_n)_{n \in N}$ for $N \subseteq \Nat$.

  It now remains to show that the expectation of $f$ is bounded by $k \in \Rext$. 
  If $k = \infty$
  then trivially $\E{\mu}{\se{f}} \leq k$.
  Hence, suppose $k \not=\infty$, and thus
  $\E{\mu_n}{f^\star}  \leq k < \infty$
  for all $n \in N$,
  i.e. $f^\star$ is integrable w.r.t. $\mu_n$.
  \Cref{l:fatou} yields
  \[
    \E{\mu}{f^\star}
    \leq \liminf_{n \to \infty, n \in N} \E{\mu_n}{f^\star} \leq k \,.
  \]
  Here, the last inequality follows as the limit inferior of a sequence uniformly bounded by $k$ is itself bounded by $k$.
\end{proof}

We now proceed to the proof of~\eqref{eq:approx-sound}.
As an auxiliary fact we will need that validity is antitone in the environment.
\begin{lemma}\label{l:valid:antitone}
  For any $\env_1 \leq \env_2$, if $\env_2 \valid \sigma$
  then $\env_1 \valid \sigma$.
\end{lemma}
\begin{proof}
  We show the lemma for $\sigma = \RHT{Z}{\ex}{\cmd}{\cmdtwo}{\extwo}$, the case where $\sigma = \RHT{Z}{\exf}{\fn}{\fntwo}{\exftwo}$ is identical.

  Assume (i)~$\env_1 \leq \env_2$ and (ii)~$\env_2 \valid \RHT{Z}{\ex}{\cmd}{\cmdtwo}{\extwo}$.
  To prove $\env_1 \valid \RHT{Z}{\ex}{\cmd}{\cmdtwo}{\extwo}$, fix $z \in Z$ and $\mem_1,\mem_2 \in Z$.
  By~(ii),
  there exists $\starcoupled{\mu_2}{\sem[\env_2]{\cmd}_{\mem_1}}{\sem[\env_2]{\cmdtwo}_{\mem_2}}$
  and
  \begin{equation}
    \label{l:valid:antitone:1}
    \ex \app z \app (\mem_1,\mem_2) \geq \E{\mu_2}{\se{\extwo} \app z}
    \,.
  \end{equation}
  Now observe that $\sem[\env_1]{\cmd}_{\mem_1}$ is a
  sub-distribution of $\sem[\env_2]{\cmd}_{\mem_1}$, i.e.,
  $\sem[\env_2]{\cmd}_{\mem_1} = \sem[\env_1]{\cmd}_{\mem_1} + d_1$
  and likewise $\sem[\env_2]{\cmdtwo}_{\mem_2} = \sem[\env_1]{\cmdtwo}_{\mem_2} + d_2$
  for some subdistributions $d_i$. This means that $\mu_2$ can be partitioned,
  \[
    \mu_2
    = \mu_{\cmd,\cmdtwo}
    + \mu_{\cmd,{-}}
    + \mu_{{-},\cmdtwo}
    + \mu_{{-},{-}} \,,
  \]
  s.t.
  $\sem[\env_1]{\cmd}_{\mem_1} = \Pi^1(\mu_{\cmd,\cmdtwo}) + \Pi^1(\mu_{\cmd,{-}})$
  and $\sem[\env_1]{\cmdtwo}_{\mem_2} = \Pi^2(\mu_{\cmd,\cmdtwo}) + \Pi^2(\mu_{{-},\cmdtwo})$.
  Let $\mu_{\cmd,\star}$ be obtained by replacing all $(\mem_1',\mem_2')$ in the support of $\mu_{\cmd,{-}}$ by $(\mem_1',\star)$,
  similar, $\mu_{\star,\cmdtwo}$ and $\mu_{\star,\star}$ are obtained this way from $\mu_{{-},\cmdtwo}$, and $\mu_{{-},{-}}$.
  Define now
  \[
    \mu_1
    \defsym
    \mu_{\cmd,\cmdtwo}
    + \mu_{\cmd,\star}
    + \mu_{\star,\cmdtwo}
    + \mu_{\star,\star} \,.
  \]
  By construction, $\starcoupled{\mu_1}{\sem[\env_1]{\cmd}_{\mem_1}}{\sem[\env]{\cmdtwo}_{\mem_1}}$
  and $\E{\mu_2}{\se{\extwo} \app z} \geq \E{\mu_1}{\se{\extwo} \app z}$.
  Using \eqref{l:valid:antitone:1} we
  conclude $\env_1 \valid \RHT{Z}{\ex}{\cmd}{\cmdtwo}{\extwo}$ as desired.
\end{proof}

We lift the semantics of statements to a distribution transformer $\sem{\cmd}_{\cdot} : \Distr(\Mem) \to \Distr(\Mem)$
in the expected way:
\[
  \sem{\cmd}_{d}(\mem) \defsym \sum_{\mathclap{\mem' \in \Mem}} d(\mem') \cdot \sem{\cmd}_{\mem'}(\mem).
\]
As a final preparatory step towards \eqref{eq:approx-sound}, we give a characterisation of validity in terms
of the lifted semantics. This lemma, in turn, will justify the composition Rule~\ref{erhls:Seq}.

\newcommand{\starit}[1]{{#1}^\star}

\begin{lemma}
  \label{l:valid-distr}
  The following are equivalent:
  \begin{enumerate}
    \item\label{l:valid-distr:1} $\env \valid \RHT{Z}{\ex}{\cmd}{\cmdtwo}{\extwo}$;
    \item\label{l:valid-distr:2}
      for every $\starcoupled{\mu}{d_1}{d_2}$
      there exists $\starcoupled{\nu}{\sem[\env]{\cmd}_{d_1}}{\sem[\env]{\cmdtwo}_{d_2}}$
      with
      $\E{\mu}{\se{\ex} \app z} \geq \E{\nu}{\se{\extwo} \app z}$.
    \end{enumerate}
\end{lemma}
\begin{proof}
  The implication $\eqref{l:valid-distr:2} \imp \eqref{l:valid-distr:1}$ is trivial, taking Dirac distributions
  $d_1$ and $d_2$.
  To prove $\eqref{l:valid-distr:1} \imp \eqref{l:valid-distr:2}$,
      Suppose \eqref{l:valid-distr:1}, and fix
      $\starcoupled{\mu}{d_1}{d_2}$ and $z \in Z$.
      By~\eqref{l:valid-distr:1}, for each pair of memories $(\mem_i,\mem_j)$, there exists
      $\starcoupled{\nu_{\mem_i,\mem_j}}{\sem[\env]{\cmd}_{\mem_i}}{\sem[\env]{\cmdtwo}_{\mem_j}}$
      with $\ex \app z \app (\mem_i, \mem_j) \geq \E{\nu_{\mem_i,\mem_j}}{\extwo \app z}$.
      Extend the family of couplings $(\nu_{\mem_1,\mem_2})_{(\mem_1,\mem_2) \in \Mem \times \Mem}$ to
      a family $(\nu_{\mem_1,\mem_2})_{(\mem_1,\mem_2) \in \Mem^\star \times \Mem^\star}$
      by letting
      \[
      \nu_{\mem_1,\star} \defsym (\sem[\env]{\cmd}_{\mem_1})^\star \times \{ \star^1 \};
      \qquad
      \nu_{\star,\mem_2} \defsym \{ \star^1 \} \times (\sem[\env]{\cmdtwo}_{\mem_2})^\star;
      \quad \text{ and } \quad
      \nu_{\star,\star} \defsym \{ (\star,\star)^1 \}.
      \]
      We claim the lemma holds for the distribution $\nu \in \Distr{\Mem^\star \times \Mem^\star}$
      \[
      \nu \defsym
      \sum_{\mathclap{\mem_1,\mem_2 \in \Mem^*}} \mu\app(\mem_1,\mem_2) \cdot \nu_{\mem_1,\mem_2},
      \]
      Note that, by definition, $\nu$ is a full distribution.
      Observe that for $\mem_1 \not = \star$,
      \begin{align*}
        \qquad\quad\Pi_1(\nu)(\mem_1)
        & = \sum_{\mem_2\in \Mem^\star} \nu \app (\mem_1, \mem_2) \\
        & = \sum_{\mem_2\in \Mem^\star}  \sum_{\mem_1',\mem_2' \in \Mem^*} \mu\app(\mem_1',\mem_2') \cdot \nu_{\mem_1',\mem_2'} \app (\mem_1,\mem_2)\\
        & \expl{$\nu_{\star,\star}\app(\mem_1,\mem_2) = \nu_{\star,\mem_2'}\app (\mem_1,\mem_2) = 0$ since $\mem_1 \not= \star$} \\
        & = \sum_{\mem_2\in \Mem^\star} \left(\sum_{\mem_1' \in \Mem} \mu\app(\mem_1',\star) \cdot \nu_{\mem_1',\star}(\mem_1,\mem_2) + \sum_{\mathclap{\mem_1',\mem_2' \in \Mem}} \mu\app(\mem_1',\mem_2') \cdot \nu_{\mem_1',\mem_2'} \app (\mem_1,\mem_2) \right)\\
        & \expl{$\nu_{\mem_1',\star}\app(\mem_1,\mem_2) = 0$ when $\mem_2 \not= \star$} \\
        & = \sum_{\mem_1' \in \Mem} \mu\app(\mem_1',\star) \cdot \nu_{\mem_1',\star}(\mem_1,\star)
          + \sum_{\mathclap{\mem_1',\mem_2' \in \Mem}} \mu\app(\mem_1',\mem_2') \cdot \sum_{\mathclap{\mem_2\in \Mem^\star}} \nu_{\mem_1',\mem_2'} \app (\mem_1,\mem_2) \\
        & \expl{definition of $\nu_{\mem_1',\star}$ and $\starcoupled{\nu_{\mem_1',\mem_2'}}{\sem[\env]{\cmd}_{\mem_1'}}{{\sem[\env]{\cmdtwo}_{\mem_2'}}}$} \\
        & = \sum_{\mem_1' \in \Mem} \mu(\mem_1',\star) \cdot \sem[\env]{\cmd}_{\mem_1'}(\mem_1) + \sum_{\mathclap{\mem_1',\mem_2'\in \Mem}} \mu(\mem_1',\mem_2') \cdot \sem[\env]{\cmd}_{\mem_1'}(\mem_1) \\
        & = \sum_{\mem_1' \in \Mem} \left(\sum_{\mem_2'\in \Mem^\star} \mu(\mem_1',\mem_2') \cdot \sem[\env]{\cmd}_{\mem_1'}(\mem_1)\right) \\
        & = \sum_{\mem_1' \in \Mem} \left(\sum_{\mem_2'\in \Mem^\star} \mu(\mem_1',\mem_2')\right) \cdot \sem[\env]{\cmd}_{\mem_1'}(\mem_1) \\
        & \expl{using $\starcoupled{\mu}{d_1}{d_2}$} \\
        & = \sum_{\mem_1' \in \Mem} d_1(\mem_1') \cdot \sem[\env]{\cmd}_{\mem_1'}(\mem_1)  = \sem[\env]{\cmd}_{d_1}(\mem_1),
      \end{align*}
      and thus, $\Pi_1(\nu) = \left(\sem[\env]{\cmd}_{d_1}\right)^\star$.
      By a similar reasoning, we have $\Pi_2(\nu) = \left(\sem[\env]{\cmdtwo}_{d_2}\right)^\star$ and conclusively $\starcoupled{\nu}{\sem[\env]{\cmd}_{d_1}}{\sem[\env]{\cmdtwo}_{d_2}}$.
      Finally, observe that
      \begin{align*}
        \E{\mu}{\se{\ex} \app z}
        = \sum_{\mathclap{\mem_1,\mem_2 \in \Mem}} \mu\app(\mem_1,\mem_2) \cdot \ex \app z \app (\mem_1,\mem_2)
        \geq \sum_{\mathclap{\mem_1,\mem_2 \in \Mem}} \mu\app(\mem_1,\mem_2) \cdot \E{\nu_{\mem_1,\mem_2}}{\extwo \app z} = \E{\nu}{\se{\extwo} \app z},
      \end{align*}
      where we use $\ex \app z \app (\mem_1,\mem_2) \geq \E{\nu_{\mem_1,\mem_2}}{\extwo \app z}$ by \eqref{l:valid-distr:1}.
      This concludes the case.
\end{proof}

We arrive at the proof of \eqref{eq:approx-sound}.
To be able to reason inductively,
we prove a more general statement that takes a logical context $\ctx$ into account.
Here, $\env<n> \valid \ctx$ means that $\env<n> \valid \RHT{Z}{\exf}{\fn}{\fntwo}{\exftwo}$ for every $(\RHT{Z}{\exf}{\fn}{\fntwo}{\exftwo}) \in \ctx$.
\begin{lemma}\label{l:soundness:approx}
  Suppose $\ctx \derivable \RHT{Z}{\ex}{\cmd}{\cmdtwo}{\extwo}$. Then
  \[
    \env<n> \valid \ctx \Imp \env<n> \valid \RHT{Z}{\ex}{\cmd}{\cmdtwo}{\extwo} \quad\text{ for all $n \in \Nat$. }
  \]
\end{lemma}
\begin{proof}
  The proof is by induction on derivation of $\ctx \derivable \RHT{Z}{\ex}{\cmd}{\cmdtwo}{\extwo}$.
  \begin{proofcases}
    \proofcase{$\ctx \derivable \RHT{Z}{\ex}{\SKIP}{\SKIP}{\ex}$ by Rule~\ref{erhls:Skip}}
    The case is trivial, taking for input memories $\mem_1$ and $\mem_2$ as witness the Dirac identity $\star$-coupling
    $\{(\mem_1,\mem_2)^1\}$
    for all $n \in \Nat$.
    \proofcase{$\ctx \derivable \RHT{Z}{\ex[\lft\vx \mapped \lft\expr,\rght\vy \mapped \rght\exprtwo]}{\vx <- \expr}{\vy <- \exprtwo}{\ex}$ by Rule~\ref{erhls:Asgn}}
    The rule is a special case of Rule~\ref{erhls:Sample}, which we consider next.
    \proofcase{$\ctx \derivable \RHT{Z}{\ex}{\vx <* \sexpr}{\vy <* \sexprtwo}{\extwo}$ by Rule~\ref{erhls:Sample}}
    In the considered case,
    \[
      \ex \app z \app (\mem_1,\mem_2) = \E[(\val_1,\val_2)]{\nu \app (\mem_1,\mem_2)}{\extwo \app z \app (\mem_1[\vx \mapped \val_1],\mem_2[\vy \mapped \val_2])}\,,
    \]
    for couplings $\coupled{\nu \app (\mem_1,\mem_2)}{\sem{\sexpr}_{\mem_1}}{\sem{\sexprtwo}_{\mem_2}}$. 
    Fix $z \in Z$, and $\mem_1,\mem_2 \in \Mem$.
    We define
    \[
      \mu \defsym \dlet{(\val_1,\val_2)}{\nu \app (\mem_1,\mem_2)}{\dunit{(\mem_1[\vx \mapped \val_1],\mem_2[\vx \mapped \val_2])}}\,,
    \]
    thus
    \begin{align*}
      \Pi_1(\mu)
      & = \dlet{\val_1}{\Pi_1(\nu \app (\mem_1,\mem_2))}{\dunit{\mem_1[\vx \mapped \val_1]}} \\
      & = \dlet{\val_1}{\sem{\sexpr}_{\mem_1}}{\dunit{\mem_1[\vx \mapped \val_1]}}
      = \sem{\vx <* \sexpr}_{\mem_1} = \sem[\env<n>]{\vx <* \sexpr}_{\mem_1} \,.
    \end{align*}
    and reasoning identically, $\Pi_2(\mu) = \sem[\env<n>]{\vy <* \sexprtwo}_{\mem_2}$.
    Since these are full distributions, we obtain
    $\starcoupled{\mu}{\sem[\env<n>]{\vx <* \sexpr}_{\mem_1}}{\sem[\env<n>]{\vy <* \sexprtwo}_{\mem_2}}$.
    As by definition
    \[
      \ex \app z \app (\mem_1,\mem_2)
      = \E[(\val_1,\val_2)]{\nu \app (\mem_1,\mem_2)}{\extwo \app z \app (\mem_1[\vx \mapped \val_1],\mem_2[\vy \mapped \val_2])}
      = \E[(\mem_1,\mem_2)]{\mu}{\extwo} \,,
    \]
    we conclude the $\env<n> \valid \RHT{Z}{\ex}{\vx <* \sexpr}{\vy <* \sexprtwo}{\extwo}$ for all $n \in \Nat$,
    independent of validity of the context.
    \proofcase{$\ctx \derivable \RHT{Z}{\ex}{\vx <* \CALL{\fn}(\expr)}{\vy <* \CALL{\fntwo}(\exprtwo)}{\extwo}$, by Rule~\ref{erhls:Call}}
    In the considered case,
    \begin{align*}
     \ex \app z \app (\mem_1,\mem_2) & = \exf \app ((\gmem_1,\sem{\expr}_{\mem_1}),(\gmem_2,\sem{\exprtwo}_{\mem_2}))\,,\text{ and} \\
     \extwo \app z \app (\mem_1,\mem_2) & = \exftwo \app ((\gmem_1,\lmem_1(\vx)),(\gmem_2,\lmem_2(\vy))) \,,
    \end{align*}
    for some $(\RHT{Z}{\exf}{\fn}{\fntwo}{\exftwo}) \in \ctx$.
    We fix $n \in \Nat$ and $\env<n> \valid \ctx$.
    To prove validity
    \[
      \env<n> \valid \RHT{Z}{\ex}{\vx <* \CALL{\fn}(\expr)}{\vy <* \CALL{\fntwo}(\exprtwo)}{\extwo}
      \,,
    \]
    fix $z \in Z$ and $\mem_1,\mem_2 \in \Mem$.
    Observe that as a consequence of $\env<n> \valid \ctx$, we have a $\star$-coupling
    \begin{equation}
      \label{eq:soundness:call}
      \tag{Ctx}
      \starcoupled{\nu}{\sem[\env<n>]{\fn}_{\gmem_1, \sem{\expr}_{\mem_1}}}{\sem[\env<n>]{\fntwo}_{\gmem_2, \sem{\exprtwo}_{\mem_2}}}\text{ s.t. }
      \exf \app z \app ((\gmem_1, \sem{\expr}_{\mem_1}), (\gmem_2, \sem{\expr}_{\mem_2})) \geq \E{\nu}{\se{\exftwo} \app z} \,.
    \end{equation}
    Unfolding the semantics,
    \begin{align*}
      \sem[\env<n>]{\vx <* \CALL{\fn}(\expr)}_{\mem_1} & = \dlet{i_1}{\sem[\env<n>]{\fn}_{\gmem_1, \sem{\expr}_{\mem_1}}}{\dunit{(i_1)_1 \uplus \lmem[\vx \mapped (i_1)_2]}}\,, \\
      \sem[\env<n>]{\vy <* \CALL{\fntwo}(\exprtwo)}_{\mem_2} & = \dlet{i_2}{\sem[\env<n>]{\fn}_{\gmem_2, \sem{\exprtwo}_{\mem_2}}}{\dunit{(i_2)_1 \uplus \lmem[\vy \mapped (i_2)_2]}}\,,
    \end{align*}
    where $(\cdot)_1$ and $(\cdot)_2$ denote the first and second projections on pairs.
    Based on the $\star$-coupling $\mu$, define
    \[
      \mu \defsym \dlet[t]{(i_1,i_2)}{\nu}{
        \dunit{
          (\begin{lalign}[t]
            \IF i_1 = \star \THEN \star \ELSE (i_1)_1 \uplus \lmem_1[\vx \mapped (i_1)_2] \FI,\\
            \IF i_2 = \star \THEN \star \ELSE (i_2)_1 \uplus \lmem_2[\vy \mapped (i_2)_2] \FI)\,.
          \end{lalign}
        }
      }
    \]
    Observe that
    \begin{align*}
      \Pi_1(\mu)
      & =
        \dlet{i_1}{\Pi_1(\nu)}
        {\dunit{ (\IF i_1 = \star \THEN \star \ELSE (i_1)_1 \uplus \lmem_1[\vx \mapped (i_1)_2] \FI) }} \\
      & \expl{by \eqref{eq:soundness:call}} \\
      & =
        \dlet{i_1}{\left(\sem[\env<i>]{\fn}_{\gmem_1, \sem{\expr}_{\mem_1}}\right)^\star}
        {\dunit{ (\IF i_1 = \star \THEN \star \ELSE (i_1)_1 \uplus \lmem_1[\vx \mapped (i_1)_2] \FI) }} \\
      & = \left(\sem[\env<n>]{\vx <* \CALL{\fn}(\expr)}_{\mem_1}\right)^\star.
    \end{align*}
    Reasoning identically,
    $\Pi_2(\mu) = \left(\sem[\env<n>]{\vy <* \CALL{\fntwo}(\exprtwo)}_{\mem_2}\right)^\star$,
    and conclusively,
    \[
      \starcoupled{\mu}{\sem[\env<n>]{\vx <* \CALL{\fn}(\expr)}_{\mem_1}}{\sem[\env<n>]{\vy <* \CALL{\fntwo}(\exprtwo)}_{\mem_2}}\,.
    \]
    The case now essentially follows from \eqref{eq:soundness:call}:
    \begin{align*}
     \ex \app z \app (\mem_1,\mem_2)
      & = \alpha \app z \app ((\gmem_1, \sem{\expr}_{\mem_1}), (\gmem_2, \sem{\exprtwo}_{\mem_2})) \\
      & \geq \E{\nu}{\se{\exftwo} \app z} \\
      & = \sum_{\mem_1,\mem_2 \in \Mem} \nu \app (\mem_1,\mem_2) \cdot \exftwo \app z \app (\mem_1,\mem_2) \\
      & = \sum_{\mem_1,\mem_2 \in \Mem} \mu \app (\mem_1,\mem_2) \cdot \exftwo  \app z \app ((\gmem_1,\lmem_1(\vx)),(\gmem_2,\lmem_2(\vy))) \\
      & = \E{\mu}{\se{\extwo} \app z}\,.
    \end{align*}
    \proofcase{
      $\ctx \derivable
      \RHT{Z}
      {\lft{\bexpr} \leftrightarrow \rght{\bexprtwo} \gd \ex}
      {\IF \bexpr \THEN \cmd_1 \ELSE \cmd_2}
      {\IF \bexprtwo \THEN \cmdtwo_1 \ELSE \cmdtwo_2}
      {\extwo}$ by Rule~\ref{erhls:If}}
    Let $\cmd = \IF \bexpr \THEN \cmd_1 \ELSE \cmd_2$ and $\cmdtwo = \IF \bexprtwo \THEN \cmdtwo_1 \ELSE \cmdtwo_2$. 
    Fix $n \in \Nat$, $\env<n> \valid \ctx$, $z \in Z$ and $\mem_1,\mem_2 \in \Mem$.
    We prove
    $\env<n> \valid \RHT{Z}
      {\lft{\bexpr} \leftrightarrow \rght{\bexprtwo} \gd \ex}
      {\cmd}
      {\cmdtwo}
      {\extwo}$.
    Note that on memories with $\sem{\bexpr}_{\mem_1} \not= \sem{\bexprtwo}_{\mem_2}$ the pre-expectation takes the value $\infty$.
    In this case, any $\star$-coupling of $\sem{\cmd}_{\mem_1}$ and $\sem{\cmdtwo}_{\mem_2}$ (e.g. taking the extension of the product coupling to a $\star$-coupling)
    witnesses validity.

    Hence, suppose $\mem_1,\mem_2$ are such that $\sem{\bexpr}_{\mem_1} = \sem{\bexprtwo}_{\mem_2}$.
    We now consider two cases. In the first, both guards $\bexpr$ and $\bexprtwo$ hold in $\mem_1$ and $\mem_2$, respectively,
    hence
    \[
      \sem[\env<n>]{\cmd}_{\mem_1} = \sem[\env<n>]{\cmd_1}_{\mem_1}
      \quad\text{ and } \quad
      \sem[\env<n>]{\cmdtwo}_{\mem_2} = \sem[\env<n>]{\cmd_2}_{\mem_2}
    \]
    The case now follows by induction hypothesis on the premise $\ctx \derivable \RHT{Z}{\lft{\bexpr} \land \rght{\bexprtwo} \gd \ex}{\cmd_1}{\cmdtwo_1}{\extwo}$,
    using that in the considered case
    \[
      (\lft{\bexpr} \leftrightarrow \rght{\bexprtwo} \gd \ex) \app z \app (\mem_1,\mem_2)
      = \ex \app z \app (\mem_1,\mem_2)
      = (\lft{\bexpr} \land \rght{\bexprtwo} \gd \ex) \app z \app (\mem_1,\mem_2) \,.
    \]
    The remaining case, where both guards fail to hold, is proven
    dual based on the premise
    and $\ctx \derivable \RHT{Z}{\lnot \lft{\bexpr} \land \lnot \rght{\bexprtwo} \gd \ex}{\cmd_2}{\cmdtwo_2}{\extwo}$.

    \proofcase{$\ctx \derivable \RHT{Z}{\ex}{\cmd_1 \sep \cmd_2}{\cmdtwo_1 \sep \cmdtwo_2}{\extwo}$ by Rule~\ref{erhls:Seq}}
    Fix again $n \in \Nat$, $\env<n> \valid \ctx$, $z \in Z$ and $\mem_1,\mem_2 \in \Mem$.
    By induction hypothesis on the first premise of the rule 
    we have $\env<n> \valid \RHT{Z}{\ex}{\cmd_1}{\cmdtwo_1}{\exthree}$, i.e.,
    there exists
    \[
      \starcoupled{\mu}{\sem[\env<n>]{\cmd_1}_{\mem_1}}{\sem[\env<n>]{\cmdtwo_1}_{\mem_2}} \quad\text{ s.t. }\quad
      \ex \app z \app (\mem_1,\mem_2) \geq \E{\mu}{\se{\exthree} \app z} \,.
    \]
    By induction hypothesis on the second premise
    we have $\env<n> \valid \ctx \derivable \RHT{Z}{\exthree}{\cmd_2}{\cmdtwo_2}{\extwo}$ and thus,
    using \Cref{l:valid-distr}, there exists
    \[
      \starcoupled{\nu}{\sem[\env<n>]{\cmd_2}_{d_1}}{\sem[\env<n>]{\cmdtwo_2}_{d_2}} \quad\text{ s.t. }\quad
      \E{\mu}{\se{\exthree} \app z} \geq \E{\nu}{\se{\extwo} \app z} \,,
    \]
    where
    $d_1 = \sem[\env<n>]{\cmd_1}_{\mem_1}$ and
    $d_2 = \sem[\env<n>]{\cmdtwo_1}_{\mem_2}$. Putting the two inequalities together, and using
    that $\sem{\cmd_1\sep\cmd_2}_{\mem_1} = d_1$ and likewise
    $\sem{\cmdtwo_1\sep\cmdtwo_2}_{\mem_2} = d_2$, we conclude
    $\env<n> \valid \RHT{Z}{\ex}{\cmd_1 \sep \cmd_2}{\cmdtwo_1 \sep \cmdtwo_2}{\extwo}$,
    as witnessed by the coupling $\nu$.
    \proofcase{$\ctx \derivable
      \RHT{Z}
      {\lft{\bexpr} \leftrightarrow \rght{\bexprtwo} \gd \ex}
      {\WHILE \bexpr \DO \cmd}
      {\WHILE \bexprtwo \DO \cmdtwo}
      {\neg \lft{\bexpr} \land \neg \rght{\bexprtwo} \gd \ex}$ by Rule~\ref{erhls:While}}
    Fix $n \in \Nat$ and $\env<n> \valid \ctx$.
    We prove, as an auxiliary step,
    \begin{equation}
      \tag{\dag}
      \label{eq:soundness:while}
      \env<n> \valid 
      \RHT{Z}
      {\lft{\bexpr} \leftrightarrow \rght{\bexprtwo} \gd \ex}
      {\WHILE<i> \bexpr \DO \cmd}
      {\WHILE<i> \bexprtwo \DO \cmdtwo}
      {\neg \lft{\bexpr} \land \neg \rght{\bexprtwo} \gd \ex}\,
    \end{equation}
    for all $i \in \Nat$.
    The proof is by induction on $i$.
    The base case $i = 0$ is trivial, hence consider the inductive step
    $i + 1$. As in the case of conditionals above,
    due to the classical pre-condition $\lft{\bexpr} \leftrightarrow \rght{\bexprtwo}$
    we can restrict our attention to $\mem_1,\mem_2$
    for which $\sem{\bexpr}_{\mem_1} = \sem{\bexprtwo}_{\mem_2}$.
    The case where the two guards fail is again trivial. In the remaining case,
    we use the (main) induction hypothesis of the premise of Rule~\ref{erhls:While}
    to obtain
    \[
      \starcoupled{\mu}{\sem[\env<n>]{\cmd}_{\mem_1}}{\sem[\env<n>]{\cmdtwo}_{\mem_2}} \quad\text{ s.t. }\quad
      \ex \app z \app (\mem_1,\mem_2) \geq \E{\mu}{\se{(\lft{\bexpr} \leftrightarrow \rght{\bexprtwo} \gd \ex)} \app z} \,.
    \]
    and the induction hypothesis on $i$ together with \Cref{l:valid-distr} to obtain
    \begin{multline*}
      \starcoupled{\nu}{\sem[\env<n>]{\WHILE<i> \bexpr \DO \cmd}_{d_1}}{\sem[\env<n>]{\WHILE<i> \bexprtwo \DO \cmdtwo}_{d_2}} \quad \text{ s.t. }\\
      \E{\mu}{\se{(\lft{\bexpr} \leftrightarrow \rght{\bexprtwo} \gd \ex)} \app z} \geq \E{\nu}{\se{(\neg \lft{\bexpr} \land \neg \rght{\bexprtwo} \gd \ex)} \app z} \,,
    \end{multline*}
    where
    $d_1 = \sem[\env<n>]{\cmd_1}_{\mem_1}$ and
    $d_2 = \sem[\env<n>]{\cmdtwo_1}_{\mem_2}$.
    As in the considered case, where $\bexpr$ and $\bexprtwo$ are true in $\mem_1$ and $\mem_2$, respectively,
    it follows that $\sem{\WHILE<i+1> \bexpr \DO \cmd}_{\mem_1} = \sem{\WHILE<i> \bexpr \DO \cmd}_{d_1}$
    and similar $\sem{\WHILE<i+1> \bexprtwo \DO \cmdtwo}_{\mem_2} = \sem{\WHILE<i> \bexpr \DO \cmdtwo}_{d_2}$,
    we conclude the claim.

    From the claim, we establish the case now by an application of \Cref{l:approx-lifting},
    using that the approximations
    $\sem{\WHILE<i> \bexpr \DO \cmd}_{\mem_1}$ and $\sem{\WHILE<i> \bexprtwo \DO \cmdtwo}_{\mem_2}$
    converge pointwise to $\sem{\WHILE \bexpr \DO \cmd}_{\mem_1}$ and $\sem{\WHILE \bexprtwo \DO \cmdtwo}_{\mem_2}$, respectively,
    as $i \to \infty$.

    \proofcase{$\ctx \derivable \RHT{Z}{\ex}{\cmd}{\cmdtwo}{\extwo}$ by Rule~\ref{erhls:Conseq}}
    Fix $n \in \Nat$, $\env<n> \valid \ctx$, $z \in Z$ and $\mem_1,\mem_2 \in \Mem$. 
    Induction hypothesis on the premise yields $\env<n> \valid \RHT{Z}{\ex}{\cmd}{\cmdtwo}{\extwo}$, thus, for every $z' \in Z'$, there exists
    \[
      \starcoupled{\mu'}{\sem[\env<n>]{\cmd}_{\mem_1}}{\sem[\env<n>]{\cmdtwo}_{\mem_2}} \quad\text{ s.t. }\quad
      \ex' \app z' \app (\mem_1,\mem_2) \geq \E{\mu'}{\se{\extwo'} \app z'}\,.
    \]
    Using this to discharge the premise in the side-condition of Rule~\ref{erhls:Conseq}
    now proves
    \[
      \starcoupled{\mu}{\sem[\env<n>]{\cmd}_{\mem_1}}{\sem[\env<n>]{\cmdtwo}_{\mem_2}} \quad\text{ s.t. }\quad
      \ex \app z \app (\mem_1,\mem_2) \geq \E{\mu}{\se{\extwo} \app z}\,,
    \]
    establishing thus $\env<n> \valid \RHT{Z}{\ex}{\cmd}{\cmdtwo}{\extwo}$.

    \proofcase{$\ctx \derivable \RHT{Z}{\ex}{\cmd}{\cmdtwo}{\extwo}$ by Rule~\ref{erhls:Ind}}
    Let $\ctxtwo = (\RHT{Z_j}{\exf_j}{\fn_j}{\fntwo_j}{\exftwo_j})_{j=1,\dots,k}$.
    Note the $k$ premises $\ctx,\ctxtwo \derivable \RHT{Z_j}{\exf_j}{\fn_j}{\fntwo_j}{\exftwo_j}$ of Rule~\ref{erhls:Ind} must have been discharged by an application of Rule~\ref{erhls:Proc},
    in particular, for declarations $\fn_j$ and $\fntwo_j$ of the form
    \[
      (\PROC{\fn_j}(\vx)\cmd_j;\RET{\expr_j})
      \text{ and } (\PROC{\fntwo_j}(\vx)\cmdtwo_j;\RET{\exprtwo_j})\,.
    \]
    a derivation of
    \[
      \ctx,\ctxtwo \derivable
      \RHT{Z}
      { \initz \gd \exf_j[\lft{\varg}\mapped\lft{\vx}, \rght{\varg}\mapped\rght{\vx}]}
      {\cmd_j}{\cmdtwo_j}
      {\exftwo_j[\lft{\vres}\mapped\lft{\expr_j}, \rght{\vres}\mapped\rght{\exprtwo_j}]}\,,
    \]
    where $\initz \app (\mem_1,\mem_2) = \forall \vy \in \LVar.\, \vy \neq \vx \imp \mem_1(\vy) = \lmemz(\vy) = \mem_2(\vy)$,
    is part of the proof. The main induction hypothesis from the $j=1,\dots,k$ first derivations states, for all $n \in \Nat$:
    \begin{align}
      \label{eq:soundness:ind-ih}
      \tag{IH${}_j$}
      & \env<n> \valid \ctx,\ctxtwo \imp
      \forall z, \mem_1,\mem_2,\ \initz(\mem_1,\mem_2) \imp \exists \starcoupled{\nu_j}{\sem[\env<n>]{\cmd_j}_{\mem_1}}{\sem[\env<n>]{\cmdtwo_j}_{\mem_2}},\\
      & \qquad
        \notag
        \exf_j \app z \app ((\gmem_1,\mem_1(x)),(\gmem_2,\mem_2(x)))
        \geq \sum_{\mathclap{\mem_1',\mem_2' \in \Mem}} \nu_j \app (\mem_1',\mem_2') \cdot \exftwo_j \app z \app ((\gmem_1',\sem{\expr_j}_{\mem_1'}),(\gmem_2',\sem{\exprtwo_j}_{\mem_1'}))\,.
    \end{align}
    Here, we've tacitly employed validity, and made use of the classical pre-condition.

    To proof the lemma,
    we now prove the intermediate claim
    \begin{equation}
      \label{eq:soundness:ind-claim}
      \tag{Claim}
      \env<n> \valid \ctx \Imp \env<n> \valid \RHT{Z_j}{\exf_j}{\fn_j}{\fntwo_j}{\exftwo_j} \quad \text{ for all $n \in \Nat$ and $j=1,\dots,k$.}
    \end{equation}
    We continue by induction on $n \in \Nat$. 
    \begin{proofcases}
      \proofcase{$0$}
      As $\sem[\env<0>]{\fn}_{(\gmem,\val)} = \dfail$ for any $\fn \in \Op$, the claim is vacuously satisfied.
      \proofcase{$n + 1$}
      Fix a logical context with $\env<n+1> \valid \ctx$, and
      consider a fixed $j=1,\dots,k$.
      To prove $\env<n+1> \valid \RHT{Z_j}{\exf_j}{\fn_j}{\fntwo_j}{\exftwo_j}$,
      we fix moreover $z \in Z_j$, initial global memories $\gmem_1,\gmem_2 \in \GMem$
      and arguments $\val_1,\val_2 \in \GMem$.
      Let $\mem_1 \defsym \gmem_1\uplus\lmemz[\vx \mapped \val_1]$, and $\mem_2 \defsym \gmem_2\uplus\lmemz[\vx \mapped \val_2]$ (thus by definition $\initz(\mem_1,\mem_2)$)
      be the initial memories upon which the bodies $\cmd_j$ and $\cmdtwo_j$ are called,
      upon invocation of $\fn_j$ and $\fntwo_j$ on $(\gmem_1,\val_1)$ and $(\gmem_2,\val_2)$, respectively.

      Note that the assumption $\env<n+1> \valid \ctx$ yields also $\env<n> \valid \ctx$ by \Cref{l:valid:antitone}, this,
      together with the induction hypothesis on $n$
      discharges the first premise $\env<n> \valid \ctx,\ctxtwo$ of \eqref{eq:soundness:ind-ih}.
      By definition,
      \begin{align*}
        \sem[\env<n+1>]{\fn_j}_{\gmem_1,\val_1}
        & = \dlet{\mem_1'}
             {\sem[{\env<n>}]{\cmd_j}_{\mem_1}}
             {\dunit{(\gmem',\sem{\expr_j}_{\mem_1'})}}\,, \\
        \sem[\env<n+1>]{\fntwo_j}_{\gmem_2,\val_2}
        & = \dlet{\mem_2'}
             {\sem[{\env<n>}]{\cmdtwo_j}_{\mem_2}}
             {\dunit{(\gmem',\sem{\exprtwo_j}_{\mem_2'})}}\,.
      \end{align*}
      Instantiating \eqref{eq:soundness:ind-ih} with $z$, $\mem_1$ and $\mem_2$ (for which $\initz(\mem_1,\mem_2)$ holds),
      yields a $\star$-coupling
      \[
        \starcoupled{\nu_j}
        {\sem[{\env<n>}]{\cmd_j}_{\mem_1}}
        {\sem[{\env<n>}]{\cmdtwo_j}_{\mem_2}}\,.
      \]
      Based on $\nu_j$, define the $\star$-coupling
      \[
        \mu \defsym
        \dlet[t]{(\mem_1',\mem_2')}{\nu_j}
        {\dunit{
            (\begin{lalign}[t]
              \IF \mem_1' = \star \THEN \star \ELSE (\gmem_1',\sem{\expr_j}_{\mem_1'}) \FI,\\
              \IF \mem_2' = \star \THEN \star \ELSE (\gmem_2',\sem{\exprtwo_j}_{\mem_2'})\FI)\,.
          \end{lalign}}
        }
      \]
      By definition, $\starcoupled{\mu}{\sem[\env<n+1>]{\fn_j}_{\gmem_1,\val_1}}{\sem[\env<n+1>]{\fntwo_j}_{\gmem_2,\val_2}}$.
      As moreover
      \begin{align*}
        \exf_j \app z \app ((\gmem_1,\val_1),(\gmem_2,\val_2))
        & = \exf_j \app z \app ((\gmem_1,\mem_1(x)),(\gmem_2,\mem_2(x)))\\
        & \geq
          \sum_{\mathclap{\mem_1',\mem_2' \in \Mem}} \nu_j \app (\mem_1',\mem_2') \cdot \exftwo_j \app z \app ((\gmem_1',\sem{\expr_j}_{\mem_1'}),(\gmem_2',\sem{\exprtwo_j}_{\mem_1'}))
        = \E{\mu}{\se{\exftwo_j} \app z}
      \end{align*}
      where the inequality is given by~\eqref{eq:soundness:ind-ih} and the equalities by definition, we conclude the inductive step.
    \end{proofcases}
    Returning to the main proof, the case is now a consequence of the induction hypothesis on $\ctx,\ctxtwo \derivable \RHT{Z}{\ex}{\cmd}{\cmdtwo}{\extwo}$,
    and
    \eqref{eq:soundness:ind-claim} to discharge the additional assumption $\env<n> \valid \ctxtwo$ ($n \in \Nat$).

    \proofcase{$\ctx \derivable \RHT{Z}{\ex}{\cmd}{\cmdtwo}{\extwo}$ via Rule~\ref{erhls:Nmod-L}}
    Fix $n \in \Nat$, $\env<n> \valid \ctx$, $z \in Z$ and $\mem_1,\mem_2 \in \Mem$,
    and let $\vx \not\in \modV{\cmd}$ as given by the side-condition of the rule. 
    By the induction hypothesis on the premise, instantiating $\val$ with $\mem_1(\vx)$, there exists
    $\starcoupled{\mu}{\sem[\env<n>]{\cmd}_{\mem_1}}{\sem[\env<n>]{\cmdtwo}_{\mem_2}}$
    with
    \begin{align*}
      \ex \app z (\mem_1,\mem_2)
      & \geq \E[(\mem_1',\mem_2')]{\mu}{\IF \mem_1' = \star \lor \mem_2' = \star \THEN \star \ELSE \extwo \app z \app (\mem_1'[\vx \mapped \mem_1(\vx)],\mem_2')} \\[1mm]
      & = \E{\mu}{\se{\extwo} \app z}\,,
    \end{align*}
    as desired.
    To see why the equality holds, observe that
    $\mem_1'(\vx) = \mem_1(\vx)$ and thus $\mem_1'[\vx \mapped \mem_1(\vx)] = \mem_1'$ for every
    $\mem_1' \in \supp(\Pi_1(\mu)) \subseteq \supp(\sem[\env<n>]{\cmd}_{\mem_1}) \cup \{\star\}$,
    as $\cmd$ does not modify $\vx$.
  \end{proofcases}
  This concludes the proof of all rules, except the one-sided rules.
  The one-sided rules are proven in a spirit similar to the two-sided rules.
  Rule~\ref{erhls:Asgn-L} and Rule~\ref{erhls:Sample-L} are treated identically to Rule~\ref{erhls:Asgn} and Rule~\ref{erhls:Sample},
  by using the product ($\star$-)coupling of the semantics.
  For Rule~\ref{erhls:Call-L} we exploit that $\SKIP$ is semantically equivalent to $\vx <* \CALL{id}(\vx)$,
  thereby the rule is provable identically to Rule~\ref{erhls:Call}.
  The case of Rule~\ref{erhls:If-L} follows, as in the proof of Rule~\ref{erhls:If}, by case analysis on the guard.
  The case of the one-sided rule for loops is more interesting.
  \begin{proofcases}
    \proofcase{$\ctx \derivable \RHT{Z}{\ex}{\WHILE \bexpr \DO \cmd}{\SKIP}{\lnot \lft{\bexpr} \gd  \ex}$ by Rule~\ref{erhls:While-L}}
    We proceed similar as in the proof for the two-sided rule, and proof the auxiliary
    statement
    \begin{equation}
      \tag{\ddag}
      \label{eq:soundness:while-l}
      \env<n> \valid
      \RHT{Z}
      {\ex}
      {\WHILE<i> \bexpr \DO \cmd}
      {\SKIP}
      {\neg \lft{\bexpr} \gd \ex}
      \text{ for all $i \in \Nat$,}
    \end{equation}
    assuming $\env<n> \valid \ctx$.
    Using that the semantics of $\WHILE<i> \bexpr \DO \cmd$ converges to that of $\WHILE \bexpr \DO \cmd$
    the case is established from this auxiliary statement by \Cref{l:approx-lifting}, as before.
    The proof of the statement is done by induction on $i$, where again, the base case is trivial.
    The inductive step $i+1$ follows by the (outer) induction hypothesis given by the premise,
    $\ctx \derivable \RHT{Z}{\lft{\bexpr} \gd \ex}{\cmd}{\SKIP}{\ex}$,
    and the induction hypothesis on $i$ together with \Cref{l:valid-distr}. 
  \end{proofcases}
  This concludes the final case. 
\end{proof}

The following refers to soundness in the logic with procedure calls, establishing \eqref{t:procs:sound} \Cref{t:procs}.
Validity under a context is defined as expected: $\ctx \valid \RHT{Z}{\ex}{\cmd}{\cmdtwo}{\extwo}$ if
$\valid \ctx$ implies $\valid \RHT{Z}{\ex}{\cmd}{\cmdtwo}{\extwo}$.
\begin{theorem}\label{t:full-sound}
  Any judgment derivable without Rule~\ref{erhls:Strassen} is valid, i.e.,
  \[
    \ctx \derivable \RHT{Z}{\ex}{\cmd}{\cmdtwo}{\extwo}
    \Imp
    \ctx \valid \RHT{Z}{\ex}{\cmd}{\cmdtwo}{\extwo}
    \,.
  \]
\end{theorem}
\begin{proof}
  Suppose $\ctx \derivable \RHT{Z}{\ex}{\cmd}{\cmdtwo}{\extwo}$.
  Consider $\env_\Prog \valid \ctx$, thus by \Cref{l:valid:antitone}
  $\env<n> \valid \ctx$, and consequently
  $\env<n> \valid \RHT{Z}{\ex}{\cmd}{\cmdtwo}{\extwo}$ by 
  \Cref{l:soundness:approx},
  for all $n \in \Nat$.
  Thus, for fixed $z \in Z$ and $\mem_1,\mem_2$, there exist couplings
  \[
    \starcoupled{\mu_n}{\sem[\env<n>]{\cmd}_{\mem_1}}{\sem[\env<n>]{\cmdtwo}_{\mem_2}} \text{ s.t. }
    \ex \app z \app (\mem_1,\mem_2) \geq \E{\mu_n}{\se{\extwo} \app z} \,,
  \]
  for all $n \in \Nat$.
  As the output distributions $\sem[\env<n>]{\cmd}_{\mem_1}$
  and $\sem[\env<n>]{\cmdtwo}_{\mem_2}$ are pointwise converging to $\sem{\cmd}_{\mem_1}$
  and $\sem{\cmdtwo}_{\mem_2}$, respectively,
  \Cref{l:approx-lifting} proves the existence of
  $\starcoupled{\mu}{\sem{\cmd}_{\mem_1}}{\sem{\cmdtwo}_{\mem_2}}$ s.t. $\ex \app z \app (\mem_1,\mem_2)\geq \E{\mu}{\se{f}}$.
\end{proof}

Finally, we consider Soundness of the Strassen rule, in the absence of procedures.
\again{t:soundness}
\begin{proof}
  Re-consider \Cref{l:soundness:approx} for the special case where $\ctx = \varnothing$. The
  proof is given by the induction on the derivation of $\RHT{Z}{\ex}{\cmd}{\cmdtwo}{\extwo}$ and proves
  \[
    \env<n> \valid \RHT{Z}{\ex}{\cmd}{\cmdtwo}{\extwo}
  \]
  for all $n \in \Nat$. As the semantics of both $\cmd$ and $\cmdtwo$ are independent of $\env<n>$,
  the conclusion is equivalent to $\RHT{Z}{\ex}{\cmd}{\cmdtwo}{\extwo}$.
  Carrying out the induction, it is thus sufficient to consider a final case, viz the one where Rule~\ref{erhls:Strassen} is applied:

  \begin{proofcases}
    \proofcase{$\RHT{z\in Z}{\ex}{\cmd}{\cmdtwo}{\cf{\setc\cond_z}}$ by Rule~\ref{erhls:Strassen}}
    Fix $\mem_1,\mem_2 \in \Mem$ and $z \in Z$.
    By assumption and induction hypothesis
    \[
      \valid \RHT{M \subseteq \Mem,\ z\in Z}
      {1 + \ex}
      {\cmd}
      {\cmdtwo}
      {\lft{\cf{M}} + \rght{\cf{\setc{(\cond_z(M))}}}}\,,
    \]
    and thus
    \[
      \forall M \subseteq \Mem,\
      1 + \ex \app z \app (\mem_1,\mem_2)
      \begin{lalign}[t]
        \geq \E[(\mem_1,\mem_2)]{\mu}{\lft{\cf{M}} + \rght{\cf{\setc{(\cond_z(M))}}}} \\
        = \sem{\cmd}_{\mem_1}(M) + \sem{\cmdtwo}_{\mem_2}(\setc{\cond_z(M)}) \\
      \end{lalign}
    \]
    for some \emph{coupling} $\mu$.
    Here, the equality follows by linearity of expectations, the coupling $\coupled{\mu}{\sem{\cmd}_{\mem_1}}{\sem{\cmd}_{\mem_2}}$ (see \Cref{l:coupling-vs-starcoupling})
    and the identity $\E{d}{\cf{\mathcal{E}}} = d(\mathcal{E})$.
    \Cref{p:Strassen} establishes a coupling $\coupled{\nu}{\sem{\cmd}_{\mem_1}}{\sem{\cmd}_{\mem_2}}$,
    thus $\star$-coupling (\Cref{l:coupling-vs-starcoupling})
    \[
      \ex \app z \app (\mem_1,\mem_2)
      \geq \nu(\setc{\cond_z(M)})
      = \E{\nu}{\se{\cf{\setc\cond_z}}}\,
    \]
    concluding validity. \qedhere
  \end{proofcases}
\end{proof}

\section{Proof of Completeness}
In this section we prove our technical completeness result, \Cref{t:completeness}.
As before, we consider procedure calls here and thereby establish \eqref{t:procs:complete} of \Cref{t:procs}.

Throughout the following, we tacitly employ that one-sided validity can be equivalently specializes to:
\begin{align*}
  & \valid \RHT{Z}{\ex}{\cmd}{\SKIP}{\extwo}
  && \Iff \forall z \in Z. \mem_1,\mem_2 \in \Mem.\ \ex \app z \app (\mem_1,\mem_2) \geq \E[\mem_1']{\sem{\cmd}_{\mem_1}}{\extwo \app z \app (\mem_1',\mem_2)} \\
  & \valid \RHT{Z}{\exf}{\fn}{\op{id}}{\exftwo}
  && \Iff \forall z \in Z. i_1,i_2 \in \Mem \times \Val.\ \exf \app z \app (i_1,i_2) \geq \E[i_1']{\sem{\fn}_{i_1}}{\exftwo \app z \app (i_1',i_2)}
\end{align*}

Following \citet{Gorelick1975}, we define a notion of
\emph{most general (left-sided) judgments} $\MGA{\fn}$ for procedures $\fn$, most general in the sense
that any valid judgment is provable from it.
Being proven mutually, we also define a notion of \emph{most general judgment} $\MGA{\cmd}$ for statements $\cmd$.

\begin{definition}[Most General Judgments]
  Given a statement $\cmd$ and procedure $\fn$, their \emph{most general (one-sided) judgments} are defined by
  \begin{align*}
    \MGA{\cmd} & \defsym
    \RHT{\Mem \to \Rext}
      {\lambda h\, (\mem,\_).\ \E{\sem{\cmd}_{\mem}}{h}}
      {\cmd}
      {\SKIP}
      {\lambda h\, (\mem,\_).\ h \app \mem}\,,
    \\
    \MGA{\fn} & \defsym
    \RHT{(\GMem \times \Val) \to \Rext}
      {\lambda h\, (i,\_).\ \E{\sem{\fn}_i}{h}}
      {\cmd}
      {\op{id}}
      {\lambda h\, (i,\_).\ h \app i}\,.
  \end{align*}
\end{definition}
\noindent
Our first lemma states that $\MGA{\cmd}$ is indeed the most general one.
\begin{lemma}\label{l:mga}
  If $\ctx \derivable \MGA{\cmd}$ then
  $\ctx \derivable \RHT{Z}{\ex}{\cmd}{\SKIP}{\extwo}$ for any
  $\ctx \valid \RHT{Z}{\ex}{\cmd}{\SKIP}{\extwo}$.
\end{lemma}
\begin{proof}
  Suppose (i)~$\ctx \derivable \MGA{\cmd}$ and
  (ii)~$\ctx \valid \RHT{Z}{\ex}{\cmd}{\SKIP}{\extwo}$.
  We conclude by one application
  of Rule~\ref{erhls:Conseq},
  using (i)~to fulfill its premise.
  To discharge the side-condition of the rule,
  fix $\mem_1,\mem_2 \in \Mem$ and $d_1,d_2 \in \Distr{\Mem}$.
  For the particular instantiation of the rule where we consider only one-sided
  judgments, the side-condition reads as
  \[
    (\forall h : \Mem \to \Rext.\ \E{\sem{\cmd}_{\mem_1}}{h} \geq \E{d_1}{h})
    \imp
    (\forall z \in Z.\ \ex \app z \app (\mem_1,\mem_2) \geq \E[\mem_1']{d_1}{\extwo \app z \app (\mem_1',\mem_2)})\,.
  \]
  To discharge it, fix $z \in Z$.
  Using validity~(ii), and instantiating $h$ with $\lambda \mem_1'.\ \extwo \app z \app (\mem_1',\mem_2)$
  in the premise
  \begin{align*}
    \ex \app z \app (\mem_1,\mem_2)
     \geq  \E[\mem_1']{\sem{\cmd}_{\mem_1}}{\extwo \app z \app (\mem_1',\mem_2)}
     \geq \E[\mem_1']{d_1}{\extwo \app z \app (\mem_1',\mem_2)}\,.
  \end{align*}
\end{proof}

The central proof step towards one-sided completeness now lies in showing that
$\MGA{\cmd}$ is derivable.
\begin{lemma}\label{l:mga-derivable:aux}
  $(\MGA{\fn})_{\fn \in \Op} \derivable \MGA{\rho}$, for any program statement or procedure $\rho$.
\end{lemma}
\begin{proof}
  We consider first the case where $\rho$ is a statement.
  Let $\ctxtwo \defsym (\MGA{\fn})_{\fn \in \Op}$, we
  have to prove
  \[
    \ctx \valid \RHT{\Mem \to \Rext}
      {\lambda h\, (\mem,\_).\ \E{\sem{\cmd}_{\mem}}{h}}
      {\cmd}
      {\SKIP}
      {\lambda h\, (\mem,\_).\ h \app \mem}\,.
    \]
    The proof is by induction on $\cmd$.
  \begin{proofcases}
    \proofcase{$\SKIP$ and $\vx <- \expr$ and $\vx <* \sexpr$}
    The judgment is an instance of Rule~\ref{erhls:Skip}, Rule~\ref{erhls:Asgn-L} and
    Rule~\ref{erhls:Sample-L}, respectively.
    \proofcase{$\vx <- \CALL\fn(\expr)$}
		\newcommand{\lvs}{\vec{l}}
		\newcommand{\gvs}{\vec{g}}
		For a memory $\mem$ and variables $\vec{\vy}$, let
		$\mem_{\vec{\vy}}$ denote
		the projection of $\mem$ to $\vec{\vy}$.
		Let $\lvs$ and $\gvs$ denote a sequence of local and global variables,
		respectively, without $\vx$,
		thus in particular every $\mem \in \Mem$ can be written as
		$\mem_{\gvs} \uplus \{ \vx \mapped \mem(\vx) \} \uplus \mem_{\lvs}$.
		Using this notation, unfolding the definition of $\sem{\vx <* \CALL\fn(\expr)}$, we have to show
    \begin{flalign*}
      \qquad
      \ctxtwo \derivable
      \RHT[t]
      {\Mem \to \Rext}
      {\lambda h\,(\mem,\_).\
        \E
        {\sem{\fn}_{\gmem,\sem{\expr}_{\mem}}}
        {\lambda (\gmem',\vres).\ h \app (\mem'_{\gvs} \uplus \{ \vx \mapped \vres \} \uplus \mem_{\lvs})}}
      {\vx <* \CALL\fn(\expr)}
      {\SKIP}
      {\lambda h\,(\mem,\_).\ h \app (\mem_{\gvs} \uplus \{ \vx \mapped \mem(\vx) \} \uplus \mem_{\lvs})}
      &&
    \end{flalign*}
    which, using $|\lvs|$ applications of Rule~\ref{erhls:Nmod-L} and its symmetric version, translates to
    \begin{flalign*}
      \qquad
      \ctxtwo \derivable
      \RHT[t]
      {\Val^{|\lvs|}\times (\Mem \to \Rext)}
      {\lambda (\vec\val,h)\,(\mem,\_).\
       \lvs = \vec\val \gd
        \E
        {\sem{\fn}_{\gmem,\sem{\expr}_{\mem}}}
        {\lambda (\mem',\vres).\ h \app (\mem'_{\gvs} \uplus \{ \vx \mapped \vres \} \uplus \{\lvs\mapped\vec\val\})}}
      {\vx <* \CALL\fn(\expr)}
      {\SKIP}
      {\lambda (\vec\val,h)\,(\mem,\_).\ h \app (\mem_{\gvs} \uplus \{ \vx \mapped \mem(\vx) \} \uplus \{\lvs\mapped\vec\val\})}
      &&
    \end{flalign*}
    and to
    \begin{flalign}
      \label{l:mga-derivable:fn:C}
      \tag{C}
      \qquad
      \ctxtwo \derivable
      \RHT[t]
      {\Val^{|\lvs|}\times (\Mem \to \Rext)}
      {\lambda (\vec\val,h)\,(\mem,\_).\
       \E
        {\sem{\fn}_{\gmem,\sem{\expr}_{\mem}}}
        {\lambda (\gmem',\vres).\ h \app (\mem'_{\gvs} \uplus \{ \vx \mapped \vres \} \uplus \{\lvs\mapped\vec\val\})}}
      {\vx <* \CALL\fn(\expr)}
      {\SKIP}
      {\lambda (\vec\val,h)\,(\mem,\_).\ h \app (\mem_{\gvs} \uplus \{ \vx \mapped \mem(\vx) \} \uplus \{\lvs\mapped\vec\val\})}
      &&
    \end{flalign}
    by weakening the pre-expectation via Rule~\ref{erhls:Weaken}.

    On the other hand, using $\MGA{\fn} \in \ctxtwo$,
    one application of Rule~\ref{erhls:Call-L} proves
    \begin{flalign}
      \label{l:mga-derivable:fn:H}
      \tag{H}
      \qquad
      \ctxtwo \derivable
      \RHT[t]
      {(\GMem\times\Val) \to \Rext}
      {\lambda f\,(\mem,\_).\
        \E{\sem{\fn}_{\gmem,\sem{\expr}_{\mem}}}{f}}
      {\vx <* \CALL\fn(\expr)}{\SKIP}
      {\lambda f\,(\mem,\_).\ f \app (\gmem,\mem(\vx))}
      &&
    \end{flalign}
    Notice how by substituting the function
    $\lambda (\gmem',\val). h \app (\mem'_{\gvs} \uplus \{ \vx \mapped \mem(\vx) \} \uplus \{\lvs\mapped\vec\val\}$ for $f$,
    our goal~\eqref{l:mga-derivable:fn:C} becomes an instance of~\eqref{l:mga-derivable:fn:H}.
    Indeed, using this substitution, one application of the derived Rule~\ref{erhls:Inst}
    finishes the case.
    \proofcase{$\cmd ; \cmdtwo$}
    By induction hypothesis on $\cmdtwo$, we have
    \begin{flalign*}
      \qquad
      \ctxtwo \derivable
      \RHT{\Mem \to \Rext}
      {\lambda h\, (\mem,\_).\ \E{\sem{\cmdtwo}_{\mem}}{h}}
      {\cmdtwo}
      {\SKIP}
      {\lambda h\, (\mem,\_).\ h \app \mem}. &&
    \end{flalign*}
    Instantiating $h$ with $\lambda \mem. \E{\sem{\cmdtwo}_{\mem}}{h}$ for $h$ via Rule~\ref{erhls:Inst}
    in the induction hypothesis on $\cmd$
    yields
    \begin{flalign*}
      \qquad
      \ctxtwo \derivable
      \RHT{\Mem \to \Rext}
      {\lambda h\, (\mem,\_).\ \E{\sem{\cmd}_{\mem}}{\lambda \mem'. \E{\sem{\cmdtwo}_{\mem'}}{h}}}
      {\cmd}
      {\SKIP}
      {\lambda h\, (\mem,\_).\ \E{\sem{\cmdtwo}_{\mem}}{h}}, &&
    \end{flalign*}
    Using one application of Rule~\ref{erhls:Seq} proves now
    \begin{flalign*}
      \qquad
      \ctxtwo \derivable
      \RHT{\Mem \to \Rext}
      {\lambda h\, (\mem,\_).\ \E{\sem{\cmd}_{\mem}}{\lambda \mem'. \E{\sem{\cmdtwo}_{\mem'}}{h}}}
      {\cmd \sep \cmdtwo}
      {\SKIP \sep \SKIP}
      {\lambda h\, (\mem,\_).\ h \app \mem}. &&
    \end{flalign*}
    Note that in the pre-expectation, the term $\E{\sem{\cmd}_{\mem}}{\lambda \mem'. \E{\sem{\cmdtwo}_{\mem'}}{h}}$ just corresponds to
    $\E{\sem{\cmd;\cmdtwo}_{\mem}}{h}$
    (see~\Cref{fig:evt}), as $\SKIP \sep \SKIP = \SKIP$ the claim follows.
    \proofcase{$\IF \bexpr \THEN \cmd \ELSE \cmdtwo$}
    Using
    \begin{align*}
      (\sem{\bexpr}_\mem \gd \E{\sem{\IF \bexpr \THEN \cmd \ELSE \cmdtwo}_{\mem}}{h}) & \geq \E{\sem{\cmd}_{\mem}}{h}
      & (\sem{\lnot \bexpr}_\mem \gd \E{\sem{\IF \bexpr \THEN \cmd \ELSE \cmdtwo}_{\mem}}{h}) & \geq \E{\sem{\cmdtwo}_{\mem}}{h}
    \end{align*}
    the induction hypothesis on $\cmd$ and $\cmdtwo$, together with Rule~\ref{erhls:Weaken},
    allows us to derive
    \begin{flalign*}
      \qquad
      & \ctxtwo \derivable \RHT{\Mem \to \Rext}
      {\lambda h\, (\mem,\_).\ \sem{\bexpr}_\mem \gd \E{\sem{\IF \bexpr \THEN \cmd \ELSE \cmdtwo}_{\mem}}{h}}
      {\cmd}
      {\SKIP}
      {\lambda h\, (\mem,\_).\ h \app \mem}\,\text{, and}
      && \\
      & \ctxtwo \derivable \RHT{\Mem \to \Rext}
      {\lambda h\, (\mem,\_).\ \sem{\lnot \bexpr}_\mem \gd \E{\sem{\IF \bexpr \THEN \cmd \ELSE \cmdtwo}_{\mem}}{h}}
      {\cmdtwo}
      {\SKIP}
      {\lambda h\, (\mem,\_).\ h \app \mem}\,,
    \end{flalign*}
    the case then follows by an application of Rule~\ref{erhls:If-L}. 
    \proofcase{$\WHILE \bexpr \DO \cmd$}
    Instantiating $h$ in the induction hypothesis with $\E{\sem{\WHILE \bexpr \DO \cmd}_{(\cdot)}}{h}$
    through the derived Rule~\ref{erhls:Inst}, and
    using the identity
    $\E[\mem']{\sem{\cmd}_{\mem}}{\E{\sem{\WHILE \bexpr \DO \cmd}_{\mem'}}{h}}
      = \E{\sem{\cmd;\WHILE \bexpr \DO \cmd}_{\mem}}{h}$
      (see \Cref{{fig:evt}}) we have
    \begin{flalign*}
      \qquad
      \ctxtwo \derivable
      \RHT{\Mem \to \Rext}
      {\lambda h\, (\mem,\_).\ \E{\sem{\cmd;\WHILE \bexpr \DO \cmd}_{\mem}}{h}}
      {\cmd}
      {\SKIP}
      {\lambda h\, (\mem,\_).\ \E{\sem{\WHILE \bexpr \DO \cmd}_{\mem}}{h}}. &&
    \end{flalign*}
    which, weakens to
    \begin{flalign*}
      \qquad
      \ctxtwo \derivable
      \RHT{\Mem \to \Rext}
      {\lambda h\, (\mem,\_).\ \sem{\bexpr}_\mem \gd \E{\sem{\WHILE \bexpr \DO \cmd}_{\mem}}{h}}
      {\cmd}
      {\SKIP}
      {\lambda h\, (\mem,\_).\ \E{\sem{\WHILE \bexpr \DO \cmd}_{\mem}}{h}}. &&
    \end{flalign*}
    with Rule~\ref{erhls:Weaken}, using the identity
    $\E{\sem{\WHILE \bexpr \DO \cmd}_{\mem}}{h} = \E{\sem{\cmd;\WHILE \bexpr \DO \cmd}_{\mem}}{h}$
    for memories $\mem$ on which the guard $\bexpr$ evaluates to true. 
    By an application of Rule~\ref{erhls:While-L}, we thus have
    \begin{flalign*}
      \qquad
      \ctxtwo \derivable
      \RHT[t]{\Mem \to \Rext}
      {\lambda h\, (\mem,\_).\ \E{\sem{\WHILE \bexpr \DO \cmd}_{\mem}}{h}}
      {\WHILE \bexpr \DO \cmd}
      {\SKIP}
      {\lambda h\, (\mem,\_).\ \sem{\lnot \bexpr}_\mem \gd \E{\sem{\WHILE \bexpr \DO \cmd}_{\mem}}{h}} &&
    \end{flalign*}
    Using that $\E{\sem{\WHILE \bexpr \DO \cmd}_{\mem}}{h} = h \app \mem$ when
    guard $\bexpr$ evaluates to false in $\mem$, we finally conclude
    \begin{flalign*}
      \qquad
      \ctxtwo \derivable
      \RHT[t]{\Mem \to \Rext}
      {\lambda h\, (\mem,\_).\ \E{\sem{\WHILE \bexpr \DO \cmd}_{\mem}}{h}}
      {\WHILE \bexpr \DO \cmd}
      {\SKIP}
      {\lambda h\, (\mem,\_).\ h \app \mem} &&
    \end{flalign*}
    with another application
    of Rule~\ref{erhls:Weaken}.
  \end{proofcases}
  \medskip
  
  This establishes the lemma for statements. For the remaining case, where $\rho$ is a procedure,
  fix $(\PROC{\fn}(\vx)\cmdtwo;\RET{\expr}) \in \prog$.
  We prove $\ctxtwo \derivable \MGA{\fn}$.
  To this end, we first apply Rule~\ref{erhls:Proc},
  leaving us with the goal
  \begin{flalign*}
    \qquad
    \ctxtwo \derivable
    \RHT[t]{(\GMem \times \Val) \to \Rext}
    {\lambda h\,(\mem_1,\mem_2).\ \initz \app (\mem_1,\mem_2) \gd \E{\sem{\fn}_{\gmem_1,\mem_1(\vx)}}{h}}
    {\cmdtwo}{\SKIP}
    {\lambda h\,(\mem_1,\_).\ h \app (\gmem_1,\sem{\expr}_{\mem_1})}
    &&
  \end{flalign*}
  where $\initz \app (\mem_1,\mem_2) = \forall \vy \in \LVar.\, \vy \neq \vx \imp \mem_1(\vy) = \lmemz(\vy) = \mem_2(\vy)$.
  Assume $\mem_1$ is such that $\initz \app (\mem_1,\mem_2)$ holds.
  Then $\mem_1 = \gmem_1 \uplus \lmemz$,
  and consequently,
  \begin{align*}
    \E{\sem{\fn}_{\gmem_1,\mem_1(\vx)}}{h}
    & = \E{\dlet{\mem_1'}
      {(\sem{\cmdtwo}_{\gmem_1\uplus\lmemz_1[\vx \mapped \mem_1(\vx)]})}
      {\dunit{(\gmem_1',\sem{\expr}_{\mem_1'})}}}{h}
    = \E[\mem_1']{\sem{\cmdtwo}_{\mem_1}}{h \app (\gmem_1',\sem{\expr}_{\mem_1'})},
  \end{align*}
  by unfolding the definition of $\sem{\fn}$, and employing \Cref{p:sem-props}.
  Our goal is thus equivalent to
  \begin{flalign*}
    \qquad
    \ctxtwo \derivable
    \RHT[t]{(\GMem \times \Val) \to \Rext}
    {\lambda h\,(\mem_1,\mem_2).\ \initz \app (\mem_1,\mem_2) \gd \E[\mem_1']{\sem{\cmdtwo}_{\mem_1}}{h \app (\gmem_1',\sem{\expr}_{\mem_1'})}}
    {\cmdtwo}{\SKIP}
    {\lambda h\,(\mem_1,\_).\ h \app (\gmem_1,\sem{\expr}_{\mem_1})}
    &&
  \end{flalign*}
  Notice that this assertion is valid, by definition.
  As we know already that $\ctxtwo \derivable \MGA{\cmdtwo}$,
  one application of \Cref{l:mga} now proves the claim.
\end{proof}

By an application of Rule~\ref{erhls:Ind}, we therefore have $\derivable \MGA{\cmd}$
for any program statement $\cmd$:
\begin{lemma}\label{l:mga-derivable}
  $\derivable \MGA{\cmd}$.
\end{lemma}
\begin{proof}
  Based on the preparatory \Cref{l:mga-derivable:aux},
  using Rule~\ref{erhls:Ind} with $\ctxtwo \defsym (\MGA{\fn})_{\fn \in \Op}$
  the lemma follows.
\end{proof}

Since $\MGA{\cmd}$ is most general, complete w.r.t. to the one-sided rules is now immediate.
\begin{lemma}\label{l:complete-onesided}
  If $\valid \RHT{Z}{\ex}{\cmd}{\SKIP}{\extwo}$ then $\derivable \RHT{Z}{\ex}{\cmd}{\SKIP}{\extwo}$.
\end{lemma}
\begin{proof}
  An immediate consequence of \Cref{l:mga} and \Cref{l:mga-derivable}.
\end{proof}
Note that through symmetry of the logic,
one-sided completeness extends to the dual assertions $\valid \RHT{Z}{\ex}{\SKIP}{\cmd}{\extwo}$.
This then almost immediately yields our main completeness result.
By symmetry, a corresponding theorem holds naturally also for the dual right-sided judgments.
\again{l:completeness}
\begin{proof}
  Observe that
  \[
    \E{\sem{\cmd}_{\mem_1} \times \sem{\cmdtwo}_{\mem_2}}{\ex \app z}
    = \E[\mem_1']{\sem{\cmd}_{\mem_1}}{
        \E[\mem_2']{\sem{\cmdtwo}_{\mem_2}}{\ex \app z \app (\mem_1',\mem_2')}\,.
    }
  \]
  Through this observation,
  the theorem is equivalent to showing
  \begin{flalign*}
    \qquad
    \derivable
    \RHT{Z}
    {\lambda z\,(\mem_1,\mem_2).\
     \E[\mem_1']{\sem{\cmd}_{\mem_1}}{
      \E[\mem_2']{\sem{\cmdtwo}_{\mem_2}}{\ex \app z \app (\mem_1',\mem_2')}}
    }
    {\cmd \sep \SKIP}
    {\SKIP \sep \cmdtwo}
    {\ex}.
    &&
  \end{flalign*}
  Applying now Rule~\ref{erhls:Seq}, using
  $\lambda z (\mem_1,\mem_2).\ \E[\mem_2']{\sem{\cmdtwo}_{\mem_2}}{\ex \app z \app (\mem_1,\mem_2')}$
  as intermediate pre/post-expectation, resulting in the two new sub-goals
  \begin{flalign*}
    \qquad & \derivable
      \RHT[t]{Z}
      {\lambda z\,(\mem_1,\mem_2).\ \E[\mem_1']{\sem{\cmd}_{\mem_1}}{\E[\mem_2']{\sem{\cmdtwo}_{\mem_2}}{\ex \app z \app (\mem_1',\mem_2')}}}
      {\cmd}
      {\SKIP}
      {\lambda z\,(\mem_1,\mem_2).\ \E[\mem_2']{\sem{\cmdtwo}_{\mem_2}}{\ex \app z \app (\mem_1,\mem_2')}}
    &&
  \end{flalign*}
  and
  \begin{flalign*}
    \qquad
    & \derivable
      \RHT{Z}
      {\lambda z\,(\mem_1,\mem_2).\ \E[\mem_2']{\sem{\cmdtwo}_{\mem_2}}{\ex \app z \app (\mem_1,\mem_2')}}
      {\SKIP}
      {\cmdtwo}
      {\ex}.
    &&
  \end{flalign*}
  As both are valid by definition, completeness of the logic for one-sided assertions
    (\Cref{l:complete-onesided}) now finishes the proof.
\end{proof}

\again{t:completeness}
\begin{proof}
  We prove the following three circular implications.
  \begin{description}
    \item[$\eqref{t:completeness:derivable} \imp \eqref{t:completeness:valid}$:]
      This is \Cref{t:soundness}.
    \item[$\eqref{t:completeness:valid} \imp \eqref{t:completeness:form}$:]
      Suppose
      $\valid
      \RHT{Z}
      {\ex}
      {\cmd}
      {\cmdtwo}
       {\lft{\extwo} + \rght{\exthree}}$, and fix $z \in Z$ and $\mem_1,\mem_2 \in \Mem$. By definition of validity, there exists
      $\starcoupled{\mu}{\sem{\cmd}_{\mem_1}}{\sem{\cmdtwo}_{\mem_2}}$ with
      \begin{align*}
        \ex \app z \app (\mem_1,\mem_2)
        & \geq \E{\mu}{\lft{\extwo} \app z \app \oplus \rght{\exthree} \app z} \\
        & \expl{losslessness} \\
        & \geq \E[(\mem_1',\mem_2')]{\mu}{\extwo \app z \app \mem_1' \oplus \exthree \app z \app \mem_2'} \\
        & \expl{commutativity property} \\
        & = \E[(\mem_1',\mem_2')]{\mu}{\extwo \app z \app \mem_1'}
          \oplus \E[(\mem_1',\mem_2')]{\mu}{\exthree \app z \app \mem_2'} \\
        & \expl{coupling} \\
        & =  \E[\mem_1']{\sem{\cmd}_{\mem_1}}{\extwo \app z \app \mem_1'}
          \oplus \E[\mem_2']{{\sem{\cmdtwo}_{\mem_1}}}{\exthree \app z \app \mem_2'},
      \end{align*}
      proving the claim.
    \item[$\eqref{t:completeness:form} \imp \eqref{t:completeness:derivable}$:]
      \Cref{l:completeness} and weakening of the pre-expectation
      $\eqref{t:completeness:form}$ via the derived Rule~\ref{erhls:Weaken}.
      \qedhere
  \end{description}
\end{proof}

\section{Soundess Proofs for Adversary Rules}

This section is devoted to showing that the adversary rules are sound.
\newcommand{\LVA}{{\LVar_{\adv}}}
\newcommand{\GVA}{{\GVar_{\adv}}}
\newcommand{\VA}{{\Var_{\adv}}}
We will prove soundness of the rules itself with \ERHL.
Throughout the following, we fix an adversary $\adv$ and an arbitrary environment with
\[
\advenv \app \adv = \PROC{\adv}(\vx) \cmd;\RET{\expr} \,.
\]
We let $\cmd_{\orcl}$ be obtained by substituting the oracle meta-variables $\vorcl$ in
$\cmd$ with $\orcl$. This way,
an instantiated adversary $\adv[\orcl]$ can be equivalently defined as
\[
  \PROC{\adv[\orcl]}(\vx) \cmd_{\orcl};\RET{\expr} \,.
\]

The first lemma states that a one-sided invariant independent on writable variables stays so,
provided invariance is preserved by the oracle.
This gives the justification for the one-sided adversary Rule~\ref{erhls:Adv-L}.
\begin{lemma}\label{l:adv-oneside}
  Let $\cmd$ be the body of an adversary $\adv$, and suppose
  $\ctx \derivable \RHT{Z}{\exftwo}{\orcl}{\op{id}}{\exftwo}$,
  where $\exftwo \perp \writeV{\adv}$ and
  $\exftwo \perp \{ \lft{\varg},\rght{\varg},\lft{\vres},\rght{\vres}\}$.
  Then
  \[
    \ctx \derivable \RHT{Z}{\exftwo}{\cmd_{\orcl}}{\SKIP}{\exftwo} \,.
  \]
\end{lemma}
\begin{proof}
  The proof is by induction on $\cmd$. All cases are standard. In the case of an
  oracle call, the assumption $\ctx \derivable \RHT{Z}{\exftwo}{\orcl}{\op{id}}{\exftwo}$ is used. 
\end{proof}

In the proof below, we perform case analysis
using the derived rule,
\[
  \Infer[erhls][Case]
  {\ctx \derivable \RHT{Z}{\ex}{\cmd}{\cmdtwo}{\extwo}}
  {
    \ctx \derivable \RHT{b \in \{0,1\}, Z}{\bexpr = b \gd \ex}{\cmd}{\cmdtwo}{\extwo}
  }
\]
which is directly obtained from Rule~\ref{erhls:Conseq}. To avoid syntactic overhead,
we may also conceive this as a binary rule
\[
  \Infer*[erhls][Case]
  {\ctx \derivable \RHT{Z}{\ex}{\cmd}{\cmdtwo}{\extwo}}
  {
    \ctx \derivable \RHT{Z}{\bexpr \gd \ex}{\cmd}{\cmdtwo}{\extwo}
    &     \ctx \derivable \RHT{Z}{\lnot \bexpr \gd \ex}{\cmd}{\cmdtwo}{\extwo}
  }
\]

The second technical lemma is the two-sided version of \Cref{l:adv-twoside}, serving as a
justification for the two-sided adversary Rule~\ref{erhls:Adv}.
\begin{lemma}\label{l:adv-twoside}
  Let $\cmd$ be the body of an adversary $\adv$, and suppose all three premises of Rule~\ref{erhls:Adv}
  have been derived:
  \begin{enumerate}
    \item\label{l:adv-twoside:1} $\ctx \derivable \RHT{Z}{\neg \BAD \land \eqmem{\varg,\GVar_{\adv}} \gd \exf}{\orcl}{\orcltwo}{\exfthree_\vres}$;
    \item $\ctx \derivable \RHT{Z}{\BAD \gd \exftwo}{\orcl}{\op{id}}{\BAD \gd \exftwo}$; and
    \item $\ctx \derivable \RHT{Z}{\BAD \gd \exftwo}{\op{id}}{\orcltwo}{\BAD \gd \exftwo}$;
  \end{enumerate}
  where $\exf,\exftwo,\BAD \perp \writeV{\adv} \cup \{ \lft{\varg},\rght{\varg},\lft{\vres},\rght{\vres} \}$,
  and $\exfthree_V \defsym \IF \BAD \THEN \exftwo \ELSE (\eqmem{V,\GVar_{\adv}} \gd \exf)$.
  Then
  \[
    \ctx \derivable \RHT{Z}{\exfthree_\LVA}{\cmd_{\orcl}}{\cmd_{\orcltwo}}{\exfthree_\LVA} \,,
  \]
\end{lemma}
\begin{proof}
  Using the identities $(\BAD \gd \exfthree_\LVA) = (\BAD \gd \exftwo)$,
  \Cref{l:adv-oneside} yields
  \[
    \ctx \derivable \RHT{Z}{\BAD \gd \exfthree_\LVA}{\cmd_{\orcl}}{\SKIP}{\BAD \gd \exfthree_\LVA}
    \text{ and dually }
    \ctx \derivable \RHT{Z}{\BAD \gd \exfthree_\LVA}{\SKIP}{\cmd_{\orcltwo}}{\BAD \gd \exfthree_\LVA}\,,
  \]
  and thus
  \[
    \ctx \derivable \RHT{Z}{\BAD \gd \exfthree_\LVA}{\cmd_{\orcl}}{\cmd_{\orcltwo}}{\BAD \gd \exfthree_\LVA}\,
  \]
  via Rule~\ref{erhls:Seq},
  and hence
  \[
    \ctx \derivable \RHT{Z}{\BAD \gd \exfthree_\LVA}{\cmd_{\orcl}}{\cmd_{\orcltwo}}{\exfthree_\LVA}\,
  \]
  by weakening the post-expectation via Rule~\ref{erhls:Weaken}, using
  that $\BAD \gd \exfthree_\LVA \geq \exfthree_\LVA$.
  
  Using the identity $(\lnot \BAD \gd \exfthree_\LVA) = (\lnot \BAD \land \eqmem{\VA} \gd \exf)$, by the derived
  Rule~\ref{erhls:Case} it remains to verify that
  \[
    \ctx \derivable \RHT{Z}{\lnot \BAD \land \eqmem{\VA} \gd \exf}{\cmd_{\orcl}}{\cmd_{\orcltwo}}{\exfthree_\LVA} \,.
  \]
  The proof is by induction on the statement $\cmd$.
  \begin{proofcases}
    \proofcase{$\SKIP$} This case follows directly from Rule~\ref{erhls:Skip}.
    \proofcase{$\vx <- \expr$ and $\vx <* \sexpr$}
    We show the more involved case of a sampling $\vx <* \sexpr$, the case $\vx <- \expr$ follows by
    identical reasoning via Rule~\ref{erhls:Asgn}.
    Using Rule~\ref{erhls:Sample} we derive
    \[
      \ctx \derivable
      \RHT{Z}
      {\E[(\val_1,\val_2)]{\mu}{\exfthree_\LVA[\lft\vx \mapped \val_1,\rght\vx \mapped \val_2]}}
      {\vx <* \sexpr}
      {\vx <* \sexpr}
      {\exfthree_\LVA}\,
    \]
    where as coupling $\mu$ (parametric in the two memories), we use the identity coupling on memories
    that agree on $\VA$, and otherwise an arbitrary one. From here, we conclude by an application of Rule~\ref{erhls:Weaken}, using that
    \begin{align*}
      (\lnot \BAD \land \eqmem{\VA} \gd \exf)
      & = (\lnot \BAD \land \eqmem{\VA} \gd \E[\val]{\lft{{\sem{\sexpr}}}}{(\eqmem{\VA} \gd \exf)[\lft\vx \mapped \val,\rght\vx \mapped \val]}) \\
      & = (\lnot \BAD \land \eqmem{\VA} \gd \E[\val]{\lft{{\sem{\sexpr}}}}{(\IF \BAD \THEN \exftwo \ELSE (\eqmem{\VA} \gd \exf))[\lft\vx \mapped \val,\rght\vx \mapped \val]}) \\
      & = (\lnot \BAD \land \eqmem{\VA} \gd \E[\val_1,\val_2]{\mu}{(\IF \BAD \THEN \exftwo \ELSE (\eqmem{\VA} \gd \exf))[\lft\vx \mapped \val_1,\rght\vx \mapped \val_2]}) \\
      & \geq \E[(\val_1,\val_2)]{\mu}{\exfthree_\LVA[\lft\vx \mapped \val_1,\rght\vx \mapped \val_2]}
    \end{align*}
    where the first two equalities follow from the assumption $\exf,\exftwo,\BAD \perp \writeV{\cmd}$,
    the thirds as $\mu$ is the identity coupling on memories satisfying $\eqmem{\VA}$, and the
    inequality is just a weakening that ignores the pre-condition.
    \proofcase{$\vx <- \CALL{\vorcl}(\expr)$}
    In this case, we first apply Rule~\ref{erhls:Ind} to internalize the assumption~\eqref{l:adv-twoside:1} in the logical context.
    The claim is then an immediate consequence of Rule~\ref{erhls:Call} tacitly employing $\exf,\exftwo \perp \{\lft{\varg},\rght{\varg},\lft{\vres},\rght{\vres}\}$.
    We also use Rule~\ref{erhls:Weaken}
    to establish the classical pre-condition $\lft{\expr} = \rght{\expr}$---stemming from $\eqmem{\varg}$---using $\eqmem{\VA}$.
    \proofcase{$\IF \bexpr \THEN \cmd \ELSE \cmdtwo$} The case follows by one application of Rule~\ref{erhls:If} and then using induction hypothesis.
    In both proofs, an implicit weakening step is used to ignore the additional pre-conditions on the guard, given by Rule~\ref{erhls:If}.
    \proofcase{$\cmd; \cmdtwo$}
    As a consequence of the induction hypothesis, as we have reasoned above in the initial application of Rule~\ref{erhls:Case}, we can prove
    \[
      \ctx \derivable \RHT{Z}{\exfthree_\LVA}{\cmd_{\orcl}}{\cmd_{\orcltwo}}{\exfthree_\LVA}
      \quad\text{ and }\quad
      \ctx \derivable \RHT{Z}{\exfthree_\LVA}{\cmdtwo_{\orcl}}{\cmdtwo_{\orcltwo}}{\exfthree_\LVA}\,,
    \]
    and thus
    \[
      \ctx \derivable \RHT{Z}{\exfthree_\LVA}{\cmd_{\orcl};\cmdtwo_{\orcl}}{\cmdtwo_{\orcltwo};\cmdtwo_{\orcltwo}}{\exfthree_\LVA} \,.
    \]
    Using
    \[
      (\lnot \BAD \land \eqmem{\VA} \gd \exf) \geq \IF \BAD \THEN \exftwo \ELSE (\eqmem{\VA} \gd \exf) = \exfthree_{\LVA}\,,
    \]
    we conclude the case by an application of Rule~\ref{erhls:Weaken}.
    \proofcase{$\WHILE \bexpr \DO \cmd$}
    The induction hypothesis establishes $\ctx \derivable \RHT{Z}{\exfthree_\LVA}{\cmd_{\orcl}}{\cmd_{\orcltwo}}{\exfthree_\LVA}$,
    and thus
    \[
      \ctx \derivable \RHT{Z}{\lft{\bexpr} \land \rght{\bexpr} \gd \exfthree_\LVA}{\cmd_{\orcl}}{\cmd_{\orcltwo}}{\exfthree_\LVA}\,,
    \]
    by one application of Rule~\ref{erhls:Weaken}.
    This establishes
    \[
      \ctx \derivable \RHT{Z}{\exfthree_\LVA}{\WHILE \bexpr \DO \cmd_{\orcl}}{\WHILE \bexpr \DO \cmd_{\orcltwo}}{\lft{\lnot \bexpr} \land \rght{\lnot \bexpr} \gd \exfthree_\LVA}
      \,
    \]
    by Rule~\ref{erhls:While}. From this, the case follows by Rule~\ref{erhls:Weaken}. 
  \end{proofcases}
\end{proof}

\again{t:adversary}
\begin{proof}
  We consider soundness of the two-sided Rule~\ref{erhls:Adv}.
  To this end, consider
  \[
    \PROC{\adv[\orcl]}(\vx) \cmd_{\orcl}; \RET \expr
    \quad\text{and}\quad
    \PROC{\adv[\orcltwo]}(\vx) \cmd_{\orcltwo}; \RET \expr\,.
  \]
  with $\cmd_{\orcl}$ and $\cmd_{\orcltwo}$ differing only in the oracle calls. 
  To prove the rule sound, we show
  \[
    \ctx \derivable
    \RHT{Z}
    {\exfthree_\varg}
    {\adv[\orcl]}{\adv[\orcltwo]}
    {\exfthree_\vres}\,,
  \]
  where $\exfthree_V \defsym \IF \BAD \THEN \exftwo \ELSE (\eqmem{V,\GVar_{\adv}} \gd \exf)$.
  Soundness of Rule~\ref{erhls:Adv} follows by soundness of the core logic \ERHLS (\Cref{t:soundness}).
  By an application of Rule~\ref{erhls:Proc}, this amounts to deriving,
  \[
    \ctx \derivable
    \RHT{Z}
    {\initz \gd \exfthree_\varg[\lft{\varg}\mapped\lft{\vx}, \rght{\varg}\mapped\rght{\vx}]}
    {\cmd_{\orcl}}{\cmd_{\orcltwo}}
    {\exfthree_\vres[\lft{\vres}\mapped\lft{\expr}, \rght{\vres}\mapped\rght{\exprtwo}]}
  \]
  Now observe that
  \begin{align*}
    (\initz \gd  \exfthree_\varg[\lft{\varg}\mapped\lft{\vx}, \rght{\varg}\mapped\rght{\vx}])
    \geq \IF \BAD \THEN \exftwo \ELSE (\eqmem{\LVar} \gd \exf)
    \geq \exfthree_\LVA\,\text{, and}\\[1mm]
    \exfthree_\LVA \geq \IF \BAD \THEN \exftwo \ELSE (\lft{\expr} = \rght{\expr} \land \eqmem{\LVA} \gd \exf) =  \exfthree_\vres[\lft{\vres}\mapped\lft{\expr}, \rght{\vres}\mapped\rght{\expr}]\,.
  \end{align*}
  Here, to resolve the substitutions, we have employed
  the assumption that variables  $\side{i}{\varg}$ and $\side{i}{\vres}$ do not occur in $\exf$ and $\exftwo$, i.e.,
  $\exf,\exftwo \perp \{ \lft{\varg},\rght{\varg},\lft{\vres},\rght{\vres} \}$,
  and that in the base $\lnot \BAD$, $\eqmem{\LVA,\GVA}$ implies $\lft{\expr} = \rght{\expr}$.
  Thereby, the above goal becomes an instance of \Cref{l:adv-twoside} through an application
  of Rule~\ref{erhls:Weaken}. The assumptions of the lemma being are discharged by the assumptions given by
  Rule~\ref{erhls:Adv}.

  Finally, the one-sided rules are proven reasoning similarly, using \Cref{l:adv-oneside}
  rather than \Cref{l:adv-twoside}.
\end{proof}

\section{Proofs of Characterisations}

\again{thm:kmdcomplete}
\begin{proof}
  Notice that due to the boundedness of $\delta$,  every $f$ that is $1$-Lipschitz, $\lft f - \rght f$ is grater than $-h$, so our post-condition is always a positive real number. This ensures that our judgment is well-defined. From soundness and \asterm assumption, we obtain that
  \[
    \forall\mem_1, \mem_2 \in \Mem. \forall f: \Mem \rightarrow \Real, ||f||_L\leq 1. \ex(\mem_1, \mem_2) + h \ge \E{\sem \cmd_{\mem_1}} f -
    \E{\sem \cmdtwo_{\mem_2}} f +h,
  \]
  so, by observing that $h$ is on both sides of the $\ge$ symbol, and by definition of $\sup$, we obtain
  \[
    \forall\mem_1, \mem_2 \in \Mem. \ex(\mem_1, \mem_2)  \ge \sup_{f: \Mem \rightarrow \Real, ||f||_L\leq 1} \E{\sem \cmd_{\mem_1}} f -
    \E{\sem \cmdtwo_{\mem_2}} f = W_1^\delta(\sem \cmd_{\mem_1},\sem \cmdtwo_{\mem_2}).
  \]

  For the opposite direction, we first apply completeness, to deduce:
  \[
    \derivable\RHT{  f: \Mem \rightarrow \Real, ||f||_L\leq 1} {\E{\sem \cmd_{\mem_1}} f -\E{\sem \cmdtwo_{\mem_2}} f +h} \cmd\cmdtwo  {\lft f - \rght f + h}.
  \]
  We conclude the proof by observing that:
  \begin{align*}
    \E{\sem \cmd_{\mem_1}} f -  \E{\sem \cmdtwo_{\mem_2}} f + h &\leq \sup_{f: \Mem \rightarrow \Real, ||f||_L\leq 1} \E{\sem \cmd_{\mem_1}} f -
                                                                  \E{\sem \cmdtwo_{\mem_2}} f +h\\
                                                                &= W_1^\delta(\sem \cmd_{\mem_1},\sem \cmdtwo_{\mem_2})+h \leq \ex(\mem_1, \mem_2) +h.
  \end{align*}
  Therefore, we conclude with an application of the \ref{erhls:Weaken} rule.
\end{proof}

\section{Encoding of cryptographic proof steps}

The embedding used in~\Cref{t:prhl-complete} suffices to show that
\ERHL proof system overcomes the limitations of \PRHL. For practical
purposes, it is however possible to use more direct embeddings. We
provide such embeddings for \emph{bridging steps}, \emph{lossy steps} and
\emph{comparison steps}, which are some of the main applications of \PRHL in
cryptographic proofs.

\subsection{Bridging steps}
Bridging steps are proofs of program equivalence. The characterization
of program equivalence in \ERHL\ is already captured by \Cref{c:equiv}.

\subsection{Lossy steps}\label{s:complete:ls}
Lossy steps are proofs of equivalence, conditioned on an event $\event\subseteq \Mem$. Such
steps are captured by judgments of the form:
\[
  \validP \RHT{}{\cond}{\cmd}{\cmdtwo}{\lft{\event} \to (=)} \,.
\]
Their validity asserts that for $\cond$ related inputs, $\cmd$ and
$\cmdtwo$ yield identical outputs, up to so some event $\event$ raised
by $\cmd$.  It is a sufficient, although not necessary, condition to
provide an upper bound on the \emph{total variation distance} on
output distribution---the maximal difference in probabilities of two
arbitrary events---in terms of the probability of the event $\event$.
\begin{corollary*}[Characterisation of PRHL's Lossy Steps]{c:condeq}
  Let $\cmd$ and $\cmdtwo$ be two \asterm programs, and let $\event \subseteq \Mem$ be an event.
  The following statements are equivalent:
  \begin{enumerate}
    \item \label{i:condeq1} \(
      \derivable \RHT{\eventm \subseteq \Mem}{\cond \gd 2} {\cmd}{\cmdtwo} {\lft{\cf{\eventm}} + \rght {\cf{\setc{\eventm}}} + \lft {\cf \event}}
      \); and
    \item \label{i:condeq2}
      \(
      \TV(\sem{\cmd}_{\mem_1}, \sem{\cmdtwo}_{\mem_2}) \leq \Prob{\sem{\cmd}_{\mem_1}}{\setc{\event}}
      \)
      for all $\mem_1\mathrel{\cond} \mem_2$.
  \end{enumerate}
\end{corollary*}
\begin{proof}
  Tacitly employing soundness (\Cref{t:soundness}) and our completeness theorem (\Cref{t:completeness}),
  \eqref{i:condeq1} is equivalent to
  \[
    \Prob{\sem{\cmd}_{\mem_1}}{\eventm} + \Prob{\sem{\cmdtwo}_{\mem_2}}{\setc{\eventm}}
    \leq 1 + \Prob{\sem{\cmd}_{\mem_1}}{\setc{\event}}
    \quad \text{for all $\eventm \subseteq \Mem$ and $\mem_1\mathrel{\cond} \mem_2$}\,,
  \]
  for \asterm program statements $\cmd$ and $\cmdtwo$.
  From here the equivalence can now be shown with the identity
  \[
    1+\TV(\sem{\cmd}_{\mem_1}, \sem{\cmdtwo}_{\mem_2})
    = \sup_{\eventm \subseteq \Mem} (\Prob{\sem{\cmd}_{\mem_1}}{\eventm}
    + \Prob{\sem{\cmdtwo}_{\mem_2}}{\setc \eventm})\,. \qedhere
  \]
\end{proof}

\subsection{Comparison steps}
Comparison steps establish the probability of an event $\event\subseteq \Mem$ on $\cmd$ is bounded by that of an event $\eventtwo \subseteq \Mem$ on $\cmdtwo$.
This
property is equivalent to validity of the following \PRHL\ judgment:
\[
  \validP \RHT {} {\cond} \cmd \cmdtwo {\lft \event \to \rght \eventtwo}
\]
Comparison steps are typically used for cryptographic reductions. For such
properties, one can give a direct encoding.
\begin{corollary*}[Characterisation of Comparison Steps]{c:lps}
  Let $\cmd$ and $\cmdtwo$ be two \asterm programs.
  Then, the following are equivalent:
  \begin{enumerate}
  \item \(
      \derivable \RHT{}{\cond \gd 1 + \ex}{\cmd}{\cmdtwo}{\lft{\cf{\event}} + \rght{\cf{\setc \eventtwo}}}
      \); and
    \item
    \(
      \Prob{\sem{\cmd}_{\mem_1}}{\event} \leq \Prob{\sem{\cmdtwo}_{\mem_2}}{\eventtwo} + \ex(\mem_1, \mem_2)
    \) for all $\mem_1 \mathrel{\cond} \mem_2$.
  \end{enumerate}
\end{corollary*}
\begin{proof}
  The two implications follow directly from soundness and our completeness theorem, respectively.
\end{proof}

\section{Missing Proof Details on Case Studies}
\label{sec:appcases}

\subsection{Total Variation}
\label{sec:prpprfapp}

Here we give some more details on the
analysis of the two inner branches of the
functions $\prp$ and $\prf$. The proof goes by
weakest precondition reasoning, and to compare the two
sampling instructions we employ the coupling
\begin{align*}
  \mu \app (\val_1,\val_2) \defsym
  \begin{cases}
    \frac{1}{|A|}
    & \text{if $\val_1 = \val_2 \text{ and } \val_1 \not \in \CALL{codom}({\vl})$,}\\
    \frac{1}{|A| \cdot |A \setminus \CALL{codom}(\vl)|}
    & \text{if $\val_2 \not \in \CALL{codom}(\vl) \text{ and } \val_1 \in \CALL{codom}(\vl)$,}\\
    0             & \text{otherwise.}
  \end{cases}
\end{align*}
Observe that, for a fixed $l$, the distribution $\mu(l)$ is a
coupling between $\ud A$ and $\ud{A\setminus \codom(l)}$.
After the application of the \ref{erhls:Sample} rule, we apply
the \ref{erhls:Weaken} rule. There, we can assume
$\lft l =\rght l, \lft x = \rght x $, $\lft \vx \notin
\CALL{codom}(\lft \vl)$, $\sz {\lft l}< Q$,  and that $\lft{l}$ is collision-free. Therefore,
in the pre-  and post-condition we can unify $\lft \vl$
and $\rght l$ with $\vl$, and
similarly for $\lft \vx$ and $\rght \vx$. The side-condition of the
rule is discharged by observing:
\begin{align*}
  \textstyle \sum_{i = 0}^{\QQ-\sz {\vl}-1} \frac {i+\sz {\vl}} {|A|}    & = \textstyle \frac{\sz{\vl}}{|A|} + \sum_{i = 1}^{\QQ-\sz {\vl}-1} \frac {i+\sz {\vl}} {|A|} \\
                                                                      & \geq \textstyle \sum_{\substack{v_1 \in A\cap \CALL{codom} (l)\\ v_2 \in A\setminus \CALL{codom} (l)}}\frac 1 {|A| \times |A\setminus \CALL{codom} (l)| } +
  \sum_{\substack{v_1 \in A\setminus \CALL{codom} (l)\\ v_2 \in A\setminus \CALL{codom} (l)\\ v_1 =v_2 }}\frac 1 {|A|}\cdot \left(\sum_{i = 0}^{\QQ-\sz {\vl}-1} \frac {i+\sz l+1} {|A|}\right)\\
                                                           & = \textstyle \E[(v_1, v_2)]{\mu(l)}{\IF v_1 \in \CALL{codom}(l) \THEN 1 \ELSE v_1 = v_2 \gd \sum_{i = 0}^{\QQ-\sz {\vl}-1} \frac {i+\sz l+1} {|A|}},
\end{align*}
where, the first equality is sound for the assumption $\sz {\lft l}< Q$ and
the last one because of the definition of $\mu$.

\subsection{Stochastic Gradient Descent}
\label{sec:sgdapp}

Our goal is to show that the Stochastic Gradient descent algorithm shown in \Cref{sec:sgd}
is $\left(\frac{2L^2}{n}\sum_{t=0}^{T-1}\alpha_t\right)$\emph{-uniformly
  stable}. We do so by establishing the following judgment:
\[
  \derivable \RHT{a \in A}{\diffone \gd \frac{2L^2}{n}\sum_{t=0}^{N-1}\alpha_t}{\sgd}{\sgd}{\lambda (\mem_1, \mem_2).\ |{\loss(a, \lft w)} - {\loss(a, \rght w)}|},
\]
where $\mem_1 \diffone \mem_2$ states that the two sets of example $\mem_1(S)$ and $\mem_2(S)$ differ only in one element. Following \cite[Theorem 3.7]{Hardt}, we make the following assumptions:

\begin{enumerate}
\item \label{sgd:h1} the function $\loss$ is $L$-Lipschitz in its
  second argument, i.e. for every $w, w'\in \Real^d$, and $a \in A$ we
  have $|\loss(a, w)-\loss(a, w')|\leq L \cdot \|w-w'\|$;
\item \label{sgd:h2} the function $\loss$ is convex, i.e.
  $\loss(a, w) \geq \loss(a, w') + \langle \grad\loss(a)(w'),
  w'-w\rangle$.
\item \label{sgd:h3} the gradient $\grad(\loss(a))$ is $\beta$-Lipschitz for every
  $a \in A$;
\item \label{sgd:h4} the step size $\alpha_t$ is such that
  $\forall t. 0\leq \alpha_t\leq \frac2\beta$.
\end{enumerate}

We start by observing that, leveraging \eqref{sgd:h1}, we can apply the
\ref{erhls:Conseq} rule to reduce the claim to:
\[
  \derivable \RHT{}{ \diffone \gd \frac{2L}{n}\sum_{t=0}^{N-1}\alpha_t}{\sgd}{\sgd}{\lambda (\mem_1, \mem_2).\ \|\lft w - \rght w\|}.
  \tag{$*$}
\]

For simplicity's sake, we call the while loop in $\sgd$  $W$, and  its body $B$. The crucial part of the proof is the application of the \ref{erhls:While} rule, where we employ the following invariant:
\[
  I \defsym \lft t =\rght t \gd \|\lft w - \rght w\| + \frac{2L} n \sum_{j = \lft t}^{T-1} \alpha_j.
\]
This rule concludes the triple:
\[
  \RHT {}{\lft t \le T \leftrightarrow \rght t \le T \gd I} {W} {W} {\lft t \ge T \land \rght t \ge T \gd I}
\]
under the premise
\[
  \RHT {}{\lft t \le T \land \rght t \le T \gd I} {B} {B} {\lft t \le T \leftrightarrow \rght t \le T \gd I}.
  \tag{\dag}
\]
Notice how, once (\dag) is established, from the conclusion of the \ref{erhls:While} rule, it is easy to conclude ($*$) by weakest precondition reasoning and by applying \ref{erhls:Weaken} to the final triple.

We show (\dag) by repeatedly applying the \ref{erhls:Asgn} rule on the instructions in $B$, and the \ref{erhls:Sample} rule to the sampling instruction. For this last rule, we employ the coupling $\mu\in \Distr (S \times S')$ that is almost everywhere the identity coupling except for the pair of distinct elements $s \in S$, $s' \in S'$ that are coupled together. This proves the following triple:
\[
  \derivable \RHT {}{\E[(v, v')]\mu{\lft t + 1 \le T \leftrightarrow \rght t + 1 \le T \gd I'}} {B} {B} {\lft t \le T \leftrightarrow \rght t \le T \gd I}
  \tag{\ddag}
\]
with
\[
  I' \defsym \lft t +1 =\rght t +1 \gd \|(\lft w -\alpha_{\lft t} \grad(\loss(v))(\lft w))- (\rght w -\alpha_{\rght t} \grad(\loss(v'))(\rght w))\| + \frac{2L} n \sum_{j =\lft t +1}^{T-1} \alpha_j.
\]
Finally, we apply the \ref{erhls:Weaken} rule to show that the precondition of ($\ddag$) is smaller than the one of ($\dag$). Concretely, this boils down to establishing:
\[
  \E[(v, v')]\mu {\|(\lft w -\alpha_{\lft t} \grad(\loss(v))(\lft w))- (\rght w -\alpha_{\lft t} \grad(\loss(v'))(\rght w))\|}
  \leq
  \|\lft w - \rght w\| + \frac{2L} n \alpha_{\lft t}.
\]
Using \eqref{sgd:h1} and the triangular inequality, we can bound the argument of the expectation for the pair $(s, s')$ with $\|\lft w-\rght w\| + 2L \alpha_{\lft t}$. For the remaining pairs, the term within the expectation is bounded by $\| \lft w - \rght w \|$, and this observation follows from assumptions \eqref{sgd:h2}, \eqref{sgd:h3} and \eqref{sgd:h4}, see~\cite[Lemma 3.6]{Hardt}. This gives precisely
\begin{multline*}
  \E[(v, v')]\mu {\|(\lft w -\alpha_{\lft t} \grad(\loss(v))(\lft w))- (\rght w -\alpha_{\lft t} \grad(\loss(v'))(\rght w))\|}\leq \\
  \frac {1} n \|\lft w_1 - \rght w\| +\frac {2L} n \alpha_{\lft t } + \frac {n-1} n \|\lft w_1 - \rght w\| = 
  \|\lft w - \rght w\| + \frac{2L} n \alpha_{\lft t}.
\end{multline*}


\end{document}